\documentclass[usenatbib]{mn2e}
\pdfoutput=1
\usepackage{graphicx}
\usepackage{mathptmx}
\usepackage{amssymb}
\usepackage[pdfpagelabels]{hyperref}
\hypersetup{pdfauthor={Michael J. Williams, M. Bureau and Michele
Cappellari},pdftitle={Kinematic constraints on the stellar and dark
matter content of spiral and S0 galaxies},breaklinks=true}

\usepackage[total={17.8cm,24.0cm},centering]{geometry}

\newcommand\aj{AJ}%
\newcommand\apj{ApJ}%
\newcommand\apjl{ApJ}%
\newcommand\apjs{ApJS}%
\newcommand\aap{A\&A}%
\newcommand\mnras{MNRAS}%
\newcommand\nat{Nature}%
\newcommand\aaps{A\&AS}%

\newcommand{\sixfigures}[6]{%
\begin{figure*}
\includegraphics[width=8.6cm]{fig/#1-bigbest.pdf}
\includegraphics[width=8.6cm]{fig/#2-bigbest.pdf} \\
\includegraphics[width=8.6cm]{fig/#3-bigbest.pdf}
\includegraphics[width=8.6cm]{fig/#4-bigbest.pdf} \\
\includegraphics[width=8.6cm]{fig/#5-bigbest.pdf}
\includegraphics[width=8.6cm]{fig/#6-bigbest.pdf} \\
\addtocounter{figure}{-1}
\caption{ --- continued}
\end{figure*}
}
\newcommand{\sixfigurescontrol}[6]{%
\begin{figure*}
\includegraphics[width=8.6cm]{fig/#1-bigbest.pdf}
\includegraphics[width=8.6cm]{fig/#2-bigbest.pdf} \\
\includegraphics[width=8.6cm]{fig/#3-bigbest.pdf}
\includegraphics[width=8.6cm]{fig/#4-bigbest.pdf} \\
\includegraphics[width=8.6cm]{fig/#5-bigbest.pdf}
\includegraphics[width=8.6cm]{fig/#6-bigbest.pdf} \\
\caption{As \reffig{fig:results} but for the control sample.}
\label{fig:control}
\end{figure*}
}
\newcommand{\refsec}[1]{Section~\ref{#1}}
\newcommand{\reffig}[1]{Fig.~\ref{#1}}
\newcommand{\refeq}[1]{equation~(\ref{#1})}
\newcommand{\ud}{\mathrm{d}}

\newcommand{\Reff}{\hbox{\ensuremath{R_\mathrm{e}}}}

\title[Constraints on stellar and dark matter in disk
galaxies]{Kinematic constraints on the stellar and dark matter content
of spiral and S0 galaxies}

\author[M. J. Williams et al.]{Michael J. Williams\thanks{E-mail:
williams@astro.ox.ac.uk}, M. Bureau and Michele Cappellari
\\Sub-Department of Astrophysics, University of
Oxford, Denys Wilkinson Building, Keble Road, Oxford OX1 3RH }

\begin{document}

\date{Accepted 2009 August 18}

\pagerange{\pageref{firstpage}--\pageref{lastpage}} \pubyear{2009}

\maketitle
\label{firstpage}

\begin{abstract}
We present mass models of a sample of 14 spiral and 14 S0 galaxies that
constrain their stellar and dark matter content. For each galaxy we
derive the stellar mass distribution from near-infrared photometry under
the assumptions of axisymmetry and a constant $K_S$-band stellar
mass-to-light ratio, $(M/L)_{K_S}$. To this we add a dark halo assumed
to follow a spherically symmetric Navarro, Frenk \& White (NFW) profile
and a correlation between concentration and dark mass within the virial
radius, $M_{\rm DM}$. We solve the Jeans equations for the corresponding
potential under the assumption of constant anisotropy in the meridional
plane, $\beta_z$. By comparing the predicted second velocity moment to
observed long-slit stellar kinematics, we determine the three
best-fitting parameters of the model: $(M/L)_{K_S}$, $M_{\rm DM}$ and
$\beta_z$. These simple axisymmetric Jeans models are able to accurately
reproduce the wide range of observed stellar kinematics, which typically
extend to $\approx$ 2--3\,\Reff{} or, equivalently, $\approx$
0.5--1\,$R_{25}$. We find a median stellar mass-to-light
ratio at $K_S$-band of $1.09\,(M/L)_{K_S,\odot}$ with an rms scatter of
0.31. We present preliminary comparisons between this large sample of
dynamically determined stellar mass-to-light ratios and the predictions
of stellar population models. The stellar population models predict
slightly lower mass-to-light ratios than we measure. The mass models
contain a median of 15 per cent dark matter by mass within an effective
radius \Reff{} (defined here as the semi-major axis of the ellipse
containing half the $K_S$-band light), and 49 per cent within the
optical radius $R_{25}$. Dark and stellar matter contribute equally to
the mass within a sphere of radius $4.1\,\Reff$ or $1.0\,R_{25}$. There
is no evidence of any significant difference in the dark matter content
of the spirals and S0s in our sample. Although our sample contains
barred galaxies, we argue a posteriori that the assumption of
axisymmetry does not significantly affect our results. Models without
dark matter are also able to satisfactorily reproduce the observed
kinematics in most cases. The improvement when a halo is added is
statistically significant, however, and the stellar mass-to-light ratios
of mass models with dark haloes match the independent expectations of
stellar population models better.
\end{abstract}


\section{Introduction}

In the dominant paradigm describing structure formation, initial
fluctuations in density are enhanced by gravity until galaxies form in
potential wells \citep{White:1978,Blumenthal:1984}. Numerical
simulations and semi-analytic models of galaxy formation are able to
reproduce many of the statistical characteristics of galaxies and some
of their detailed features
\citep[e.g.][]{Kauffmann:1993,Cole:1994,Benson:2003,Baugh:2006,Somerville:2008}.
A detailed and consistent theory remains elusive, however, and there are
apparent contradictions between the predictions of models and the
observations \citep[see, e.g.][and references
therein]{Baugh:2006,Mayer:2008}.

The main stumbling blocks are the significant uncertainties and
computational difficulties involved in capturing the baryonic physics
that is crucial on sub-Mpc scales. Progress can be made in two ways.
Firstly, using observational constraints which are thought to be
insensitive to baryonic physics, one can circumvent these uncertainties
and directly test the predictions of large-scle structure formation,
which is in itself crucially important. Alternatively, one can test
formation models with galaxy-scale observations. The outcomes of these
tests can be used to refine and improve the models. Perhaps one of the
most useful observational constraints for either approach is an
understanding of the relative distribution of dark and luminous matter.
In this work we aim to measure the radial dark matter distribution in a
sample of 28 edge-on disc galaxies. We make cosmologically-motivated
assumptions, which allow us to lift certain modelling degeneracies. 

It is well-established that, within the optical disc, the kinematics of
high surface brightness spiral galaxies can be reproduced by maximal
disc models, in which luminous material contributes the maximum amount
consistent with the observed rotation curve
\citep[e.g.][]{van-Albada:1986,Persic:1996,Palunas:2000}. The stellar
components of maximal disc models typically contribute 75--95 per cent
of the rotational velocity at $2.2\,R_{\rm disc}$, where $R_{\rm disc}$
is the scale length of the exponential disc and $2.2\,R_{\rm disc}$ is
the radius where the rotational velocity of the exponential disc peaks
\citep{Sackett:1997}. This permits dark haloes that comprise 10--45 per
cent of the total mass within 2.2 $R_{\rm disc}$. In a Freeman disc
\citep{Freeman:1970}, $R_{25}$, the radius of the 25\,mag\,arcsec$^{-2}$
isophote occurs at $\approx 3\,R_{\rm disc}$.

Side-stepping the maximal disc assumption, \cite{Ratnam:2000}
demonstrate that rotation curves can be adequately fitted without dark
matter. \cite{Kassin:2006} avoided the maximal disc assumption entirely
by using independent estimates of the stellar mass-to-light ratio
inferred from a relationship with colour \citep{Bell:2001,Bell:2003}.
Most of their models were consistent with a maximal disc. Further
evidence is provided by model-independent analysis of rotation curves at
intermediate radii \citep{McGaugh:2007}.

In the case of elliptical and S0 galaxies, evidence from dynamical
modelling and gravitational lensing studies suggests that dark matter
makes only a small contribution to the total mass within \Reff{}, the
radius of the elliptical isophote containing half the light. For
example, from the dynamical studies, \cite{Gerhard:2001} find that
10--40 per cent of the mass within \Reff{} is dark in their sample of 21
ellipticals, \cite{Borriello:2003} find 30 per cent dark matter within
\Reff{} by constraining a sample of 221 ellipticals to lie on a
fundamental plane, \cite{Cappellari:2006} find a median of 30 per cent
dark matter within \Reff{} in their sample of 25 ellipticals and S0s,
\cite{Thomas:2007} find 10--50 per cent dark matter within \Reff{} for a
sample of 17 ellipticals and S0s, and \cite{Weijmans:2008} find 55 per
cent dark matter within 5\,\Reff{} in NGC~2974. Lensing studies find
similar results. For example, \cite{Rusin:2003} find 22 per cent dark
matter within 2 \Reff{} of 22 elliptical lenses, \cite{Koopmans:2006}
find 25 per cent dark matter within the Einstein radius of 15 elliptical
lenses (the Einstein radius is approximately equal to the effective
radius for galaxy scale lenses), and \cite{Bolton:2008} find 38 per cent
dark matter inside \Reff{} for 53 elliptical lenses.

However, the total absence of dark matter in ellipticals is sometimes
excluded with only low significance due to intrinsic degeneracies in the
dynamical models (see, e.g. \citealt{Romanowsky:2003} and the response
by \citealt{Dekel:2005}) and the lack of kinematic tracers at large
radii. Moreover, in all cases $\Reff \ll R_{25}$. The
constraints on both the mass of the halo and its extent are therefore
less strong for ellipticals than for spirals.  

Perhaps the most compelling evidence for the dominance of stellar matter
in either ellipticals or high surface brightness disc galaxies comes
from the analysis of the bars and spiral arms often present in disc
galaxies. These non-axisymmetric features lift the degeneracy between
the contributions from luminous and dark matter. This is done by
assuming that dark matter is axisymmetric, so all non-circular motions
can be attributed to the non-axisymmetric luminous component. This
approach has been applied to both bars
\citep[e.g.][]{Englmaier:1999,Weiner:2001} and spiral arms
\citep[e.g.][]{Kranz:2003} and provides results consistent with maximal
disc studies. Further constraints come from $N$-body simulations of
bars, which imply that a significant central dark component would slow
or even destroy bars \citep[e.g.][]{Debattista:2000}, while observations
systematically indicate that bars are fast
\citep[e.g.][]{Aguerri:2003,Gerssen:2003}. 

To summarize, observational evidence indicates that the kinematics of
both spiral and elliptical galaxies can often be reproduced by mass
models that include a sub-dominant contribution from dark matter in the
optical or central regions. The specific amount required and
correlations between halo and luminous galaxy properties are, however,
unclear. Note that we have not mentioned low surface brightness and
dwarf galaxies, which are dark matter-dominated at all radii
\citep[e.g.][]{Persic:1996,de-Blok:1997,Verheijen:1997,Swaters:1999} and
may provide the most stringent constraints on halo shapes, once
observational uncertainties are resolved \citep[e.g.][and references
therein]{de-Blok:2001,Swaters:2003}. Such galaxies are, however,
unrepresentative and extrapolating their results to systems with greater
stellar masses may introduce biases. It is therefore crucial to place
accurate constraints on the dark matter content of giant, high surface
brightness spiral, S0 and elliptical galaxies.

This work uses a modelling technique which is different but
qualitatively similar to traditional mass decomposition and rotation
curve analyses of spiral galaxies. The key differences are (i) the
stellar components of our mass models are based on deep near-infrared
photometry, which accurately traces the smooth stellar potential of the
galaxy (see \refsec{sec:mge}), (ii) we lift a degeneracy by making
assumptions about the shape of the dark halo that are motivated by the
results of cosmological simulations and observational constraints (see
\refsec{sec:halo}) (iii) we lift a further degeneracy (and account for
pressure support) by comparing the predicted second velocity moment
rather than the rotational velocity to the observed kinematics (see
\refsec{sec:jeans}). 

We apply this technique to a sample of 28 edge-on spiral and S0
galaxies. We use edge-on galaxies because, under the assumption of
axisymmetry, the deprojection is unique. Moreover, when a galaxy is
close to edge-on ($i = 90^\circ$), an inclination error does not
propagate on to significant uncertainties in the kinematics in the plane
of the galaxy (which is proportional to $\sin i$), and thus on to
significant mass uncertainties. The mass models consist of a dark halo
component and an unusually detailed parametrization of the projected
light. Under justifiable assumptions, the mass model has three free
parameters, the stellar mass-to-light ratio, the mass of the dark halo
and the velocity anisotropy. We place constraints on these parameters by
adjusting them so that the mass model predicts stellar kinematics that
closely match those observed. We hope that these simple but powerful
quantitative statements will be of use in constraining models of galaxy
formation and evolution.

In addition to the halo masses inferred, the constraints we place on the
near-infrared stellar mass-to-light ratios are themselves of great
interest. There are significant difficulties in modelling the spectral
energy distribution of stellar populations in the near-infrared, due to
the importance of the complex thermally pulsing asymptotic giant branch
at these wavelengths \citep{Maraston:2005}. Moreover, the normalization
of the models is uncertain due to a lack of knowledge about the initial
mass function of stars. Dynamical measures of the near-infrared stellar
mass-to-light ratio like ours, that do not depend on population models
and seek to correctly account for dark matter, therefore provide
important independent tests of these models.

A final goal of the study is to provide detailed constraints on dark
matter in S0 galaxies. The dark matter content of S0s is an important
quantity to constrain because the dominant model of S0 formation as
faded spirals predicts a
simple and verifiable relation between the Tully-Fisher relations of
spirals and S0s \citep[e.g.][and references therein]{Bedregal:2006}.
Galaxies that lie on a single Tully-Fisher relation linking their
luminosities and rotational velocities are believed to have equal
dynamical mass-to-light ratios. This will be investigated in more detail
in a future paper (Williams, Bureau \& Cappellari, in preperation).

This paper is organised as follows: in \refsec{sec:methods} we describe
the multi-Gaussian expansion used to
model the luminous and dark mass distribution of the galaxies and the
dark halo model adopted. We then
give an overview of the Jeans modelling technique used to model the
stellar kinematics. In \refsec{sec:sample} we present the sample of
edge-on galaxies under consideration and describe the photometric and
kinematic data. In \refsec{sec:results} we present the results of our
modelling of both the mass distribution and stellar kinematics of the
sample, which we discuss in depth in \refsec{sec:discussion}. Finally,
we summarise our conclusions and discuss possibilities for future work
in \refsec{sec:conclusion}.

\section{Methods}
\label{sec:methods}

\subsection{Modelling the luminous mass distribution}
\label{sec:mge}

We use multi-Gaussian expansions (MGEs) to create mass models that have
a simple analytic form whose kinematics can easily be solved using the
Jeans equations. MGE is a method of parametrizing an image of a galaxy
as the sum of a finite number of two-dimensional Gaussian functions. The
original application of two-dimensional Gaussians to galaxy images
\citep{Bendinelli:1991} was extended to the non-circular case and to
more general point spread functions (PSFs) by \cite{Monnet:1992} and
\cite{Emsellem:1994}. Here we briefly summarize the method using a
formalism due to \cite{Cappellari:2002}, which we simplify for the
special case of axisymmetric edge-on disc galaxies with a luminous
component with a constant mass-to-light ratio.

In a coordinate system ($x$, $y$, $z$), where $x$ and $z$ are centred
on the galaxy nucleus and point along the major and minor axes in the
plane of the sky while the $y$-axis points away from the observer,
$\tilde{\Sigma}_X(x, z)$ the apparent surface brightness of a galaxy at
an arbitrary waveband $X$ can be written as a sum of $N$ two-dimensional
Gaussians of apparent width $\tilde{\sigma}_i$ in the $x$-direction
and $\tilde{\sigma}_i \tilde{q}_i$ in the $z$-direction:
\begin{equation}
\label{eqn:obsmge}
\tilde{\Sigma}_X(x, z) = \sum_{i=1}^N 
\frac{L_i}{2\pi\tilde{\sigma}^2_i \tilde{q}_i}
\exp \left[ - \frac{1}{2 \tilde{\sigma}^2_i} 
\left(x^2 + \frac{z^2}{\tilde{q}^2_i} \right) \right],
\end{equation}
where $L_i$ is the total luminosity of the $i$th Gaussian component.
We further model the PSF as a circular Gaussian of width
$\sigma_{\mathrm{PSF}}$ such that the intrinsic projected light
distribution, deconvolved from seeing effects, is
\begin{equation} 
\label{eqn:sb}
\Sigma_X(x, z) = \sum_{i=1}^N \frac{L_i}{2\pi\sigma^2_i q_i} 
\exp \left[ 
- \frac{1}{2\sigma^2_i} 
\left( x^2 + \frac{z^2}{q^2_i} \right) 
\right],
\end{equation}
where $\sigma_i$, the intrinsic width of the $i$th Gaussian in the
$x$-direction, and $\sigma_i q_i$, the intrinsic width in the
$z$-direction, are given by
\begin{equation}
\sigma^2_i = \tilde{\sigma}^2_{i} - \sigma^2_\mathrm{PSF},
\end{equation}
\begin{equation}
\sigma^2_i q^2_i = \tilde{\sigma}^2_{i} \tilde{q}^2_{i} 
- \sigma^2_\mathrm{PSF}.
\end{equation}

Once the projected light distribution is expressed in this simple
analytic form, it can be deprojected straightforwardly to give a full
three-dimensional model of the light. In this work we deproject by
assuming that the galaxy is axisymmetric, but other assumptions about
the geometry are possible. Assuming axisymmetry and an edge-on view
allows us to trivially transform from the ($x$, $y$, $z$) Cartesian
coordinates of the Gaussians on the sky to the ($R$, $\phi$, $z$)
cylindrical system of the galaxy, where $R$ is the galactocentric
radius, $\phi$ the azimuthal angle and $z$ the axis of symmetry of the
galaxy. We also assume a constant stellar mass-to-light ratio 
$(M/L)_X$ to transform the stellar light at waveband $X$ to a stellar
mass distribution. 

Together these assumptions of axisymmetry, an edge-on view and a
constant mass-to-light ratio imply that the intrinsic mass distribution
of the luminous component of the galaxy can be written as
\begin{equation}
\label{eqn:density}
\rho(R, z) = (M/L)_X \sum^N_{i=1}a_i
\exp \left[ - \frac{1}{2\sigma_i^2} \left(
R^2 + \frac{z^2}{q^2_i}
\right) \right],
\end{equation}
where $a_i = L_i/(\sqrt{2\pi}\sigma_i)^3 q_i$.

In this work we determined an optimal MGE parametrization of the images
described in Section~\ref{sec:sample} using the public fitting
routine of
\cite{Cappellari:2002}\footnote{\url{http://www-astro.physics.ox.ac.uk/~mxc/idl/}},
which minimizes the quantity
\begin{equation}
\label{eqn:chi2mge}
\chi^2_\mathrm{MGE} \equiv \sum_{j=1}^M \left( \frac{C_j(x, z) -
\tilde{\Sigma}(x,z)}{C_j(x,z)}
\right)^2
\end{equation}
for a given set of $\sigma_i$, $q_i$ and $L_i$, where $M$ is the number
of photometric data points $C_j$ and $\tilde{\Sigma}$ is the apparent
surface brightness of the model. 

A MGE is simply a sum of Gaussians reproducing the observed surface
brightness. As such, the amplitude of each individual Gaussian, $L_i$,
is not constrained to be positive as long as the total luminosity and
density are positive. Our sample consists of edge-on disc galaxies,
which are particularly difficult to fit with solely positive Gaussian
terms. As was demonstrated by \cite{Bureau:2006}, the major axis surface
brightness profiles exhibit significant plateaus and secondary maxima.
These cannot be fitted by a sum of concentric Gaussians with positive
amplitudes, which is necessarily monotonically decreasing with radius.
The complex rectilinear or concave two-dimensional structures visible in
the isophotes of boxy and peanut-shaped (B/PS) bulges are similarly
challenging.

\begin{figure}
\includegraphics[width=8.4cm]{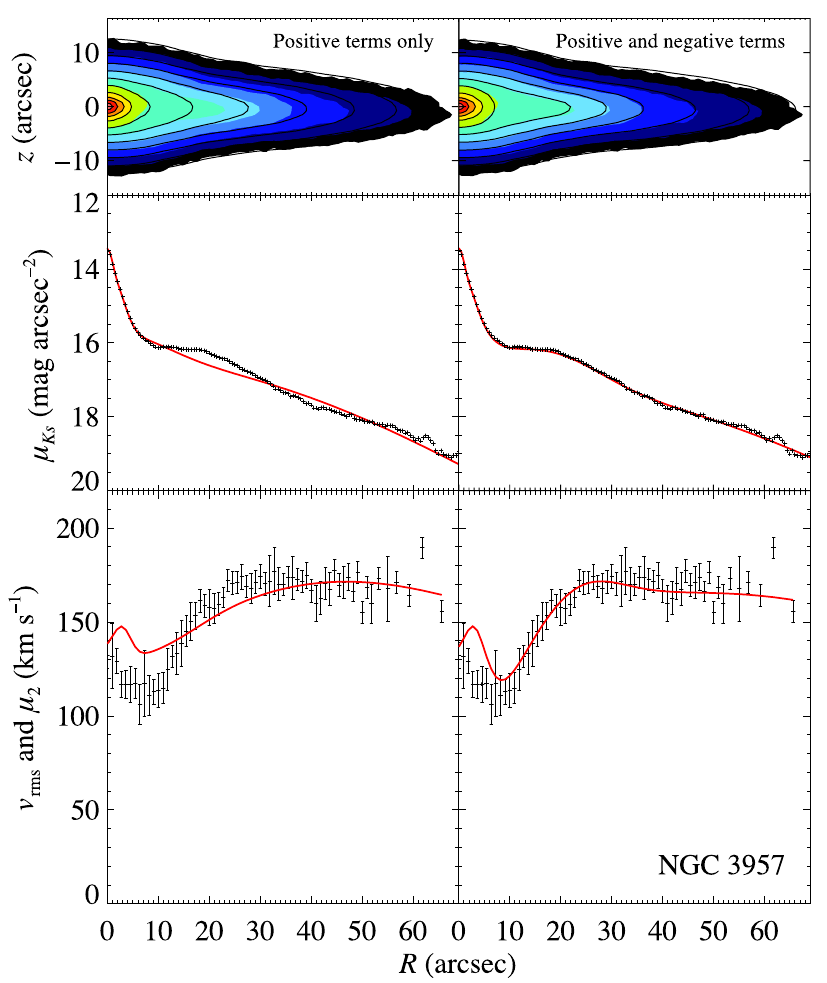}
\caption{A comparison of the best MGE model with only positive amplitude
Gaussians (left-hand panels) and both positive and negative amplitude
Gaussians (right-hand panels) for the same object, NGC 3957. The top
plots show the $K_S$-band surface brightness of the image (filled contours) and
model isophotes (solid lines). Contours are separated by 0.5 mag
arcsec$^{-2}$. The middle plots show the major axis surface brightness
profiles of the image (points) and models (solid lines). The bottom
panels show $v_\mathrm{rms}$ (points), the observed root mean square
velocity, and $\mu_2$, the best-fitting second moment found by solving
the Jeans equations for the MGE model (solid lines). Note that allowing
terms in the Gaussian expansion to have a negative amplitude improves
both the fit to the photometry and the accuracy with which the
kinematics can be reproduced. In order to make the effect clear, the mass
model shown here does not include a dark halo.}
\label{fig:negfit}
\end{figure}

We therefore lifted the positivity constraint on $L_i$ when modelling
the luminous mass. In doing so, we encountered numerical issues with
both the accuracy and stability of the fitting algorithm when using
large numbers of Gaussians ($\approx 30$). We solved these by reducing
the maximum number of Gaussians in the sum to a relatively small number
($\approx 10$), enabling double precision arithmetic to avoid
cancellation errors, and finally, where necessary, tweaking the minimum
surface brightness level down to which the fit was constrained by the
photometry. We show a typical example of the improvement that is
possible when terms with negative amplitude are allowed in
\reffig{fig:negfit}. These improved mass models resulted in small but
systematic improvements in the accuracy of the modelled kinematics.
That a more accurate description of the light gives a more accurate
model of the kinematics gives us some confidence that our kinematic
modelling methods and assumptions (described in \refsec{sec:jeans}) are not
significantly flawed.

The pixel-by-pixel absolute deviation from the photometry of our MGEs is
typically 2--4 per cent. For a given constant $(M/L)_X$ these small
errors are propogated linearly into the mass model. In most previous
work, the observed surface brightness distribution of disc galaxies is
parametrized using fits to azimuthally averaged radial profiles
\citep[e.g.][]{Ratnam:2000,Kassin:2006} or two-dimensional Sersic and/or
exponential decompositions \citep[e.g.][]{Gentile:2004}. While the terms
in our MGEs lack any direct physical association with intrinsic
components of the galaxies (bulge, disc, etc.), they reproduce the
observed surface photometry more accurately than the simpler
parametrizations, and their form is mathematically convenient for Jeans
modelling (see \refsec{sec:jeans}).

\begin{figure}
\includegraphics[width=8.4cm]{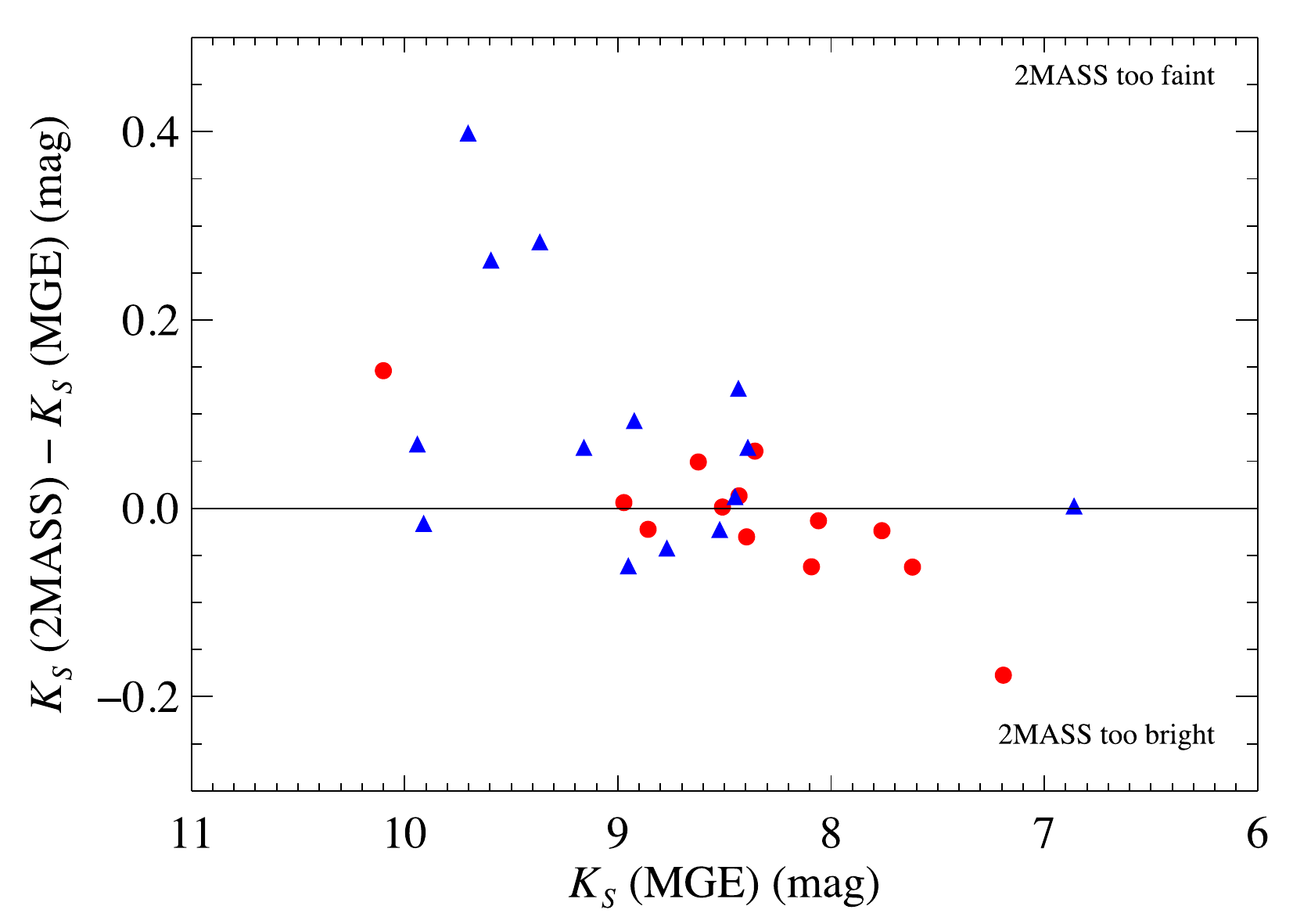}
\caption{Comparison of the apparent magnitudes of the sample galaxies
computed by summing the light of all Gaussians in the best-fitting MGE
models with the apparent magnitudes published in the 2MASS Extended
Source Catalog. The 2MASS values are derived in the usual way using a
growth curve. The methods typically agree to within 0.2\,mag. S0s are
shown as red circles and spirals as blue triangles.}
\label{fig:appmag}
\end{figure}

At large radii, where the model is only weakly constrained by
photometric data, models including negative Gaussians can look a little
unphysical to the eye (see, e.g., NGC~1886 in \reffig{fig:results}).
Because there is so little light at these radii, however, we are
confident that this does not affect the results significantly. We
confirmed this by summing the light of all Gaussians in the MGE models,
which can of course be computed analytically, to derive estimates of the
total apparent magnitudes of the galaxies.
As is shown in \reffig{fig:appmag}, these match the total apparent
magnitudes presented in the 2MASS Extended Source Catalog
\citep{Jarrett:2000} to within $0.2$\,mag. This demonstrates that the
excess light sometimes present in the outer regions of MGEs including
negative terms is not significant, and that the photometric calibration
described in Appendix~\ref{app:2mass} is reliable. \reffig{fig:appmag}
also hints that the application of a growth curve method to the
relatively shallow 2MASS images of faint objects results in total
magnitudes that are systematically too faint \citep[see
also][]{Noordermeer:2007}.

\subsection{Modelling the dark halo mass distribution}
\label{sec:halo}

If mass models are derived under the assumption that mass follows light
then that implicitly assumes that dark matter is not a significant
component by mass within the radius probed by the model. Strictly
speaking, such models also admit the possibility that the spatial
distribution of dark matter closely matches that of the luminous matter,
but we know of no physically motivated model of galaxy formation that
predicts this behaviour. 

We extend our models by explicitly including a dark halo. Although the
long-slit kinematic data for each galaxy typically reach $\approx$
2--3\,\Reff{} or, equivalently, $\approx$ 0.5--1\,$R_{25}$, they are
insufficient to allow a us to constrain the shape of the halo of giant,
high surface brightness galaxies like these \citep[see,
e.g.][]{Banerjee:2008}. We therefore emphasise that we are not able to
constrain the shapes of the dark haloes. Rather we attempt to probe the
more limited question of what fraction by mass of each galaxy is dark
\emph{assuming} the haloes follow the particular one-parameter density
profile described below.

Our halo model is motivated by the results of
$N$-body simulations of cold dark matter (CDM) in an
up-to-date cosmology. In such simulations the dark matter halo density
$\rho_{\rm DM}$ is given by the spherically averaged NFW profile:
\begin{equation}
\rho_{\rm DM}(r) = \frac{\rho_{\rm s}}{(r/r_{\rm s})(1+r/r_{\rm s})^2}
\label{eqn:rhonfw}
\end{equation}
\citep[e.g.][]{Navarro:1997,Bullock:2001,Wechsler:2002}, were $r_{\rm s}$
is a scale radius characterising the location of the `break' in the
profile and $\rho_{\rm s}$ is a corresponding inner density.

Rewriting \refeq{eqn:rhonfw} in terms of $M_{\rm DM}$, the total dark
matter mass enclosed within the virial radius $r_\mathrm{vir}$, and defining a
concentration parameter $c_\mathrm{vir} \equiv r_\mathrm{vir}/r_{\rm s}$
implies
\begin{equation}
\rho_{\rm DM}(r)=\frac{M_{\rm DM}}{4 \pi A(c_{\rm
vir})}\frac{1}{r(r_\mathrm{s}+r)^2} ,
\label{eqn:rhonfw2}
\end{equation}
where $M_{\rm DM} = 4 \pi \rho_\mathrm{s} r_\mathrm{s}^3 A(c_{\rm vir})$ and 
\begin{equation}
A(x) = \ln (1+x) - \frac{x}{1+x}.
\label{eqn:a}
\end{equation}
The virial radius is defined as the radius within which the mean density
is $\Delta \rho_\mathrm{crit}$ and $\rho_\mathrm{crit,0} = 3H_0^2/8\pi
G$ now. Throughout this work we adopt the cosmological parameters found
by the \emph{Wilkinson Microwave Anisotropy Probe} five-year results
(WMAP5), i.e. $\Delta = 95.1$ and the Hubble constant $H_0 =
100\,h$\,km\,s$^{-1}$\,Mpc$^{-1}$ where $h = 0.70$ \citep{Komatsu:2009}.

Following \cite{Napolitano:2005}, we use a key result from simulations
of $\Lambda$CDM to eliminate one of the free parameters from
\refeq{eqn:rhonfw2}. The simulations demonstrate that concentration and
halo mass are correlated: low mass haloes are more concentrated and the
concentration--mass relation is well described by a
single power law with a slope $\approx -0.1$
\citep[e.g.][]{Navarro:1997,Bullock:2001,Eke:2001,Kuhlen:2005,Neto:2007}.
We choose the fitting formula presented by \cite{Maccio:2008} that is
consistent with their WMAP5 simulations over the range
$10^{10}\,M_\odot < M_\mathrm{vir} < 10^{15}\,M_\odot$: 
\begin{equation}
c_{\rm vir}(M_{\rm vir})\approx 9.354 \left( \frac{M_{\rm vir}}{h^{-1} 10^{12}
\,M_{\odot}}\right )^{-0.094},
\end{equation}
where $M_\mathrm{vir}$ is the \emph{total} mass enclosed within the
virial radius. 

We then rewrite this relationship in terms of the dark
mass inside the virial radius, rather than the total mass using the
equation $M_\mathrm{vir} = \alpha\,M_\mathrm{DM}$. A number of
plausible choices are available to us for the constant $\alpha$: 
to adopt the cosmological value, to neglect the contribution of baryons
entirely, or to adopt some intermediate value. The cosmological
value is derived from the ratios of matter density to critical density
$\Omega_\mathrm{m} = 0.258$ and baryon density to critical density
$\Omega_\mathrm{b} = 0.0438$, which together imply
\begin{equation}
\alpha = \frac{\Omega_\mathrm{m}}{\Omega_\mathrm{m} - \Omega_\mathrm{b}} =
1.20.
\end{equation}
The most extreme 'missing baryon' scenario, which neglects the
contribution of baryons to the total viral mass, implies $\alpha = 1$.
An intermediate possibility is motivated by the results of observationally
constrained halo occupation distribution methods and semi-analytic
models, which imply that the stellar mass of a galaxy is typically
around 3 per cent of that of its halo, i.e. $\alpha = 1.03$.
In this work we follow \cite{Napolitano:2005} and adopt the
cosmological value, $\alpha = 1.20$, which gives
\begin{equation}
\label{eqn:cmd}
c_\mathrm{vir}(M_{\rm DM}) \approx 
9.195 \left( \frac{M_{\rm DM}}{h^{-1} 10^{12}\,M_{\odot}} \right)^{-0.094}.
\label{eqn:cmdm}
\end{equation}

However, we note that the choice of $\alpha$ makes almost no different
to our results. This is because $\alpha$ is immediately raised to the
power $-0.094$ in order to define the concentration of the halo; a 20
per cent change in $\alpha$ changes $c_\mathrm{vir}$ at the 2 per cent
level. If our choice is wrong and results in the introduction of such a
small systematic error in $c_\mathrm{vir}$, then the consequences for
the parameters of the best-fitting mass models are in any case
negligible compared to the observational errors.

The dark halo density profile may therefore be written as a function of
a single parameter, $M_{\rm DM}$, using equations~(\ref{eqn:rhonfw2}),
(\ref{eqn:a}) and (\ref{eqn:cmdm}). We perform a multi-Gaussian
expansion of this one-dimensional, single parameter profile to allow us
to easily include it in the potential for which we derive model
kinematics. 

We refrain from including prescriptions for the effects of baryonic
contraction on our halo model
\citep[e.g.][]{Blumenthal:1986,Gnedin:2004,Abadi:2009}. We do this for
two reasons. Firstly, our goal is not to determine the detailed shapes
of dark haloes but rather the total dark and stellar masses in the optical
parts of galaxies. Secondly, while our kinematic data do not allow us to
constrain the halo shape, other observational evidence suggests that
contracted NFW haloes do not reproduce observed kinematics.
\citep[e.g.][]{Gentile:2004,Kassin:2006,Thomas:2007}.

\subsection{Modelling the stellar kinematics}
\label{sec:jeans}

The most general dynamical methods are particle-based
\citep[e.g.][]{de-Lorenzi:2007} or orbit-based
\citep[e.g.][]{Schwarzschild:1979,Cappellari:2007,van-den-Bosch:2008,Thomas:2009}.
These powerful methods are so general, however, that a wide range of
unrealistic models can be made to fit long-slit data, providing only
weak constraints on the parameters of the mass models \citep[see, e.g.,
fig. 2 of][]{Cappellari:2005}. In fact, the stellar kinematics of
early-type fast-rotator galaxies are well-described by Jeans models with
a cylindrically-aligned velocity ellipsoid with a constant flattening in
the $z$-direction \citep{Cappellari:2007,Thomas:2009}. This has lead to
the use of a Jeans modelling technique in which such a velocity
ellipsoid is assumed. By varying the flattening of the velocity
ellipsoid, the Jeans modelling approach has been shown to reproduce a
wide range of two-dimensional observed kinematics in the central regions
of early-type fast-rotators in the SAURON survey
\citep{Cappellari:2008,Scott:2009} and out to $5\,\Reff$ in the case of
the edge-on S0 NGC~2549 \citep{Weijmans:2009}. The large size of our
sample (28 galaxies) allows us to test whether the assumptions provide a
good description of our galaxies, but we note that our galaxies are at
least as rotationally supported as the SAURON galaxies, so we
expect to be even less vulnerable to assumptions about anisotropy. 

For convenience we provide an overview of the derivation and solution of
the Jeans equations. This is essentially a summary of Sections~2 and 3.1
of \cite{Cappellari:2008}. Under the assumptions required for the
collisionless Boltzmann equation to hold (a smooth potential and 
steady state), and the further assumption of axisymmetry, the Jeans
equations may be written in cylindrical coordinates as 
\begin{eqnarray}
    \frac{\rho\overline{v_R^2}-\rho\overline{v_\phi^2}}{R}
    + \frac{\partial(\rho\overline{v_R^2})}{\partial R}
    + \frac{\partial(\rho\overline{v_R v_z})}{\partial z}
    & = & -\rho\frac{\partial\Phi}{\partial R},
    \label{eqn:jeans_cyl_R}\\
    \frac{\rho\overline{v_R v_z}}{R}
    + \frac{\partial(\rho\overline{v_z^2})}{\partial z}
    + \frac{\partial(\rho\overline{v_R v_z})}{\partial R}
    & = & -\rho\frac{\partial\Phi}{\partial z}
    \label{eqn:jeans_cyl_z}
\end{eqnarray}
(\citealt{Jeans:1922}; \citealt{Binney:2008}). Here $\rho$ is the
density, $\Phi$ is the gravitational potential and we use the usual
notation 
\begin{equation}
    \rho\overline{v_k v_j}\equiv\int v_k v_j f\; \ud^3 \mathbf{v}.
\end{equation}

Even if $\rho$ and $\Phi$ are known (as is the case for our mass
models), equations (\ref{eqn:jeans_cyl_R}) and (\ref{eqn:jeans_cyl_z})
are still two equations with four unknowns, $\overline{v_R^2}$,
$\overline{v_z^2}$, $\overline{v_\phi^2}$ and $\overline{v_R v_z}$, so
they do not specify a unique solution. In order to close the equations
one can assume a particular anisotropy, i.e. a relationship between the
lengths of the axes of the velocity ellipsoid. We assume here that the
velocity ellipsoid is aligned with the cylindrical coordinate system of
the galaxy and that the anisotropy in the meridional plane is constant
($\overline{v_R^2} = b \overline{v_z^2}$ where $b$ is a constant). Under
these assumptions, equations~(\ref{eqn:jeans_cyl_R}) and
(\ref{eqn:jeans_cyl_z}) reduce to:
\begin{eqnarray}
    \frac{b\,\rho\overline{v_z^2}-\rho\overline{v_\phi^2}}{R}
    + \frac{\partial(b\,\rho\overline{v_z^2})}{\partial R}
    & = & -\rho\frac{\partial\Phi}{\partial R},
    \label{eq:jeans_beta_R}\\
    \frac{\partial(\rho\overline{v_z^2})}{\partial z}
    & = & -\rho\frac{\partial\Phi}{\partial z}.
    \label{eq:jeans_beta_z}
\end{eqnarray}
These equations are solved for the case of a density and potential
described by a sum of Gaussians in \cite{Cappellari:2008}. Once the
observable quantities in the solutions have been projected along the
line-of-sight, a single integration gives a model of the second velocity
moment, $\mu_2$ (see equation~[28] of \citealt{Cappellari:2008}). We use
the Jeans Anisotropic MGE (\textsc{jam}) routines to perform the
calculation.$^1$ The second velocity moment is compared to the observed
root mean square velocity, $v_\mathrm{rms} \equiv (v^2 +
\sigma^2)^{1/2}$, where $v$ is the observed line-of-sight velocity and
$\sigma$ the observed line-of-sight velocity dispersion.
$v_\mathrm{rms}$ is a good approximation to the true second moment. The
second moment $\mu_2$ is perhaps a less familiar quantity to work with
than the circular or line-of-sight velocities, but it has the distinct
advantage that it is not necessary to assume a particular `splitting'
between ordered and random motions. We discuss its physical meaning in
more detail in \refsec{sec:modeluncertainties}.

The quantity $b$ is often expressed as an anisotropy parameter $\beta_z
= 1 - 1/b$, where $\beta_z = 0$ corresponds to isotropy. To give an idea
of the kinds of anisotropies observed in real galaxies,
\cite{Shapiro:2003} found flattened velocity ellipsoids with $0.35
\lesssim \beta_z \lesssim 0.75$ in the discs of Sa and Sb galaxies.
\cite{Cappellari:2007} and \cite{Thomas:2009} found more circular
ellipsoids for which $\beta_z \lesssim 0.4$ in fast-rotator ellipticals
and S0s. 

If the anisotropy $\beta_z$ is free then the predicted kinematics for
each galaxy are a function of three parameters: $(M/L)_X$, $M_{\rm DM}$
and $\beta_z$. We find, however, that for our galaxies, which are
rotation dominated, the predicted kinematics along the slit are not very
sensitive to the particular choice of $\beta_z$. This means that we are
unable to place stringent constraints on anisotropy using these data,
but it also means that we are not vulnerable to serious systematic
errors due to our assumptions about anisotropy.

\subsection{Summary of methods and assumptions}

To sum up our methods section, we find the stellar component of each
mass model by assuming axisymmetry and a constant stellar mass-to-light
ratio. To each model we add a dark halo that follows a spherically
symmetric NFW profile and assumes the correlation between halo
concentration and halo mass defined by \refeq{eqn:cmdm}. The total mass
model is a function of two parameters, the stellar mass-to-light ratio
$(M/L)_X$ and the dark halo mass within the virial radius $M_{\rm DM}$. It is
expressed as a sum of Gaussians.

The mass model is used to calculate an estimate of the observed stellar
kinematics. This is done by solving the Jeans equations under the
assumption of constant anisotropy in the meridional plane, yielding the
second velocity moment, $\mu_2$. This predicted quantity is then
compared to the observed root mean square velocity $v_\mathrm{rms}
\equiv (v^2 + \sigma^2)^{1/2}$. The two parameters of the mass model
$(M/L)_X$ and $M_{\rm DM}$, and the velocity anisotropy $\beta_z$ are
then adjusted until the predicted stellar kinematics match the
observations, placing constraints on those parameters.

Having determined the parameters for each galaxy, we also compute
circular the velocities $v_\mathrm{c}$ of the stellar, dark and total
mass distributions using the numerical techniques described in
\cite{Cappellari:2002}. $v_\mathrm{c}$ provides an intuitive measure of
the mass enclosed as a function of radius, which is proportional to
$v_\mathrm{c}^2$. The ratio of the squares of the stellar and dark
circular velocities is therefore equal to the ratio of the stellar and
dark mass enclosed. $v_\mathrm{c}$ will also be useful in a forthcoming
paper on the Tully-Fisher relation (Williams, Bureau \& Cappellari, in
preparation).

All mass-to-light ratios in this work are given in solar units, i.e. the
mass-to-light ratio at waveband $X$, $(M/L)_X$, is in units of
$(M/L)_{X,\odot}$.

\section{Sample and observations}
\label{sec:sample}

We apply the methods described above to a sample of 28 edge-on disc
galaxies selected by \cite{Bureau:1999}. 14 of the 28 galaxies are
classified as S0s and the remaining 14 are Sa--Sb galaxies (see
Table~\ref{tab:sample}).

The galaxies in the sample were originally selected to investigate the
nature of boxy and peanut-shaped (B/PS) bulges. Of the 28 galaxies, 22
have a B/PS bulge and 6 form a control sample with spheroidal bulges.
Galaxies with a B/PS bulges are edge-on disc systems with bulge
isophotes above and below the centre of the disc that are either
horizontal (boxy) or concave (peanut-shaped). B/PS bulges are thought to
simply be bars viewed edge-on. Many $N$-body simulations have shown that
bars buckle and thicken soon after formation, resulting in an object
that appears peanut-shaped when viewed side-on and boxy when viewed at
intermediate angles
\citep[e.g.][]{Combes:1981,Combes:1990,Raha:1991,Athanassoula:2002}.
\cite{Lutticke:2000} found that 45 per cent of edge-on disc galaxies have a
B/PS bulge, which is roughly consistent with the fraction of bars in
face-on spirals. Moreover, several studies have identified the kinematic
signatures of bars in systems with a B/PS bulge
\citep[e.g.][]{Kuijken:1995,Merrifield:1999,Bureau:1999,Chung:2004} and
a recent study demonstrates the converse by observing kinematic evidence
for a B/PS bulge in a face-on barred system \citep{Mendez-Abreu:2008}.
The sample therefore contains galaxies whose central regions are
probably non-axisymmetric. If there is a direct correspondence between
B/PS bulges and bars, then the bar fraction in our sample ($\approx 75$
per cent) is representative of that in all disc galaxies ($\approx 65$
per cent, e.g. \citealt{Sheth:2008}). In fact, we will argue in
\refsec{sec:discussion} the these non-axisymmetries do not affect our
results, despite our axisymmetric modelling techniques.

%
%
\begin{table*}
\caption{Galaxy sample}
\label{tab:sample}
\begin{tabular*}{\textwidth}{@{\extracolsep{\fill}}llccccccccc}
\hline
Galaxy & Type  &  D & $R_{25}$ & \Reff{} & $R_\mathrm{max}/R_{25}$ &
$R_\mathrm{max}/\Reff{}$ & $K_S$ & $B$ & $M_{K_S}$ & $M_{B}$ \\
       &       &  (Mpc) & (arcsec) & (arcsec) &  & & (mag) &  (mag) & (mag) & (mag) \\
(1)    & (2)   &  (3)   & (4) & (5) & (6) & (7) & (8) & (9) & (10) & (11) \\
\hline
\multicolumn{11}{c}{B/PS bulges} \\
\\
     NGC 128 &       S0 pec & $ 59.2$ &  95.3 & 18.1 & 0.49 & 2.58 &  8.51 & 12.46 & -25.35 & -21.40 \\
ESO 151-G004 &     S0$^{0}$ & $106.9$ &  42.2 & 14.1 & 0.71 & 2.13 & 10.29 & 14.73 & -24.85 & -20.41 \\
    NGC 1381 &          SA0 & $ 16.8$ &  76.2 & 20.1 & 0.92 & 3.47 &  8.36 & 12.35 & -22.76 & -18.77 \\
    NGC 1596 &          SA0 & $ 15.3$ & 116.7 & 21.7 & 0.39 & 2.11 &  8.06 & 11.94 & -22.86 & -18.98 \\
    NGC 1886 &          Sab & $ 24.5$ &  96.2 & 31.1 & 0.71 & 2.20 &  9.16 & 12.22 & -22.79 & -19.73 \\
    NGC 2310 &           S0 & $ 15.2$ & 135.6 & 39.2 & 0.64 & 2.21 &  8.43 & 11.48 & -22.48 & -19.43 \\
ESO 311-G012 &        S0/a? & $ 14.5$ & 122.8 & 29.7 & 0.49 & 2.04 &  7.76 & 10.89 & -23.05 & -19.92 \\
    NGC 3203 & SA(r)0$^{+}$ & $ 35.4$ &  96.6 & 22.1 & 0.64 & 2.80 &  8.86 & 12.76 & -23.89 & -19.99 \\
    NGC 3390 &           Sb & $ 44.0$ &  93.8 & 28.4 & 0.92 & 3.03 &  8.39 & 11.73 & -24.83 & -21.49 \\
    NGC 4469 &    SB(s)0/a? & $ 16.5$ & 143.3 & 39.1 & 0.60 & 2.21 &  8.09 & 12.07 & -23.00 & -19.02 \\
    NGC 4710 & SA(r)0$^{+}$ & $ 16.5$ & 131.6 & 44.0 & 0.73 & 2.19 &  7.62 & 11.69 & -23.47 & -19.40 \\
   PGC 44931 &          Sbc & $ 61.2$ &  91.4 & 32.3 & 0.66 & 1.86 &  9.60 & 12.60 & -24.34 & -21.33 \\
ESO 443-G042 &           Sb & $ 35.8$ &  84.6 & 35.9 & 0.86 & 2.02 &  9.37 & 12.56 & -23.40 & -20.21 \\
    NGC 5746 &    SAB(rs)b? & $ 30.4$ & 217.3 & 51.0 & 0.67 & 2.84 &  6.86 & 10.11 & -25.55 & -22.30 \\
     IC 4767 &        S pec & $ 53.3$ &  57.8 & 21.4 & 0.64 & 1.72 &  9.94 & 13.91 & -23.69 & -19.72 \\
    NGC 6722 &           Sb & $ 70.9$ &  87.9 & 22.4 & 0.81 & 3.19 &  8.95 & 12.28 & -25.30 & -21.97 \\
    NGC 6771 & SA(r)0$^{+}$ & $ 64.6$ &  98.7 & 15.5 & 0.52 & 3.29 &  8.97 & 13.24 & -25.08 & -20.81 \\
ESO 185-G053 &       SB pec & $ 66.9$ &  37.4 & 12.3 & 0.77 & 2.34 &  9.91 & 14.04 & -24.22 & -20.09 \\
     IC 4937 &           Sb & $ 72.4$ &  51.5 & 27.7 & 0.87 & 1.63 &  9.70 & 13.84 & -24.60 & -20.46 \\
ESO 597-G036 & S0$^{0}$ pec & $123.5$ &  43.3 & 18.5 & 0.87 & 2.04 & 10.10 & 14.76 & -25.36 & -20.70 \\
     IC 5096 &           Sb & $ 47.0$ & 105.7 & 27.6 & 0.56 & 2.14 &  8.52 & 12.15 & -24.84 & -21.21 \\
ESO 240-G011 &           Sb & $ 41.4$ & 167.2 & 41.5 & 0.73 & 2.92 &  8.43 & 11.53 & -24.65 & -21.56 \\
\\
\multicolumn{11}{c}{Control sample} \\
\\
    NGC 1032 &         S0/a & $ 37.0$ & 106.0 & 21.0 & 0.58 & 2.91 &  8.40 & 12.13 & -24.44 & -20.71 \\
    NGC 3957 &    SA0$^{+}$ & $ 24.1$ & 100.7 & 31.6 & 0.65 & 2.08 &  8.62 & 12.71 & -23.29 & -19.20 \\
    NGC 4703 &           Sb & $ 72.0$ &  50.9 & 28.0 & 1.33 & 2.43 &  8.92 & 13.37 & -25.36 & -20.92 \\
    NGC 5084 &           S0 & $ 24.2$ & 297.9 & 27.0 & 0.30 & 3.26 &  7.19 & 10.07 & -24.73 & -21.85 \\
    NGC 7123 &           Sa & $ 55.3$ &  80.2 & 17.1 & 0.60 & 2.83 &  8.45 & 12.90 & -25.26 & -20.81 \\
     IC 5176 &    SAB(s)bc? & $ 24.8$ & 135.2 & 29.6 & 0.46 & 2.10 &  8.77 & 12.02 & -23.20 & -19.95 \\
\hline
\end{tabular*}
\begin{minipage}{\textwidth}
\emph{Notes}. (1) Galaxy name. To ensure continuity with previous
studies, the sample is subdivided into B/PS bulges and control galaxies and
arranged in order of increasing right ascension. (2)
Morphological type taken from \cite{Jarvis:1986}, \cite{de-Souza:1987},
\cite{Shaw:1987} and \cite{Karachentsev:1993}. (3) Distance. The
distance of NGC~1381 is taken from \citet{Jensen:2003} and that of
NGC~1596 from \citet{Tonry:2001}. NGC~4469 and NGC~4710 are members of
the Virgo cluster so we adopt the cluster distance derived by
\cite{Mei:2007}. For all other galaxies we use distances from the
NASA/IPAC Extragalactic Database (NED) calculated assuming a WMAP5
cosmology (\citealt{Komatsu:2009}, $H_0 = 70$\,km\,s$^{-1}$\,Mpc$^{-1}$)
and a Virgocentic flow model \citep{Mould:2000}. (4) Radius of
the 25\,mag\,arcsec$^{-2}$ $B$-band isophote taken from HYPERLEDA
\citep{Paturel:2003}. (5) Semi-major axis of the ellipse
containing half the light at $K_S$-band, measured using the
near-infrared photometry presented in \protect{\cite{Bureau:2006}} and
the method described in \refsec{sec:results}. (6) and (7) 
Radius of the last stellar kinematic data point used in this work,
expressed as a fraction of $R_{25}$ and \Reff. (8) Apparent
magnitude at $K_S$-band derived from the MGE parametrization of the $Kn$-band
image \citep{Bureau:2006} calibrated using the 2MASS $K_S$-band image
\protect{\citep{Skrutskie:2006}} corrected for Galactic extinction. 
(9) Apparent magnitude at $B$-band corrected for internal and Galactic
extinction taken from HyperLEDA. (10) and (11) Absolute magnitudes
derived from the distance and apparent magnitudes in columns (3), (8)
and (9). 
\end{minipage} \end{table*}

To parametrize the projected light distribution we use the relatively
deep, high-resolution $K$n-band images presented by \cite{Bureau:2006}.
In the course of this work we discovered that the photometric
calibration of the images was incorrect (they were too faint by $\approx
0.8$\,mag), so we used shallower 2MASS images to recalibrate them,
transforming from $K$n to $K_S$-band in the process. We describe this
procedure in detail in Appendix~\ref{app:2mass}. We also correct the
2MASS images and therefore our models for the effects of foreground
Galactic extinction using the dust maps of \cite{Schlegel:1998}. Because
the apparent surface brightness distributions that we parametrize are at
$K_S$-band, the mass-to-light ratios we measure are also at $K_S$-band,
and therefore denoted $(M/L)_{K_S}$. Throughout this work we adopt a
value for the absolute magnitude of the Sun at $K_S$-band of
$M_{K_S,\odot} = 3.29$ \citep{Blanton:2007}.

We compare the predicted kinematics of the mass models to major-axis
long-slit stellar kinematics presented in \cite{Chung:2004}. We use
stellar kinematics rather than gas kinematics because gas is not present
and/or not extended in many cases. Where gas is present, it is strongly
affected by non-circular motions and shocks in the inner parts of many
of the objects \citep[see][]{Bureau:1999a,Athanassoula:1999}.
Line-of-sight velocity distributions were extracted using the Fourier
Correlation Quotient algorithm \citep{Bender:1990}; the $v$ and $\sigma$
used are those of the best-fitting Gaussian. 

\section{Results}
\label{sec:results}

\subsection{Exploring parameter space}
\label{sec:paramspace}

\begin{figure*}
\includegraphics[width=17.5cm]{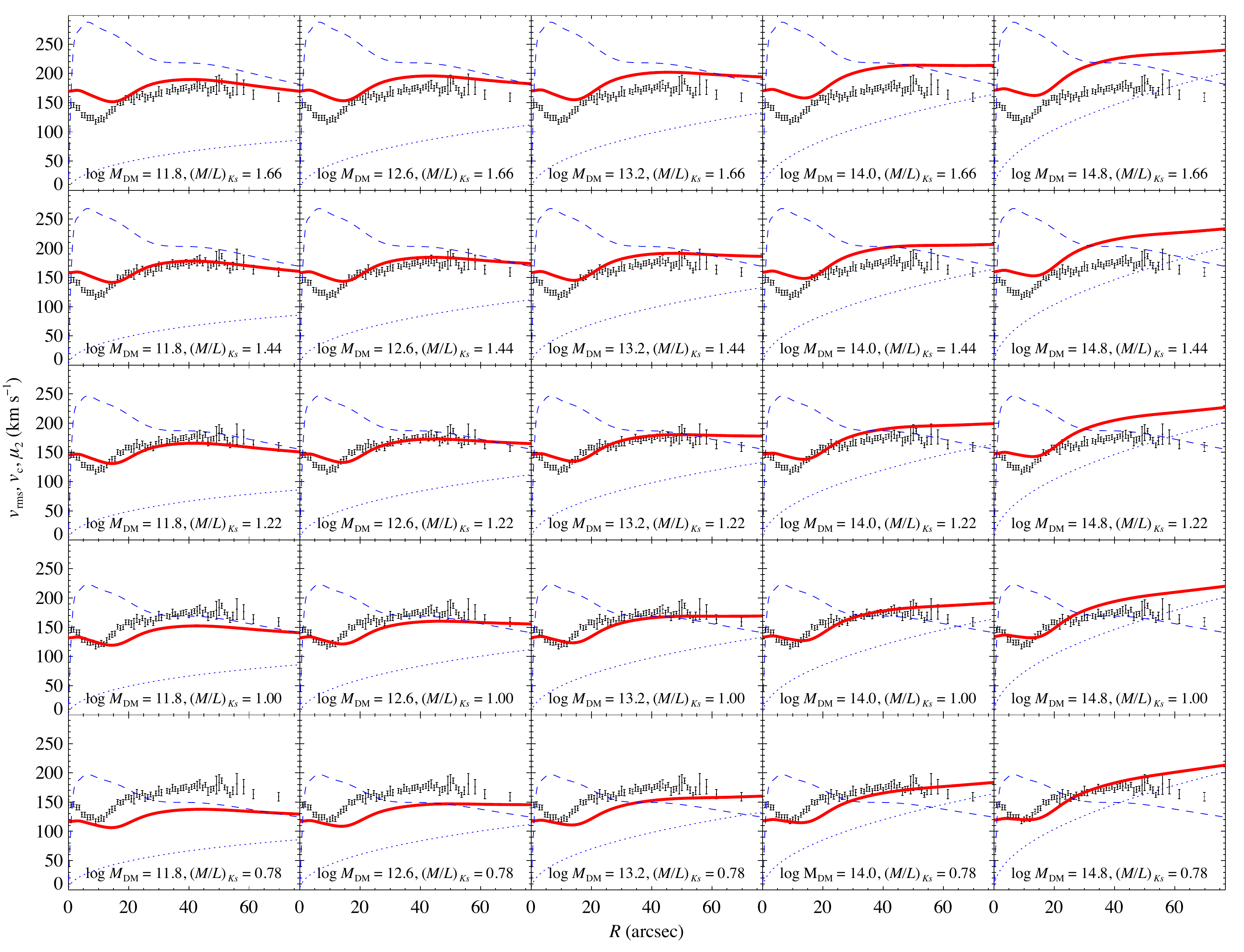}
\caption{Predicted kinematics for several mass models of NGC~1381,
demonstrating the effects of varying the model parameters
$M_\mathrm{DM}$ and $(M/L)_{K_S}$ Each plot shows the observed
$v_\mathrm{rms}$ (points with error bars) and predicted $\mu_2$ (solid
red line). We also show the circular velocity of the luminous (dashed
blue line) and dark (dotted blue line) components. The text at the
bottom of each plot gives the parameters of the mass model in solar
units at $K_S$-band. The best fitting mass model for this galaxy has
$(M/L)_{K_S} = 1.22$, $M_\mathrm{DM} = 10^{13.21}$ and $\beta_z = 0.14$.
For this figure we assumed isotropy, i.e. $\beta_z = 0.0$.}
\label{fig:jamparam}
\end{figure*}

Before presenting our best models, we give an overview of how the
changes in $(M/L)_{K_S}$ and $M_{\rm DM}$ affect the predicted second
velocity moment $\mu_2$, and what kinds of constraints this allows us to
place on these parameters. In general, increasing $(M/L)_{K_S}$ shifts
the model $\mu_2$ to greater velocities at all radii. Increasing
$M_\mathrm{DM}$ increases $\mu_2$ at large radii more than it does at
small radii. \reffig{fig:jamparam} demonstrates this behaviour for
NGC~1381.

\begin{figure*}
\includegraphics[width=17cm]{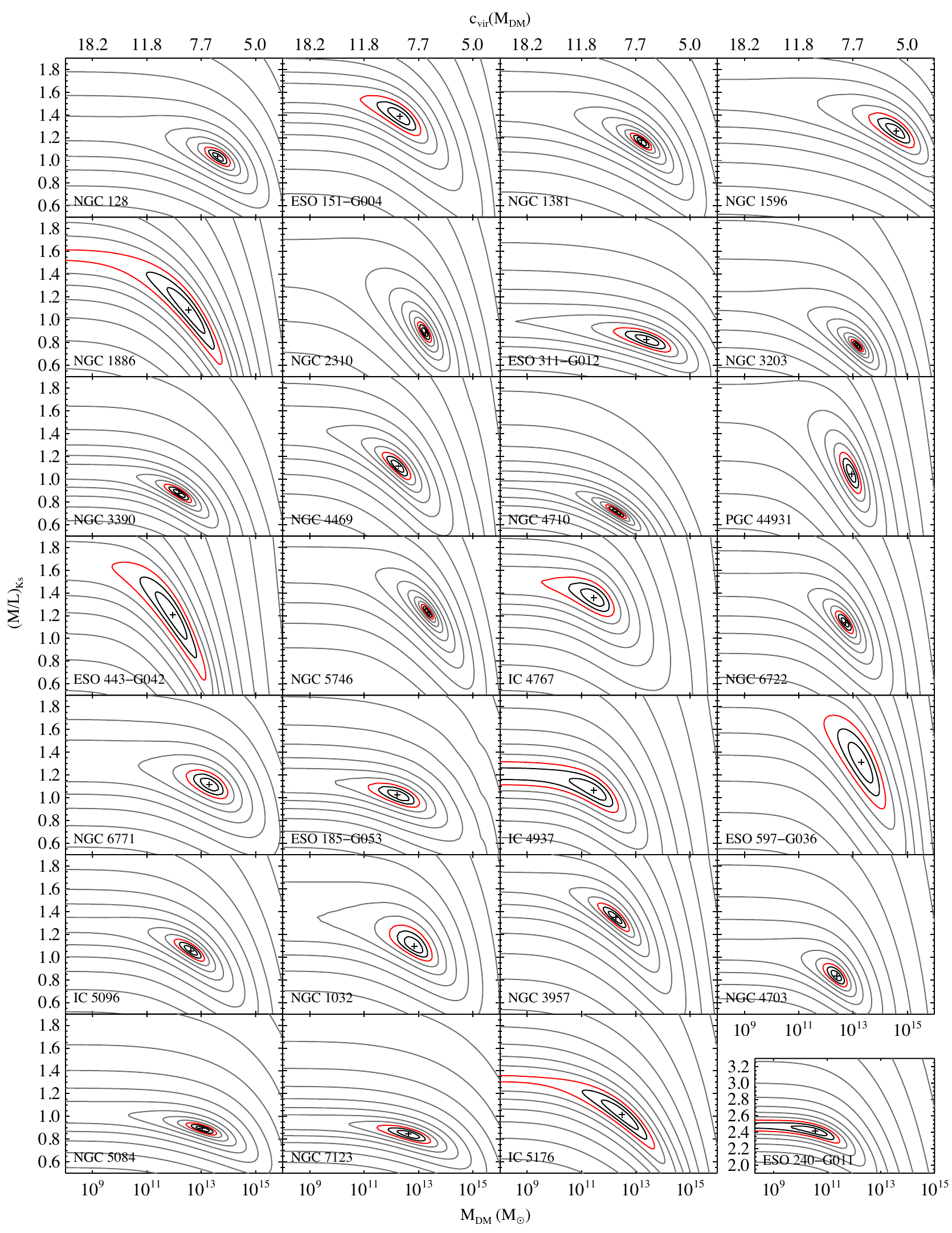}
\caption{$\chi^2$ contour plots for the complete sample showing fit
quality as a function of dark halo mass $M_{\rm DM}$ and stellar
mass-to-light ratio $(M/L)_{K_S}$. The third parameter, the anisotropy
$\beta_z$, has been marginalized out for this figure, as described in
\refsec{sec:paramspace}. The red contours are the $3\,\sigma$ confidence
levels. The inner black contours are at $1\,\sigma$ and $2\,\sigma$. The
cross in each plot shows the location of minimum $\chi^2$. The upper
horizontal axis shows the halo concentrations $c_{\rm vir}(M_{\rm DM})$
corresponding to $M_{\rm DM}$, as defined by \refeq{eqn:cmd}. With the
exception of ESO~240-G011, we arrange galaxies in the same order in
which we present them in Table~\ref{tab:sample}. The parameters of the
best-fitting model for ESO~240-G011 are outliers, so this galaxy is
shown over a different range of $(M/L)_{K_S}$.}
\label{fig:jamchi2}
\end{figure*}

\begin{figure*}
\includegraphics[width=8.4cm]{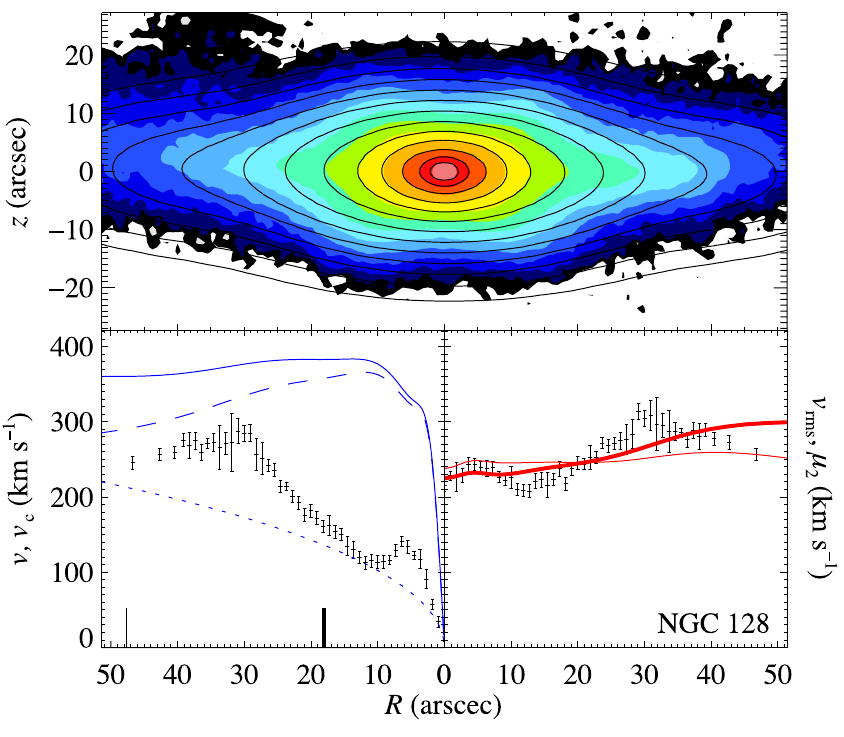}
\includegraphics[width=8.4cm]{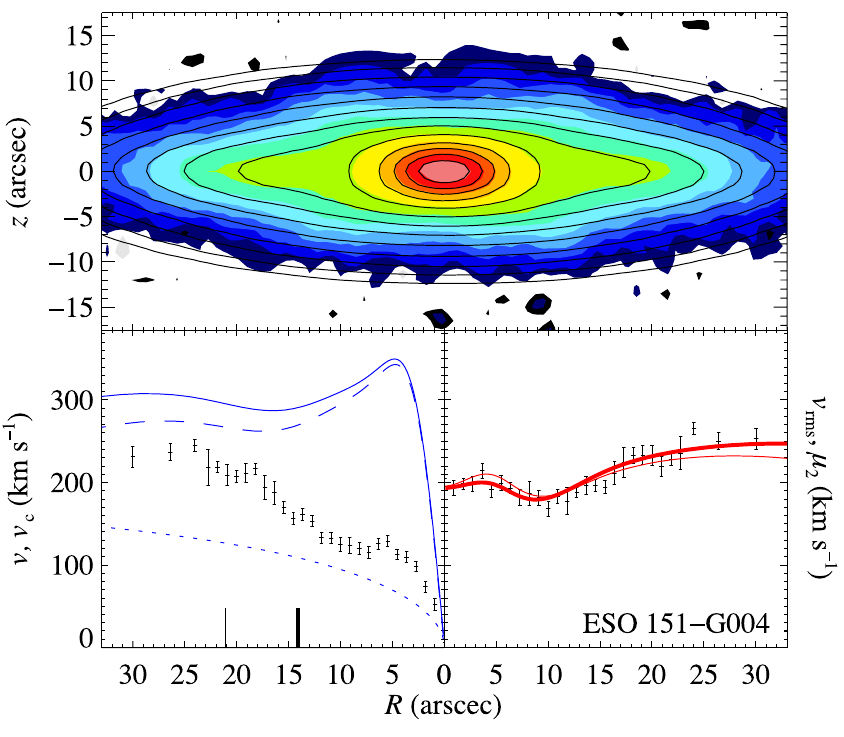} \\
\includegraphics[width=8.4cm]{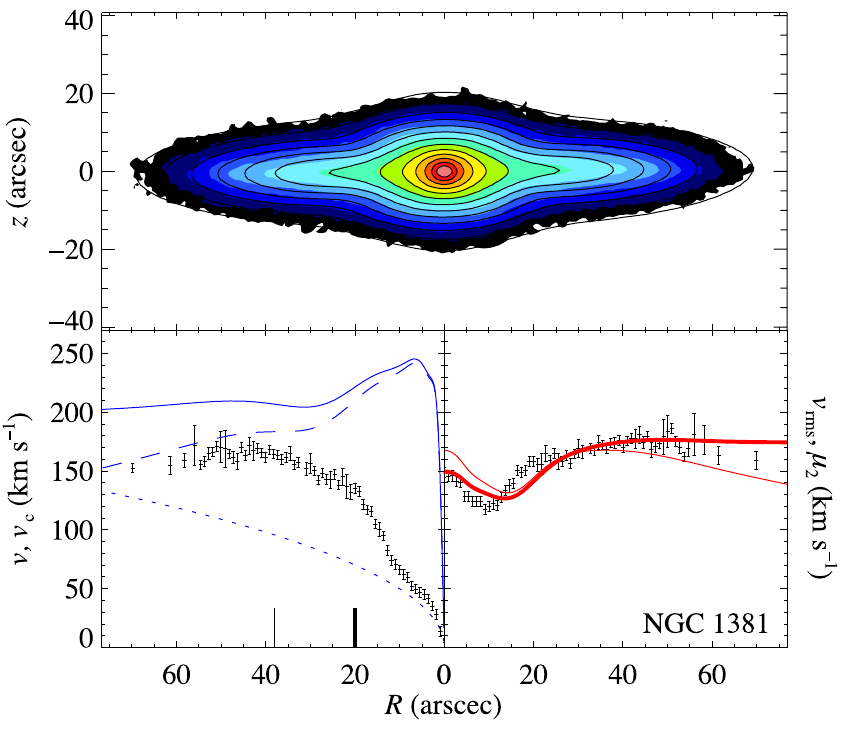}
\includegraphics[width=8.4cm]{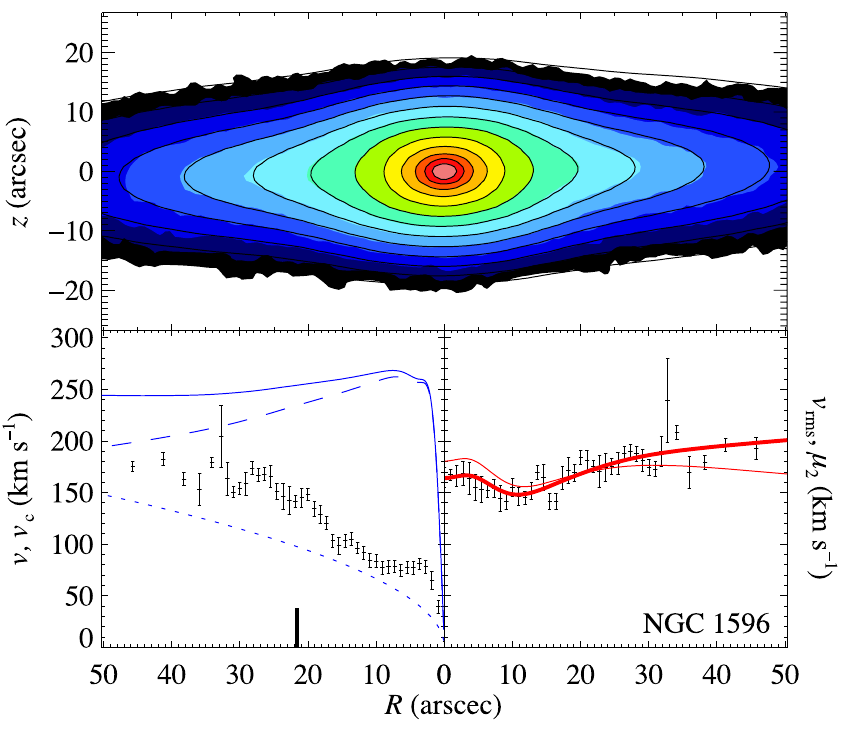} \\
\caption{Complete results of the mass and dynamical
modelling described in Section~\ref{sec:methods} for the
sample of galaxies with a B/PS bulge. Briefly, the top plot of each
panel shows the isophotes of the image (filled contours) and the MGE
parametrization of this luminous component (solid black lines). The bottom
left plot shows $v$, the observed line-of-sight velocity (points),
$v_\mathrm{c}$, the total circular velocity (solid blue line), and the
circular velocities of the luminous component (dashed blue line) and
dark halo (dotted blue line). The thick notch on the $R$-axis shows
\Reff{}, the $K_S$-band effective radius. The thin notch shows
$R_{25}/2$, half the $B$-band optical radius. The bottom right plot
shows $v_\mathrm{rms}$, the observed root mean square line-of-sight
velocity (points), $\mu_2$, the best-fitting model second velocity
moment (thick solid red line) and the best-fitting second moment with no
dark halo (thin solid red line). See Section~\ref{sec:bestfit} for a
more detailed explanation of these figures.}
\label{fig:results}
\end{figure*}
\sixfigures{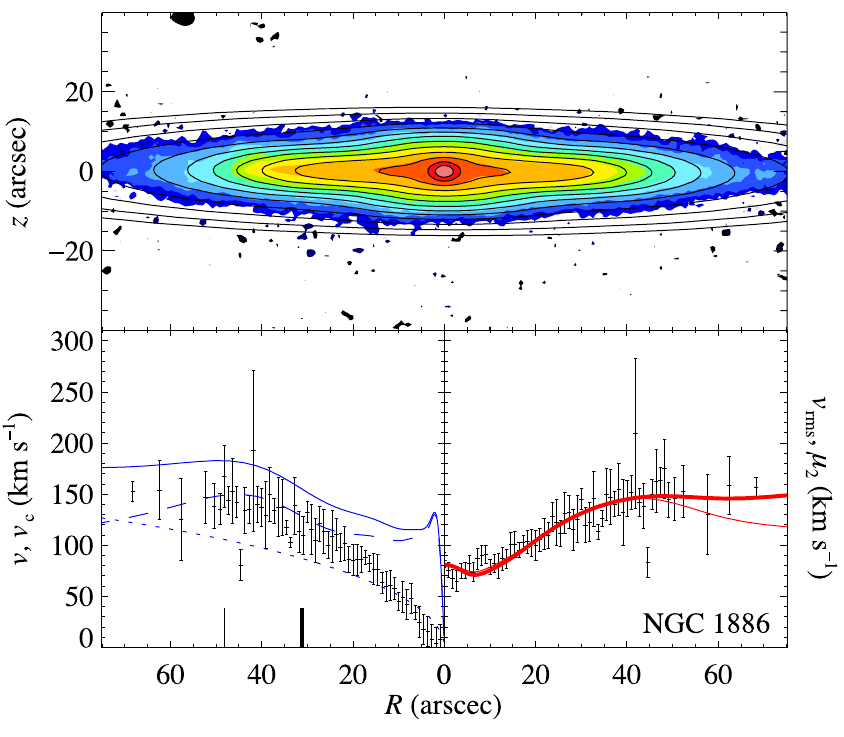}{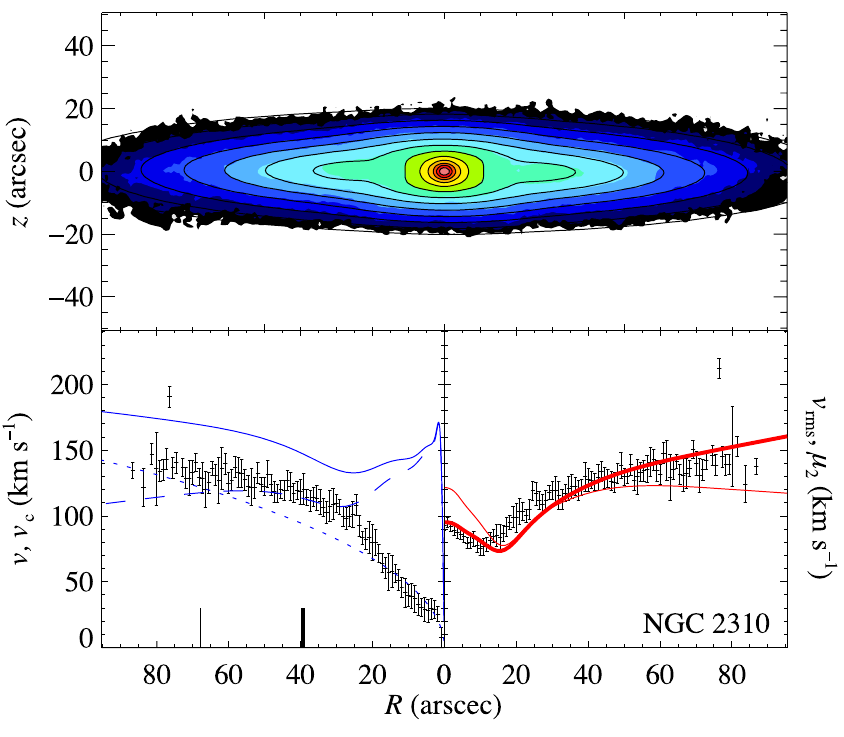}{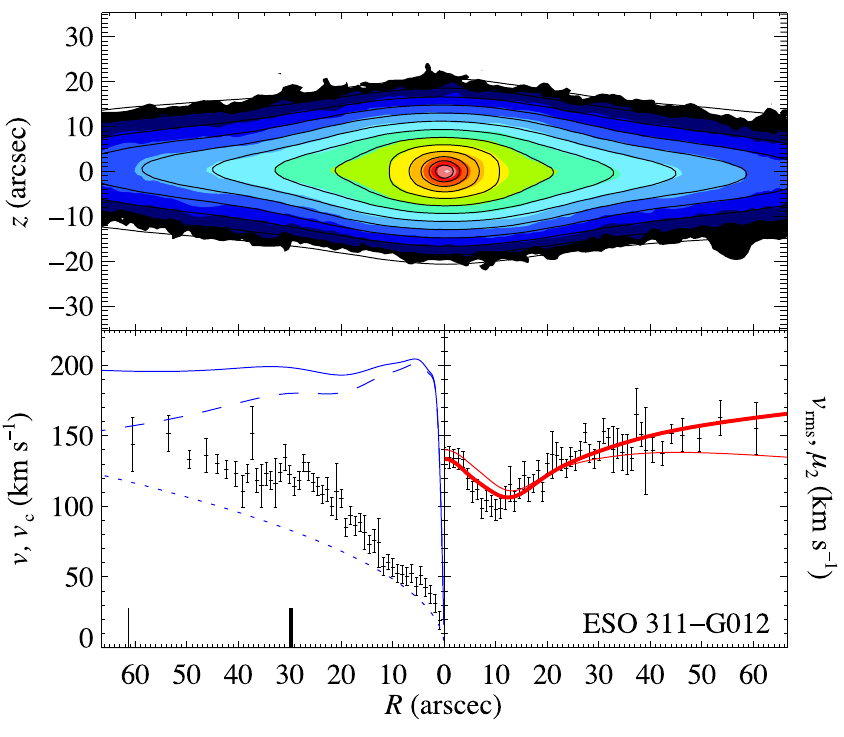}{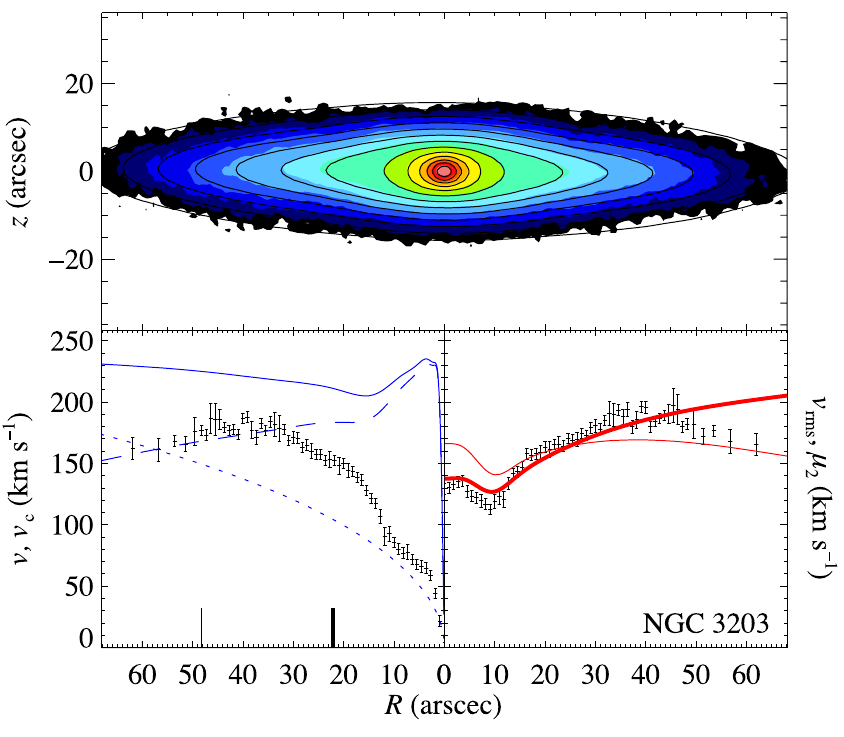}{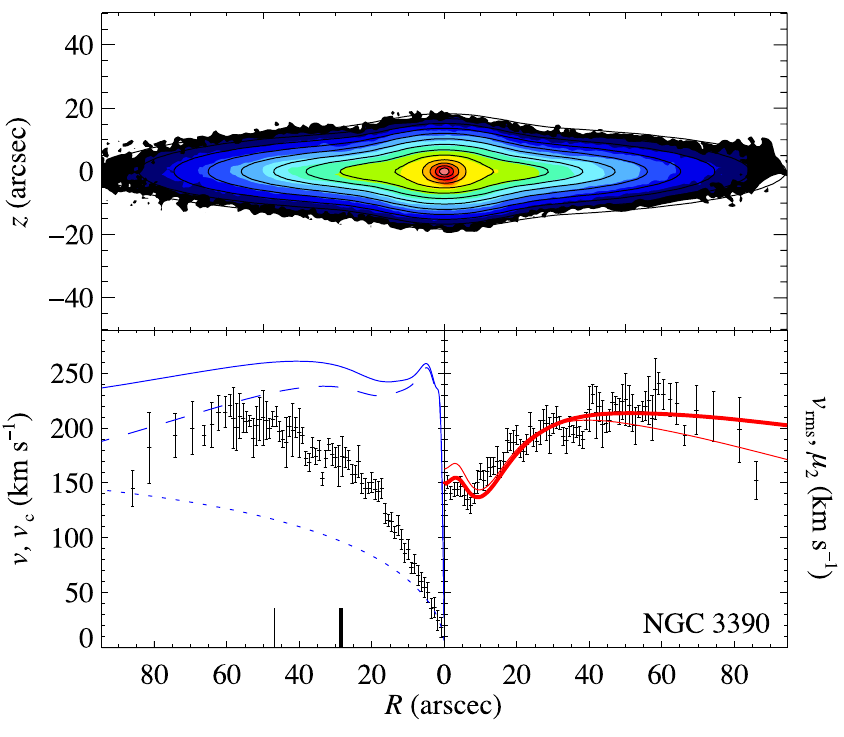}{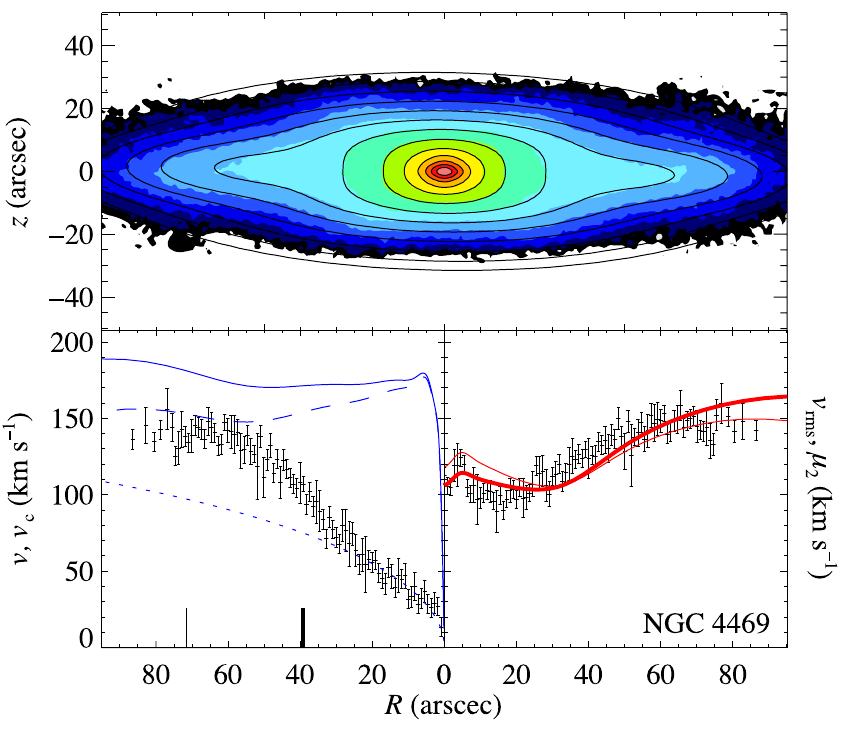}
\sixfigures{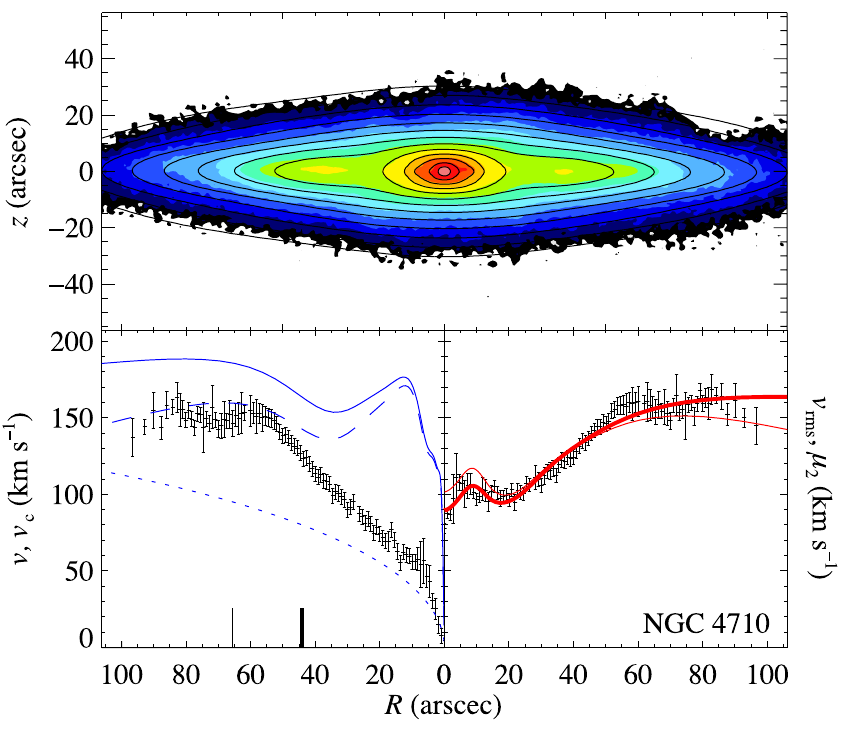}{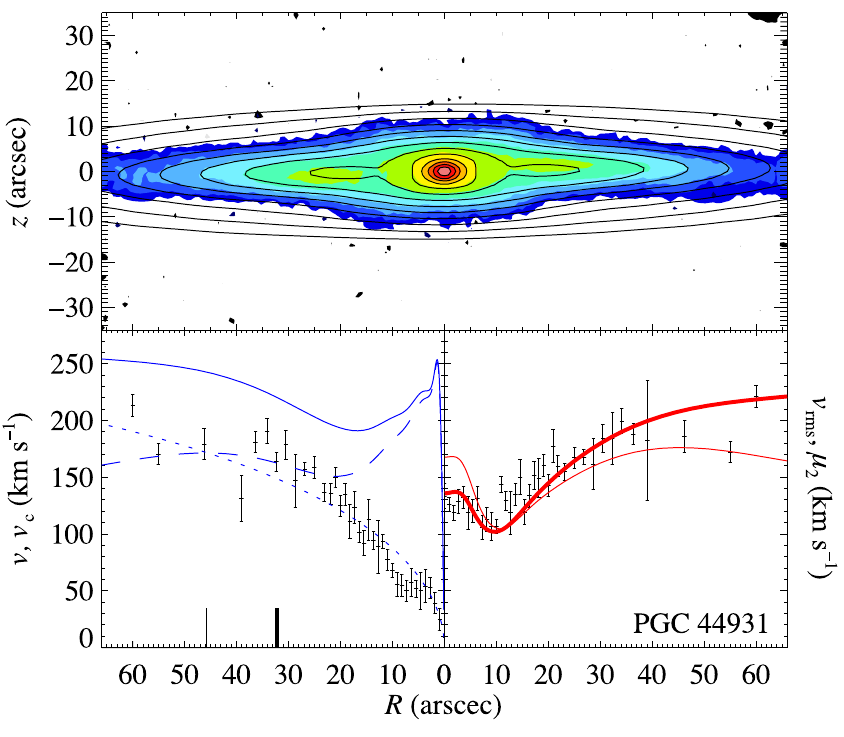}{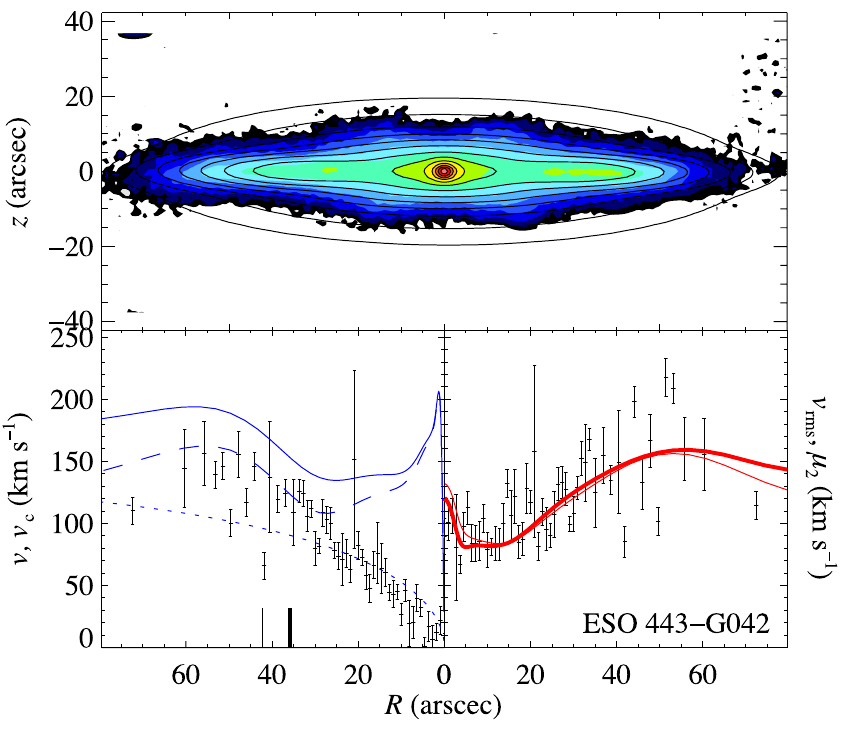}{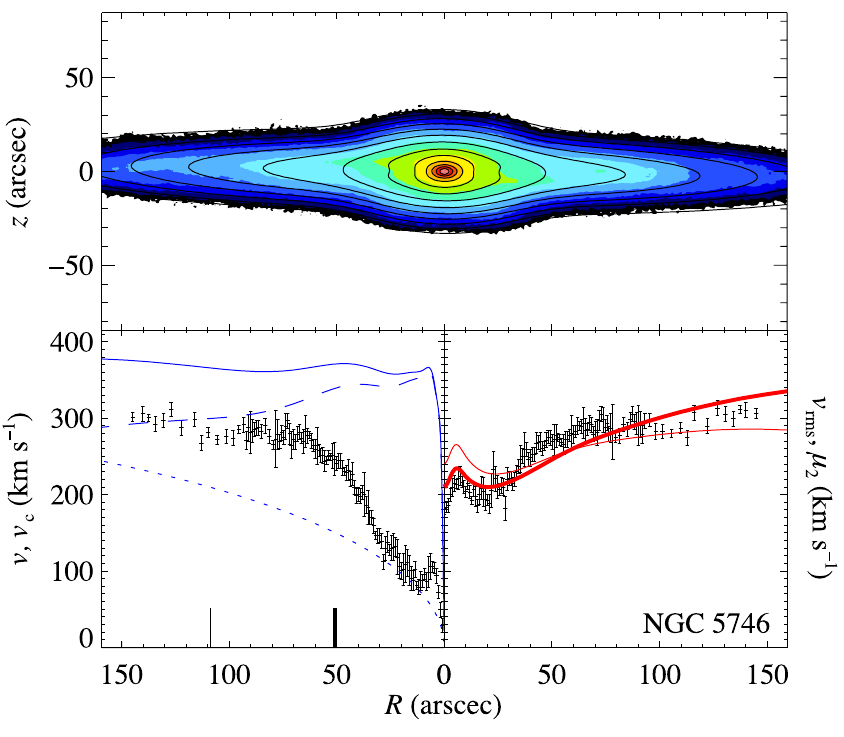}{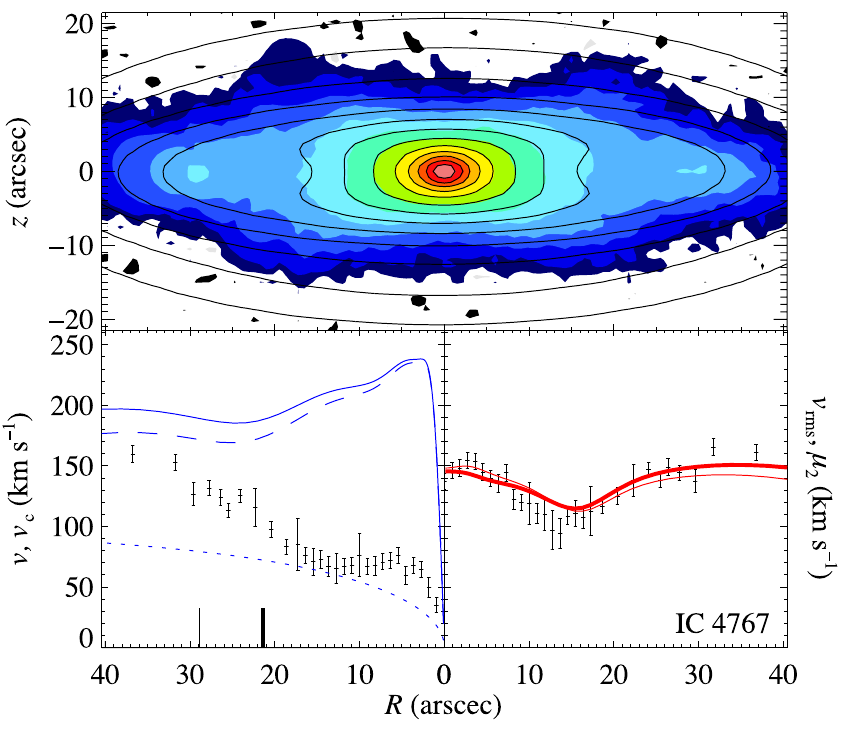}{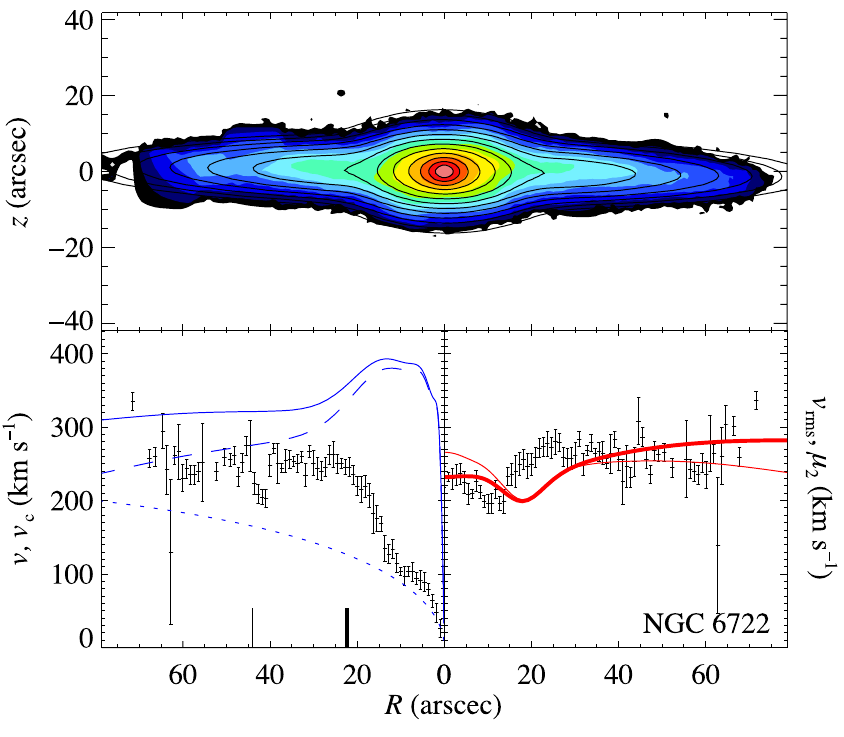}
\sixfigures{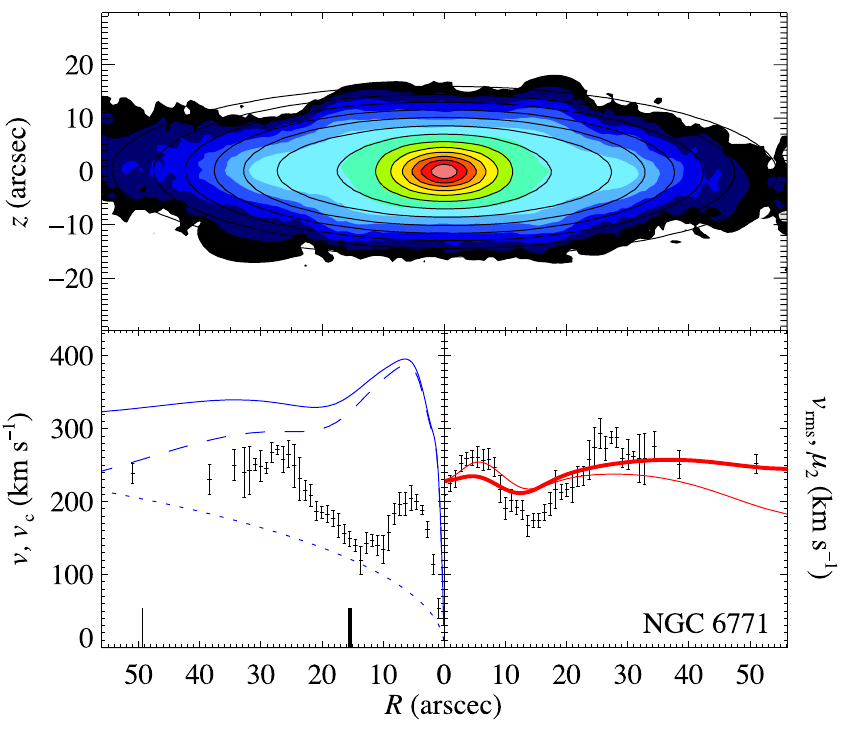}{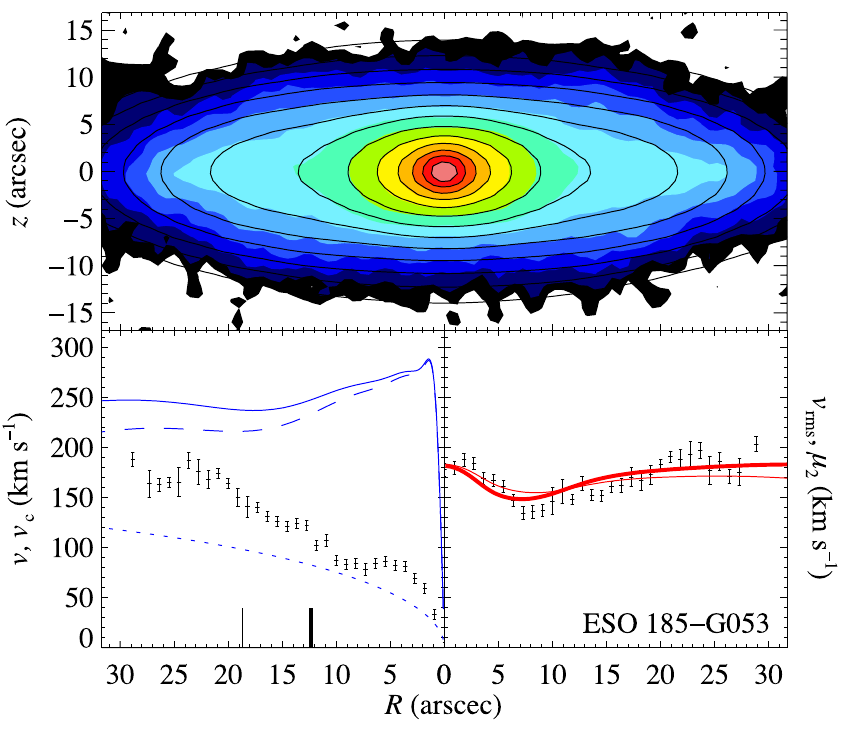}{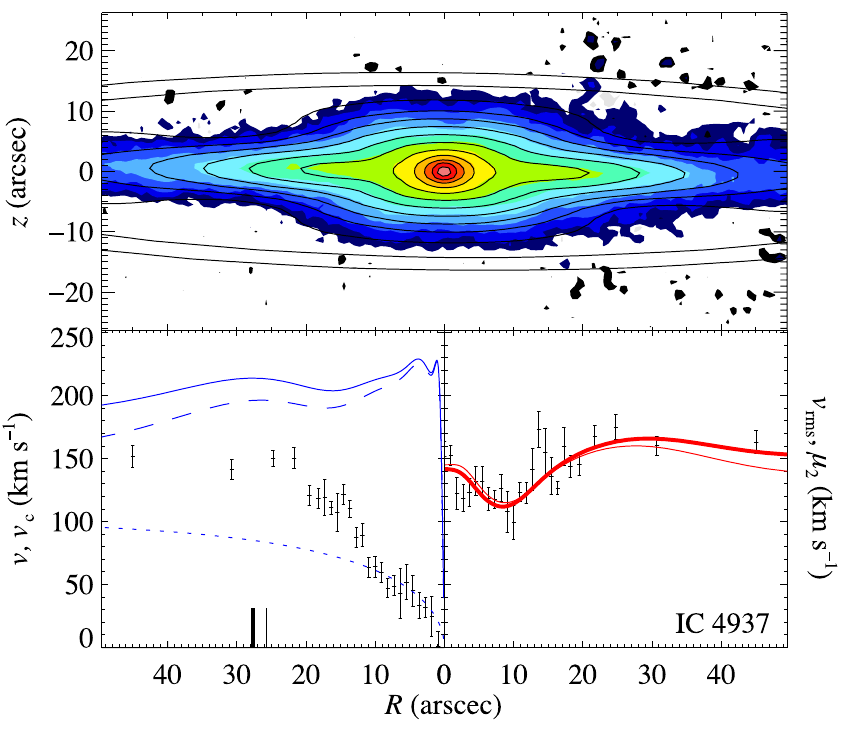}{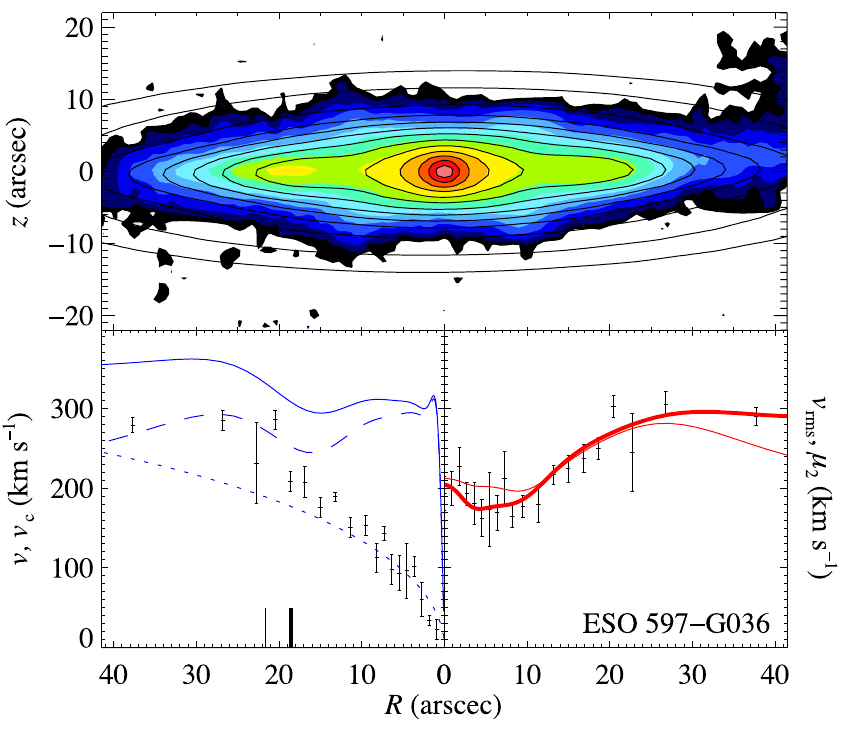}{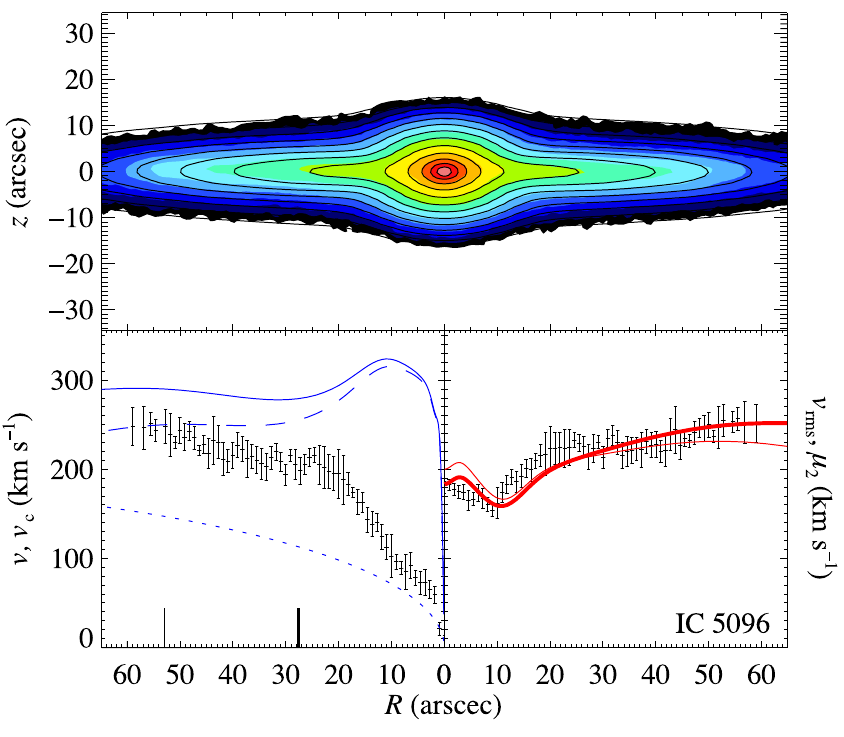}{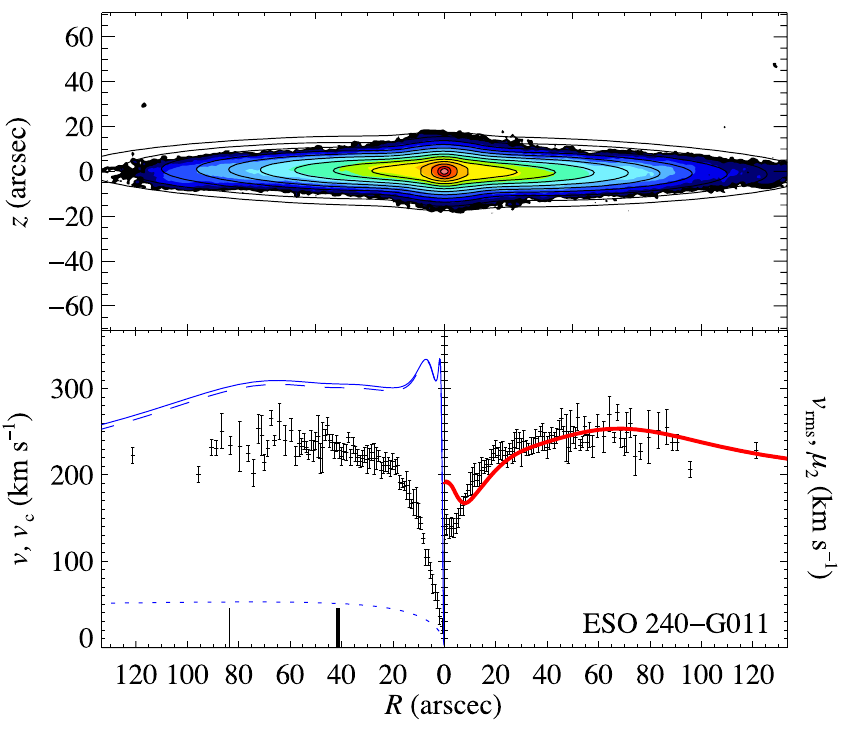}
\sixfigurescontrol{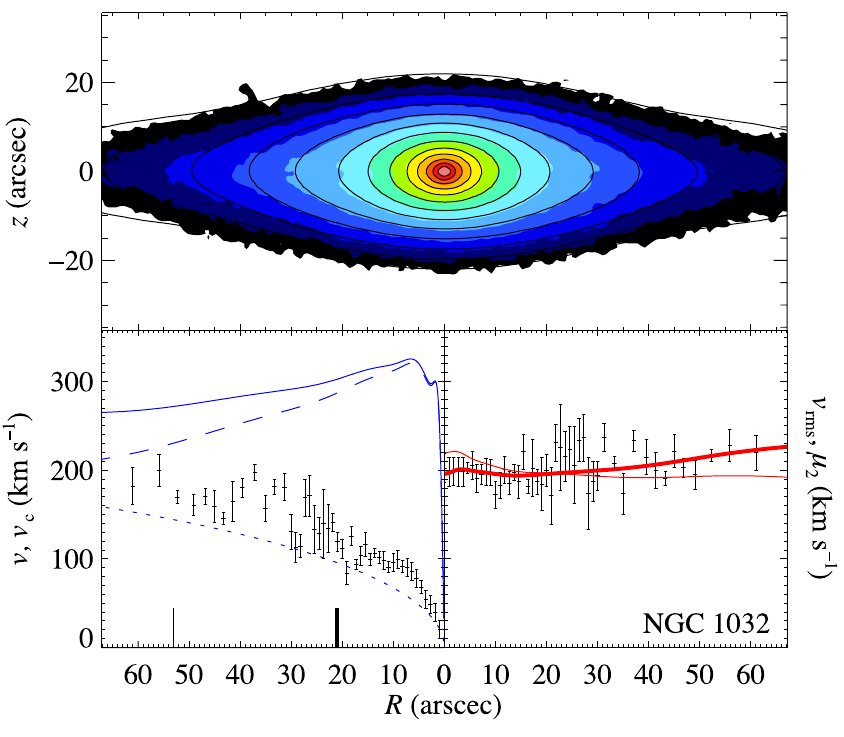}{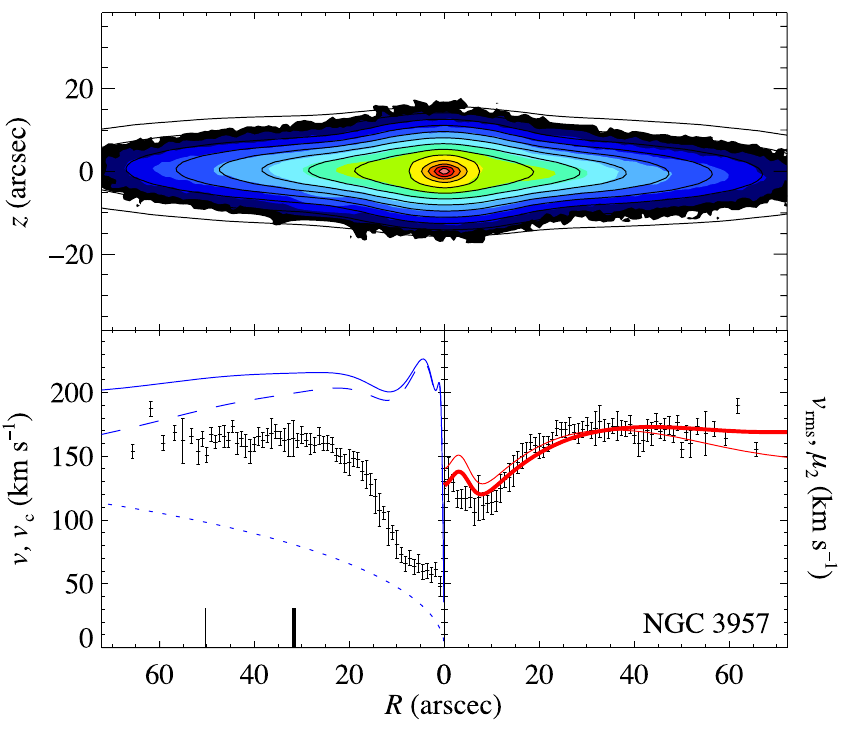}{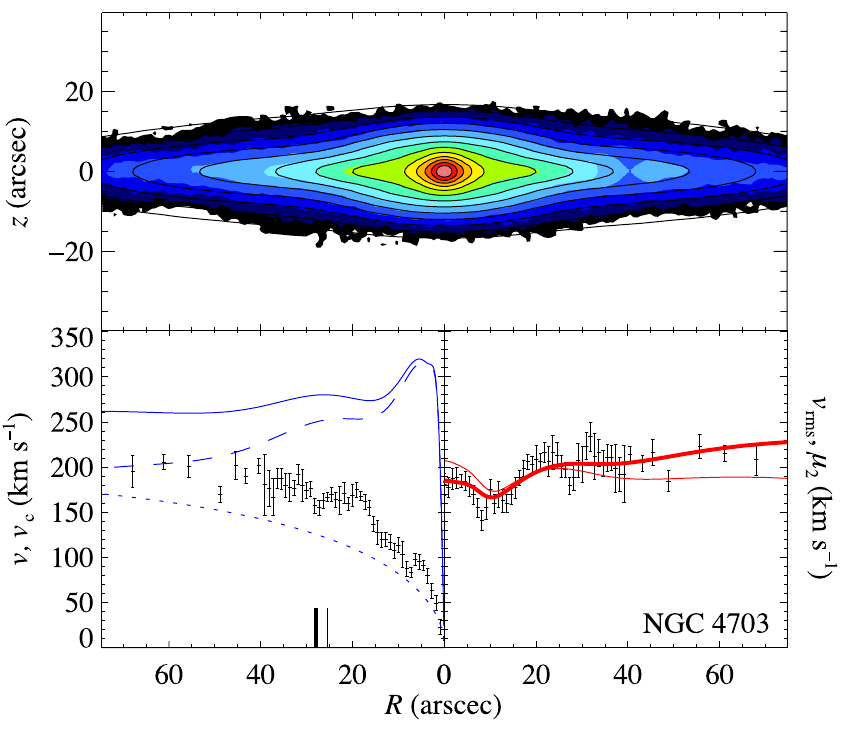}{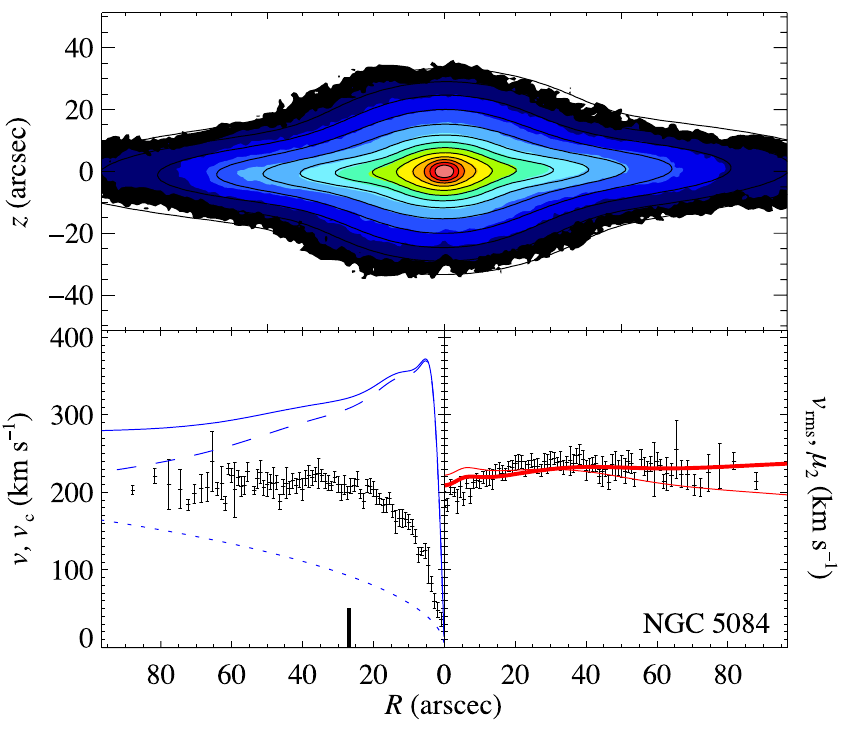}{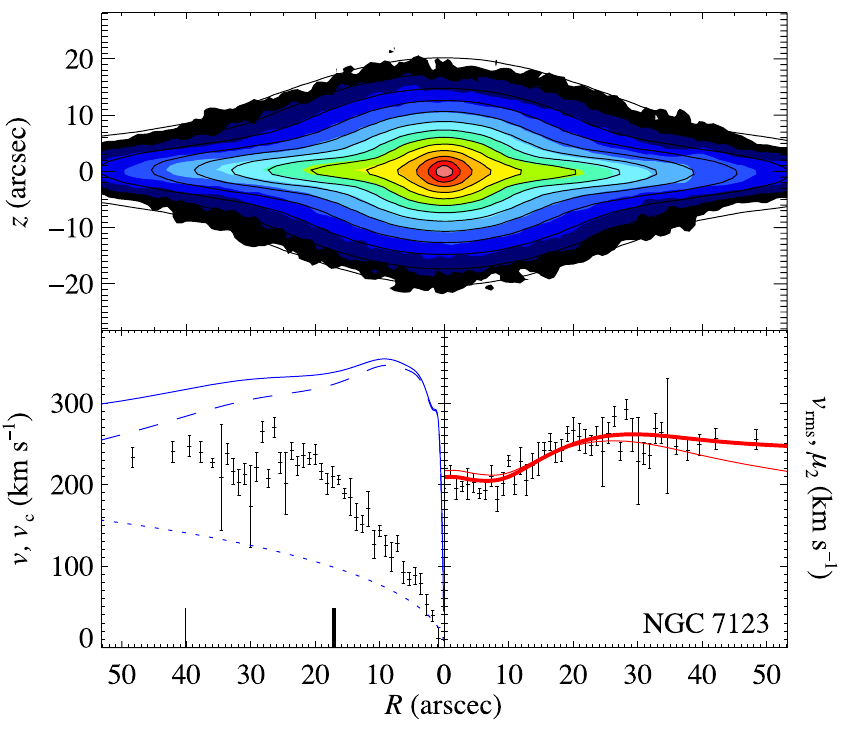}{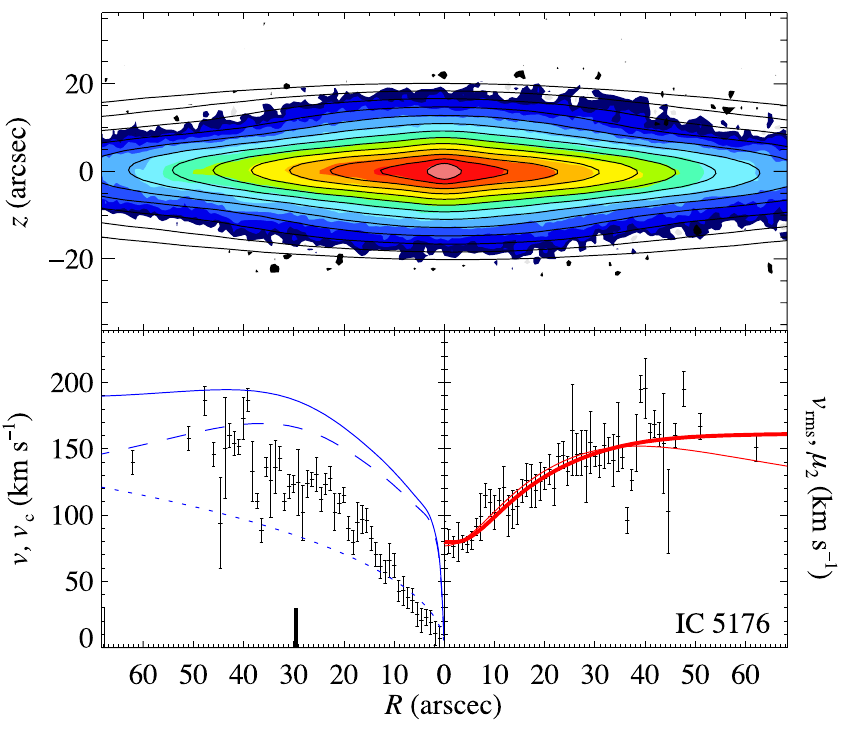}

We define the best-fitting model for each galaxy to be the one with the
combination of parameters $(M/L)_{K_S}$, $M_\mathrm{DM}$ and $\beta_z$ that
results in a predicted $\mu_2$ that most closely matches the observed
$v_\mathrm{rms}$. The figure of merit, $\chi^2$, is therefore defined as 
\begin{equation}
\label{eqn:chi2}
\chi^2 \equiv \sum \frac{\{v_\mathrm{rms} - \mu_2[M_{\rm DM},
(M/L)_{K_S}, \beta_z]\}^2}{\Delta v_\mathrm{rms}^2},
\end{equation}
where $\Delta v_\mathrm{rms}$ is the error in $v_\mathrm{rms}$ and the
summation is over all kinematic data points.

We show contour plots of $\chi^2$ as a function of these parameters for
the complete sample in \reffig{fig:jamchi2}. For clarity, these plots
are marginalized over the anisotropy $\beta_z$. This means that the
value of $\chi^2$ used at each point in $M_{\rm DM}$--$(M/L)_{K_S}$
space is set to $\min(\chi^2[M_{\rm DM}, (M/L)_{K_S}, \beta_z])$ for $0
< \beta_z < 0.75$. The contours become horizontal for $M_{\rm DM}
\lesssim 10^8 M_\odot$ because very low mass haloes do not affect the
dynamics of the galaxies. There are two classes of $\chi^2$ plot: those
where we can rule out $M_{\rm DM} = 0$ at at least the $3\,\sigma$
confidence level (i.e. where the red contour in \reffig{fig:jamchi2}
fully encircles the minimum in $\chi^2$, e.g. NGC~128 and NGC~1381) and
those where we cannot (where the red contour forms an open horizontal
`tongue', e.g. NGC~1886 and IC~4937). All galaxies therefore have a
best-fitting combination of mass model parameters and allow us to place
upper and lower bounds on $(M/L)_{K_S}$ and at least an upper limit on
$M_\mathrm{DM}$. Not all galaxies allow us to place a lower limit on
$M_\mathrm{DM}$.

\subsection{Best-fitting models}
\label{sec:bestfit}

The parameters of the best-fitting mass models are shown in
Table~\ref{tab:upsilons}. The MGE parametrizations of the luminous
component of each mass model are shown in Table~\ref{tab:mge}.
The complete results of the mass modelling and dynamical modelling
procedures described in Section~\ref{sec:methods} are shown in
\reffig{fig:results} (B/PS bulges) and \reffig{fig:control} (control
sample). In each panel:

\begin{enumerate}

\item The top plot shows the image and the luminous
component of the mass model. The filled contours are the observed
light distribution while the solid lines are the isophotes of the
best-fitting MGE model, projected and convolved with the PSF. Contours
are separated by 0.5 mag arcsec$^{-2}$. The lowest contour is 7 mag
arcsec$^{-2}$ below the highest. The horizontal axes of these images are
registered with those of the kinematic data below, which usually
requires that the image be cropped. The full extent of the photometric
data is presented in fig.~1 of \cite{Bureau:2006}.

\item The bottom left plot shows the observed line-of-sight velocity $v$
(points with error bars) and the circular velocities of the total mass
model (solid blue line), the luminous component (dashed blue line) and
the dark halo (dotted blue line) of the best-fitting mass model (see
below). We remind the reader that the ratio of the squares of the dark
and luminous circular velocities at a particular radius gives the
approximate ratio of dark to luminous matter enclosed within a sphere of
that radius (this statement is strictly true for spherical models only).

\item The bottom right plot shows the observed root mean square velocity
$v_\mathrm{rms}$ (points with error bars), the second velocity moment
$\mu_2$ of the best-fitting mass model with a dark halo (thick solid red
line) and the second velocity moment of the best model without a dark
halo (thin solid red line). 

\end{enumerate}

We mark the radial axis of the bottom left plot with $R_{25}/2$ and
\Reff{}, two distance scales used for spiral and elliptical galaxies,
respectively. We place a mark at $R_{25}/2$ because $R_{25}$ is always
beyond the extent of the stellar kinematic data. Showing both scales aids
comparison with previous studies of the influence of dark matter in both
classes of galaxies. This is especially interesting because our sample
contains many S0s, which represent a transition between spirals and
ellipticals. 

\begin{figure}
\includegraphics[width=8.4cm]{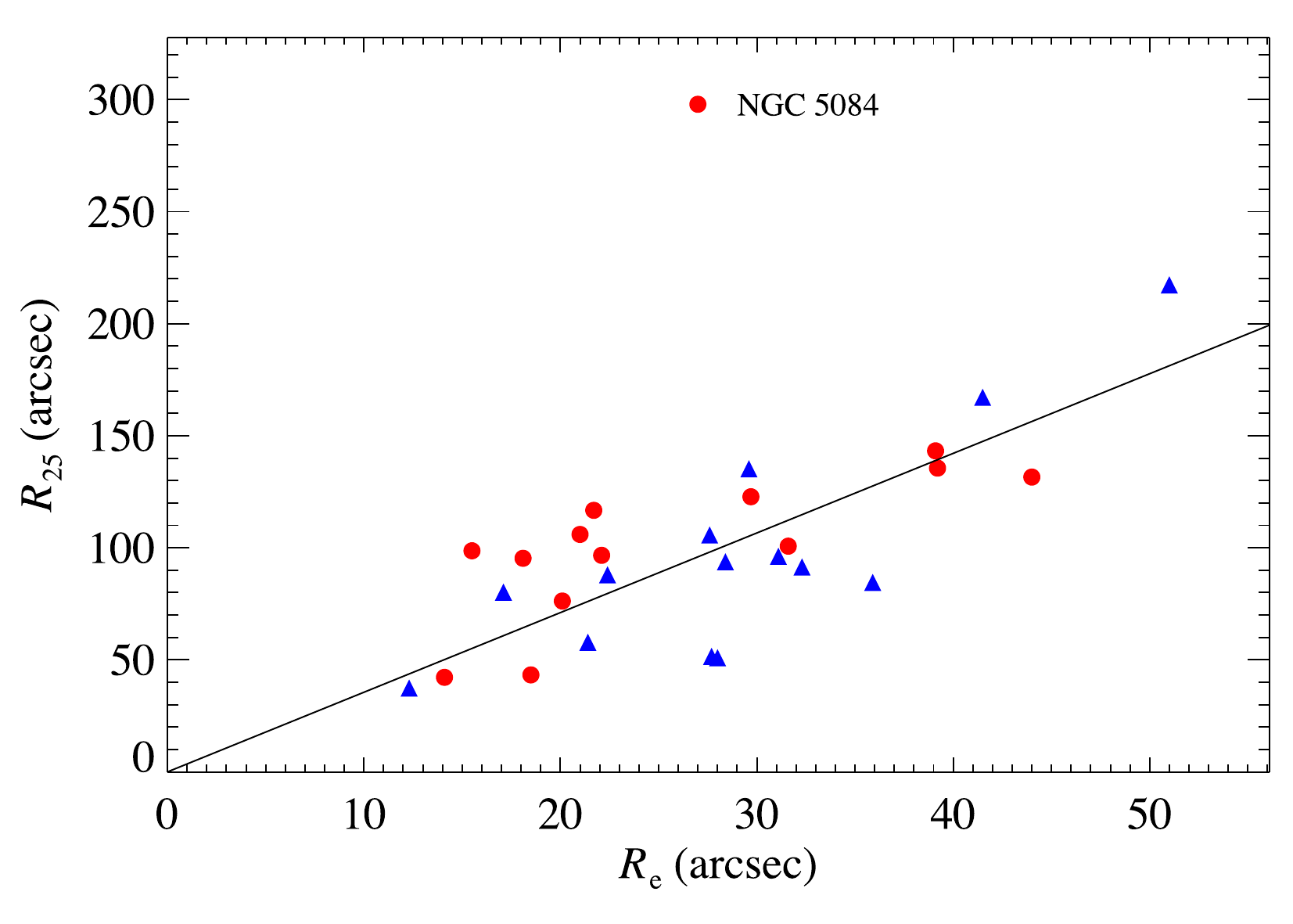}
\caption{The $B$-band optical radius $R_{25}$ shown as a function of the
$K_S$-band effective radius \Reff{} (as defined in \refsec{sec:bestfit})
for our sample. The line is a linear fit to the data with the intersect
constrained to be 0 and the outlier NGC~5084 excluded. The best fit is
$R_{25} = 3.6\,\Reff$. Symbols are as in \reffig{fig:appmag}.}
\label{fig:reffr25}
\end{figure}

We note that the effective radius of very flat objects is particularly
sensitive to its definition. We define it here to be the semi-major axis of the
ellipse containing half the integrated $K_S$-band light, which we determine
directly from the photometry. We use the integrated light of the MGE model to
determine the half-light radius rather than the more usual growth curve method.
Using our MGE models, we tested that this definition yields a value similar to
that which would obtained from an ideal observation with the galaxy oriented
face-on. We found \Reff (face-on) $= 0.9 \Reff$ (edge-on) $\pm 0.1$. For this
definition of \Reff{} and for our sample of edge-on spirals and S0s, $R_{25}
\approx 3.6\,\Reff$ (see \reffig{fig:reffr25}). This empirical finding will be
used in this paper, and is probably true to within approximately 50 per cent
for other edge-on objects, but it should obviously not be applied to less
inclined spirals.

An alternative classic definition of \Reff{} is to measure the radius of
the circle containing half the light, independent of the apparent
ellipticity \citep{Burstein:1987,Roman:1991,Jorgensen:1996}. We measured
this value for our edge-on galaxies and found that it was typically
smaller than the semi-major axis value by a factor $2.0\pm0.2$. It is
therefore rather poorly related to the `true' face-on value and we prefer
to integrate within ellipses. It is important, however, to keep our
definition of \Reff{} in mind when comparing our results to other
studies.

%
%
\begin{table*}
\caption{Mass and dynamical modelling results}
\label{tab:upsilons}
\begin{tabular*}{\textwidth}{l@{\extracolsep{8em}}c@{\extracolsep{\fill}}ccc@{\extracolsep{8em}}c@{\extracolsep{\fill}}cc}
\hline
& \multicolumn{4}{c}{Best-fitting models with halo} & \multicolumn{3}{c}{Best-fitting models without halo} \\
Galaxy & $(M/L)_{K_S}$ & $\log (M_\mathrm{DM}/M_\odot)$ & $\beta_z$ & $\chi_\mathrm{red}$ &
$(M/L)_{K_S,\mathrm{nohalo}}$ & $\beta_{z,\mathrm{nohalo}}$ & $\chi_\mathrm{red,\,nohalo}$ \\ 
\hline
\\
\multicolumn{8}{c}{B/PS bulges} \\
\\
     NGC 128 &    $1.04^{+0.07}_{-0.08}$&  $13.61^{+ 0.45}_{- 0.45}$&  $0.23^{+0.06}_{-0.06}$&  1.86  &  $1.28^{+0.06}_{-0.06}$&  $0.17^{+0.08}_{-0.09}$&  2.61 \\[2pt]
ESO 151-G004 &    $1.39^{+0.18}_{-0.16}$&  $12.31^{+ 0.77}_{- 1.44}$&  $0.32^{+0.12}_{-0.20}$&  1.16  &  $1.59^{+0.11}_{-0.11}$&  $0.21^{+0.15}_{-0.21}$&  1.38 \\[2pt]
    NGC 1381 &    $1.18^{+0.06}_{-0.08}$&  $13.21^{+ 0.36}_{- 0.40}$&  $0.14^{+0.05}_{-0.11}$&  1.89  &  $1.48^{+0.04}_{-0.04}$&  $0.17^{+0.06}_{-0.08}$&  2.85 \\[2pt]
    NGC 1596 &    $1.26^{+0.16}_{-0.14}$&  $14.59^{+ 0.66}_{- 0.99}$&  $0.14^{+0.11}_{-0.14}$&  1.17  &  $1.58^{+0.11}_{-0.11}$&  $0.12^{+0.11}_{-0.12}$&  1.80 \\[2pt]
    NGC 1886 &    $1.09^{+0.53}_{-0.47}$&  $12.53^{+ 1.22}_{-\ldots}$&  $0.43^{+0.26}_{-0.43}$&  1.04  &  $1.57^{+0.16}_{-0.15}$&  $0.00^{+0.37}_{-0.00}$&  1.12 \\[2pt]
    NGC 2310 &    $0.88^{+0.09}_{-0.08}$&  $13.26^{+ 0.18}_{- 0.27}$&  $0.00^{+0.18}_{-0.00}$&  1.55  &  $1.49^{+0.07}_{-0.06}$&  $0.00^{+0.03}_{-0.00}$&  3.09 \\[2pt]
ESO 311-G012 &    $0.82^{+0.10}_{-0.10}$&  $13.39^{+ 0.86}_{- 1.17}$&  $0.32^{+0.11}_{-0.15}$&  1.00  &  $0.99^{+0.07}_{-0.07}$&  $0.24^{+0.12}_{-0.20}$&  1.40 \\[2pt]
    NGC 3203 &    $0.77^{+0.05}_{-0.05}$&  $13.17^{+ 0.18}_{- 0.23}$&  $0.00^{+0.15}_{-0.00}$&  1.92  &  $1.17^{+0.03}_{-0.03}$&  $0.00^{+0.05}_{-0.00}$&  3.94 \\[2pt]
    NGC 3390 &    $0.87^{+0.07}_{-0.08}$&  $12.18^{+ 0.45}_{- 0.54}$&  $0.00^{+0.09}_{-0.00}$&  1.33  &  $1.06^{+0.04}_{-0.04}$&  $0.00^{+0.05}_{-0.00}$&  1.63 \\[2pt]
    NGC 4469 &    $1.12^{+0.11}_{-0.11}$&  $12.27^{+ 0.40}_{- 0.59}$&  $0.05^{+0.17}_{-0.05}$&  1.44  &  $1.46^{+0.05}_{-0.06}$&  $0.00^{+0.09}_{-0.00}$&  1.81 \\[2pt]
    NGC 4710 &    $0.71^{+0.06}_{-0.06}$&  $12.31^{+ 0.31}_{- 0.40}$&  $0.00^{+0.06}_{-0.00}$&  1.08  &  $0.95^{+0.03}_{-0.02}$&  $0.00^{+0.02}_{-0.00}$&  1.72 \\[2pt]
   PGC 44931 &    $1.04^{+0.19}_{-0.16}$&  $12.94^{+ 0.36}_{- 0.41}$&  $0.00^{+0.17}_{-0.00}$&  1.75  &  $1.66^{+0.13}_{-0.12}$&  $0.00^{+0.06}_{-0.00}$&  2.79 \\[2pt]
ESO 443-G042 &    $1.21^{+0.45}_{-0.57}$&  $11.95^{+ 1.22}_{- 2.21}$&  $0.47^{+0.28}_{-0.47}$&  2.05  &  $1.68^{+0.19}_{-0.18}$&  $0.29^{+0.29}_{-0.29}$&  2.10 \\[2pt]
    NGC 5746 &    $1.23^{+0.07}_{-0.05}$&  $13.35^{+ 0.18}_{- 0.22}$&  $0.00^{+0.02}_{-0.00}$&  2.13  &  $1.64^{+0.04}_{-0.03}$&  $0.00^{+0.00}_{-0.00}$&  3.08 \\[2pt]
     IC 4767 &    $1.36^{+0.18}_{-0.16}$&  $11.45^{+ 0.72}_{- 1.89}$&  $0.24^{+0.14}_{-0.23}$&  1.23  &  $1.51^{+0.13}_{-0.12}$&  $0.17^{+0.17}_{-0.17}$&  1.39 \\[2pt]
    NGC 6722 &    $1.14^{+0.10}_{-0.09}$&  $12.72^{+ 0.27}_{- 0.36}$&  $0.14^{+0.09}_{-0.14}$&  2.22  &  $1.47^{+0.06}_{-0.06}$&  $0.20^{+0.08}_{-0.11}$&  2.68 \\[2pt]
    NGC 6771 &    $1.12^{+0.12}_{-0.12}$&  $13.30^{+ 0.68}_{- 0.81}$&  $0.29^{+0.11}_{-0.18}$&  2.05  &  $1.31^{+0.10}_{-0.10}$&  $0.12^{+0.17}_{-0.12}$&  2.34 \\[2pt]
ESO 185-G053 &    $1.03^{+0.10}_{-0.10}$&  $12.22^{+ 0.81}_{- 1.04}$&  $0.34^{+0.09}_{-0.14}$&  1.73  &  $1.15^{+0.08}_{-0.08}$&  $0.21^{+0.11}_{-0.18}$&  1.93 \\[2pt]
     IC 4937 &    $1.07^{+0.25}_{-0.19}$&  $11.45^{+ 0.90}_{-\ldots}$&  $0.18^{+0.21}_{-0.18}$&  1.35  &  $1.21^{+0.13}_{-0.11}$&  $0.08^{+0.28}_{-0.08}$&  1.38 \\[2pt]
ESO 597-G036 &    $1.31^{+0.41}_{-0.41}$&  $13.30^{+ 0.86}_{- 1.35}$&  $0.37^{+0.26}_{-0.37}$&  0.96  &  $1.87^{+0.23}_{-0.20}$&  $0.00^{+0.40}_{-0.00}$&  1.44 \\[2pt]
     IC 5096 &    $1.06^{+0.10}_{-0.09}$&  $12.62^{+ 0.50}_{- 0.59}$&  $0.00^{+0.06}_{-0.00}$&  1.28  &  $1.29^{+0.05}_{-0.06}$&  $0.00^{+0.05}_{-0.00}$&  1.70 \\[2pt]
ESO 240-G011 &    $2.41^{+0.14}_{-0.14}$&  $10.56^{+ 0.90}_{-\ldots}$&  $0.00^{+0.06}_{-0.00}$&  1.76  &  $2.49^{+0.08}_{-0.08}$&  $0.00^{+0.08}_{-0.00}$&  1.76 \\[2pt]
\\
\multicolumn{8}{c}{Control sample} \\
\\
    NGC 1032 &    $1.10^{+0.18}_{-0.12}$&  $12.85^{+ 0.63}_{- 0.90}$&  $0.00^{+0.21}_{-0.00}$&  0.98  &  $1.34^{+0.16}_{-0.11}$&  $0.08^{+0.18}_{-0.08}$&  1.42 \\[2pt]
    NGC 3957 &    $1.35^{+0.12}_{-0.12}$&  $12.27^{+ 0.49}_{- 0.63}$&  $0.00^{+0.21}_{-0.00}$&  1.22  &  $1.65^{+0.07}_{-0.06}$&  $0.00^{+0.17}_{-0.00}$&  1.63 \\[2pt]
    NGC 4703 &    $0.83^{+0.10}_{-0.09}$&  $12.40^{+ 0.36}_{- 0.49}$&  $0.09^{+0.23}_{-0.09}$&  1.07  &  $1.06^{+0.08}_{-0.07}$&  $0.12^{+0.21}_{-0.12}$&  1.65 \\[2pt]
    NGC 5084 &    $0.89^{+0.04}_{-0.05}$&  $13.03^{+ 0.49}_{- 0.59}$&  $0.00^{+0.06}_{-0.00}$&  1.18  &  $1.02^{+0.04}_{-0.03}$&  $0.00^{+0.08}_{-0.00}$&  1.68 \\[2pt]
    NGC 7123 &    $0.84^{+0.07}_{-0.08}$&  $12.62^{+ 0.81}_{- 1.12}$&  $0.09^{+0.15}_{-0.09}$&  1.18  &  $0.95^{+0.05}_{-0.05}$&  $0.06^{+0.14}_{-0.06}$&  1.43 \\[2pt]
     IC 5176 &    $1.02^{+0.35}_{-0.29}$&  $12.49^{+ 1.17}_{-\ldots}$&  $0.34^{+0.23}_{-0.34}$&  1.38  &  $1.33^{+0.11}_{-0.10}$&  $0.06^{+0.29}_{-0.06}$&  1.44 \\[2pt]
\hline
\end{tabular*}
\begin{minipage}{\textwidth}
\emph{Notes}. All values are computed assuming the distances in
Table~\ref{tab:sample} and are presented in solar units at $K_S$-band,
\hbox{\ensuremath{(M/L)_{K_S,\odot}}} Errors are the formal fitting errors at the $3\,\sigma$
confidence level and neglect the uncertainties discussed in
\refsec{sec:uncertainties}. If the lower error on halo mass $M_{\rm DM}$
is denoted with ellipsis, e.g. $\log M_\mathrm{DM} = 12.53^{+1.22}_{ -
\ldots}$, then for this galaxy there is no lower bound on $M_\mathrm{DM}$
because $M_\mathrm{DM} = 0$ cannot be ruled out at the $3\,\sigma$
level. In such cases the quoted value for $M_\mathrm{DM}$ is highly
unconstrained up to some upper limit, and should be used with caution.
$\chi_\mathrm{red}$ and $\chi_\mathrm{red,\,nohalo}$ are the reduced $\chi$
for the models with and without dark haloes, i.e.
$(\chi^2/\mathrm{DOF})^{1/2}$ where $\chi^2$ is defined in
\refeq{eqn:chi2} and $\mathrm{DOF}$ is the number of degrees of
freedom. 
\end{minipage}
\end{table*}

\section{Discussion}
\label{sec:discussion}

The form of the observed $v_\mathrm{rms}$ as a function of radius is
usually a double-humped curve (with the inner hump reaching a smaller
speed), but sometimes a monotonically rising curve reaching a plateau
toward the edge of the disc (rather like a typical observed rotation
curve). For example, the observed second moment rises monotonically in
NGC~1886 and IC~5176, is almost constant in NGC~1032 and NGC~5084, falls
before rising again in NGC~1381 and NGC~2310, and rises then falls then
rises again in NGC 3203 and NGC~6771. The position, height, depth and
number of bumps is different for each galaxy. The first thing to note,
therefore, is that the models are able to reproduce this whole range of
behaviours. This is demonstrated by $\chi_\mathrm{red}$, the square root
of the reduced $\chi^2$ of the best-fitting models (see
Table~\ref{tab:upsilons}). The median $\chi_\mathrm{red}$ is 1.35 with
an rms scatter of 0.40.

These low figures of merit are neither trivial given the wide range of
kinematics observed, nor necessarily expected given the simplicity of
the models. Although the MGE technique allows for accurate
parametrization of the $K_S$-band photometry, the deprojected mass
models are axisymmetric descriptions of objects that, in most cases, we
have good reason to believe are in fact barred
\citep[see][]{Bureau:1999,Chung:2004}. Our one-parameter description of
the dark halo is model dependent, and our total model contains only
three free parameters. Despite this, we are able to accurately reproduce
the wide range of observed second velocity moment profiles as far as
outermost data point, which is typically at $R_\mathrm{max} \approx$
2--3\,\Reff{} or, equivalently, $R_\mathrm{max} \approx$
0.5--1\,$R_{25}$ (see Table~\ref{tab:sample}). Because of the lack of
freedom that we have to vary the \emph{shape} of the predicted kinematics, it
seems that a great deal of information about the kinematical structure
of these early-type disc galaxies is contained in their photometry
alone. This result is consistent with that of \cite{Cappellari:2008}: 

Given the extent to which our models reproduce the observed kinematics,
we are justified to consider that they correspond in some way to the
intrinsic properties of the galaxies. We defer our discussion of the
possible systematics introduced by our assumptions to
\refsec{sec:modeluncertainties}.

\begin{figure}
\includegraphics[width=8.4cm]{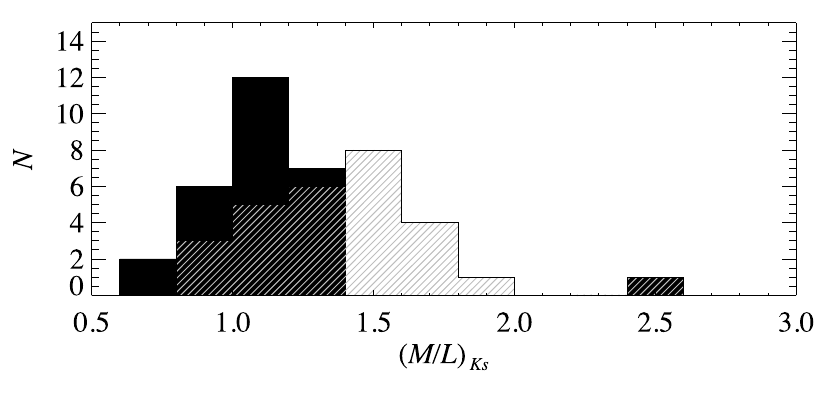}
\includegraphics[width=8.4cm]{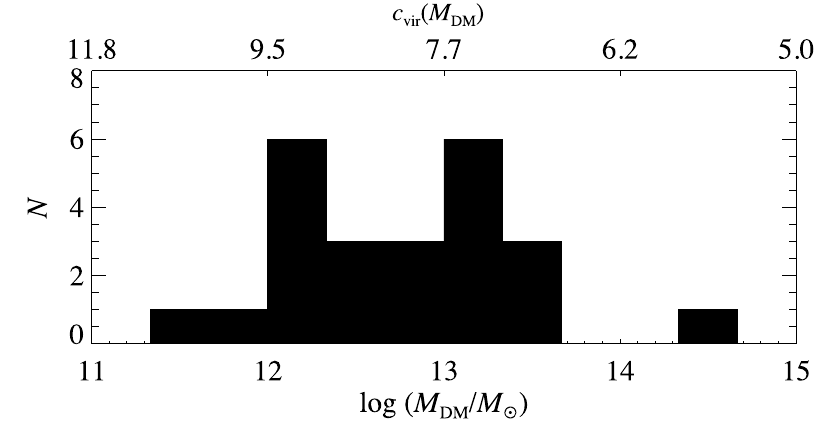}
\includegraphics[width=8.4cm]{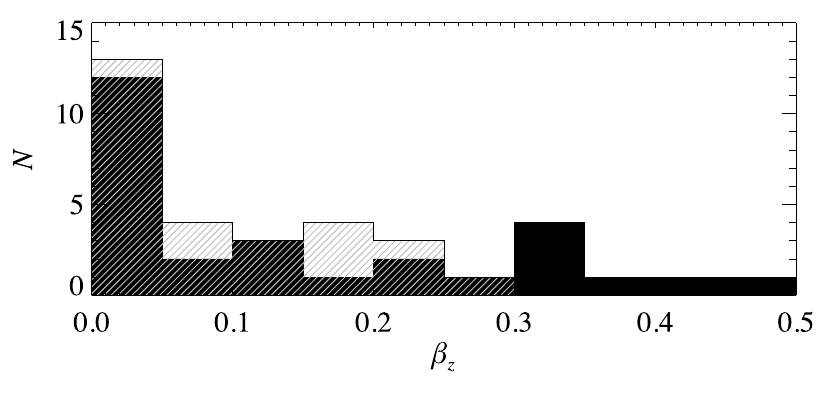}
\caption{Distributions of the parameters of the best-fitting models with
a dark halo (solid black) and without a dark halo (hatched white) in
Table~\ref{tab:upsilons}. We do not show halo masses where there is no
lower mass limit. Corresponding values of concentration
$c_\mathrm{vir}(M_{\rm DM})$ are shown on the upper horizontal axis of
the halo mass histogram.}
\label{fig:bestparams}
\end{figure}

The following discussion of the model parameters is separated into two
parts. In the first, we discuss the properties of the best-fitting mass
models that include a dark halo. In the second, we discuss those
without. We present the three parameters, $(M/L)_{K_S}$, $M_\mathrm{DM}$
and $\beta_z$, of the models with dark haloes and the two parameters,
$(M/L)_{K_S}$ and $\beta_z$, of the models without in
Table~\ref{tab:upsilons} and \reffig{fig:bestparams}. We precede this
discussion by noting than none of them appears to correlate with the
absolute magnitude $M_{K_S}$, the stellar mass or the Hubble type.

\subsection{Models with dark haloes: mass-to-light ratios}

We find a median $K_S$-band mass-to-light ratio $(M/L)_{K_S} = 1.09$
with an rms scatter of 0.31. As we noted in the introduction, the
mass-to-light ratios are themselves interesting because there are
relatively few dynamical measurements of stellar $K$-band mass-to-light
ratios and they are important for normalizing and constraining stellar
population models in the near-infrared. We therefore compare our values
to other available dynamical and stellar population-based measures. Of
course there have been many measurements of $(M/L)$ ratio at wavebands
other than $K_S$ that can be transformed to $K_S$ using a global colour.
To minimize the uncertainties associated with such transformations, we
restrict our discussion to direct determinations at $H$, $K$ or, almost
equivalently, $K_S$. $K - K_S \lesssim 0.03$\,mag (the exact value
depends on which $K$-band filter is used). For a given stellar mass,
this is equivalent to a difference between $(M/L)_{K_S}$ and $(M/L)_K$
of less than 3 per cent. We neglect this difference in the discussion
that follows.

\subsubsection{Comparison to previous dynamical measures of stellar $(M/L)_K$}

\cite{Devereux:1987} made among the earliest determinations of
near-infrared mass-to-light ratios using nuclear dispersions and
1.65\,$\mu$m (i.e. $H$-band) photometry. For 72 bright spiral, S0 and
elliptical galaxies, they found a mean $(M/L)_H$ = 0.94 (where we have
updated the value they adopted for $M_{\odot,H}$ with a more recent
determination). For a constant characteristic $H - K_S$ color of
0.28\,mag \citep{Jarrett:2003}, this implies $(M/L)_{K_S} \approx 0.7$.
\cite{Kranz:2003} determined maximal disc $(M/L)_K$ ratios by comparing
the results of hydrodynamical simulations to the observed spiral
patterns in spiral galaxies. For a sample of five high surface
brightness late-type spiral galaxies, they find a mean $(M/L)_K \approx
0.6$. In the case of the small elliptical galaxy Cen~A ($\sigma_{\rm e}
\approx 140$\,km\,s$^{-1}$), there are two independent dynamical
measurements of $(M/L)_K$ \citep{Silge:2005,Cappellari:2009}, both of
which are $\approx 0.7$. We therefore find that our direct dynamical
measures of $(M/L)_{K_S}$, which we believe accurately account for dark
matter, are somewhat larger than most previous dynamical estimates at
around the 1--$2\,\sigma$ significance level. Our method depends on an
assumed single-parameter NFW dark halo, rather than the two-parameter
isothermal halo used by \cite{Kranz:2003}. It is however somewhat more
direct and avoids the maximal disc assumption.

\subsubsection{Comparison to previous stellar population measures of
$(M/L)_K$}

\begin{figure}
\includegraphics[width=8.4cm]{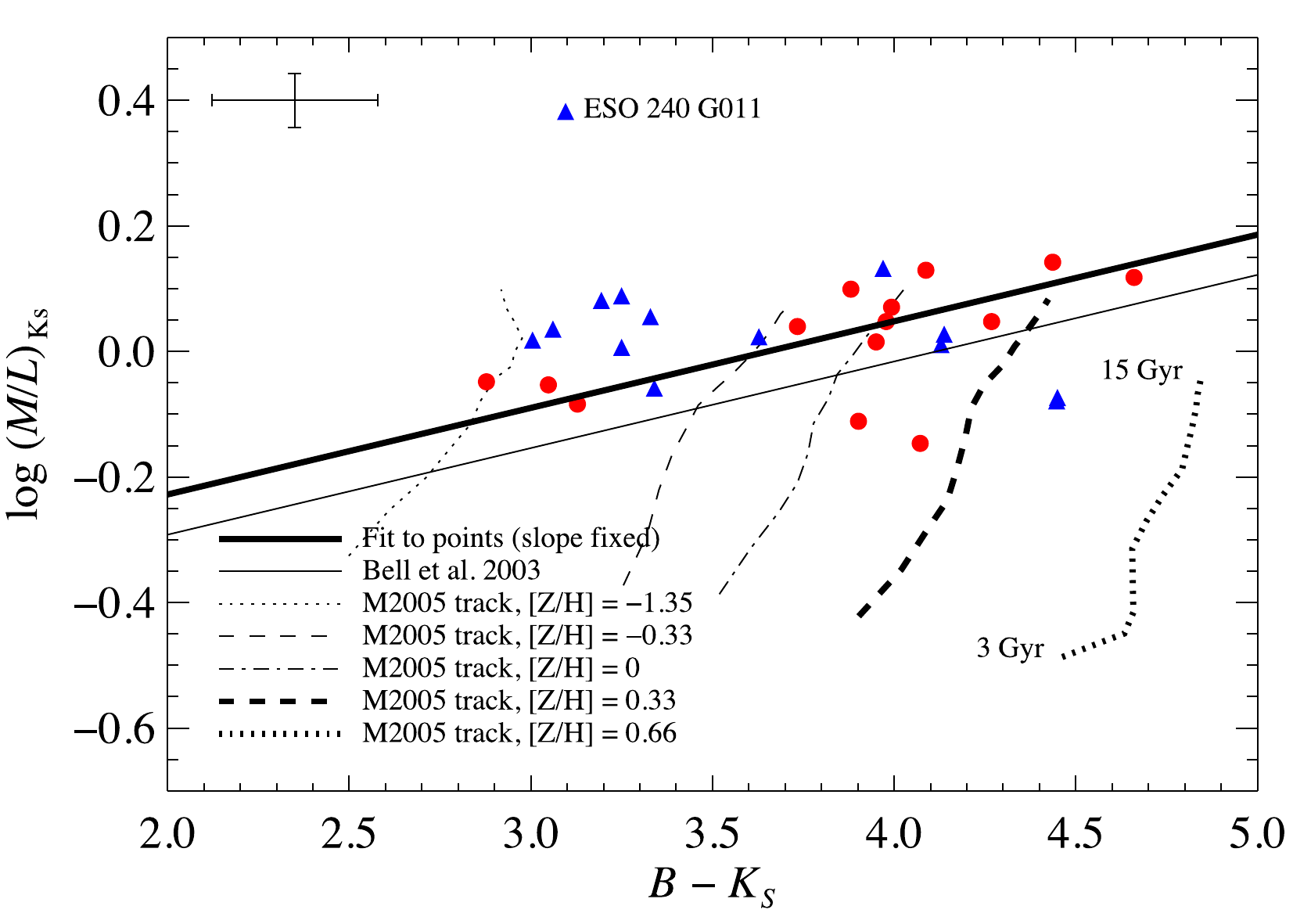}
\caption{Stellar $(M/L)_{K_S}$ against $B-K_S$ colour. The data points 
use the $K_S$-band mass-to-light ratios of the stellar components of the
best-fitting mass-models including a dark halo, i.e. column (2) of
Table~\ref{tab:upsilons} and $B$-band magnitudes, corrected for Galactic
and internal extinction, are taken from HyperLEDA. $K_S$-band magnitudes
are derived from our MGE models which incorporate a correction for
Galactic extinction. Internal extinction at $K$-band is neglected. A
median error bar that does not include the uncertainties introduced by
the extinction corrections is shown in the upper left corner (see text).
Symbols are as in \reffig{fig:appmag}. The thin straight line is the
prediction of the spectrophotometric galaxy evolution models of
\protect\cite{Bell:2003}. The thick straight line is a fit to our data
constrained to the slope of the dashed line. The evolutionary tracks for
a range of metallicities are from the models of
\protect\cite{Maraston:2005} and run from 3\,Gyr (low mass-to-light
ratio) to 15\,Gyr (high mass-to-light ratio). Metallicity increases from
left to right for these tracks.} \label{fig:colorml}
\end{figure}

If our modelling assumptions are justified, then because mass is
correctly shared between the luminous and dark components, the
$(M/L)_{K_S}$ ratios that we measure should agree with those predicted
by well-calibrated stellar population models. A forthcoming paper will
present an extensive stellar population analysis of the galaxies in our
sample using absorption line strengths and a more complete comparison to
the predictions of models. For now though, we perform a preliminary
check by comparing our results to the predictions of two models. In
\reffig{fig:colorml} we show the dynamically determined stellar
mass-to-light ratios of our sample galaxies as a function their $B -
K_S$ colour. We also show the predictions of two independent stellar
population models.

We use the $B$-band magnitudes given in HYPERLEDA, which are corrected
for both Galactic and internal extinction. We adopt the uncertainties in
the uncorrected $B$-band magnitude, which neglect random and systematic
errors in the extinction corrections and are therefore lower limits. The
Galactic correction is derived from the dust maps of
\cite{Schlegel:1998}. The median Galactic correction at $B$ is 0.2\,mag
and the uncertainties are probably rather smaller than this.
\mbox{ESO~311-G012} is close to the equatorial plane of the Galaxy
(Galactic latitude $b =
-8.0^\circ$) so the Galactic correction for that galaxy is very large
(1.7\,mag) and is derived from a region of the \cite{Schlegel:1998} maps
where extinction corrections are unreliable ($b \lesssim 5^\circ$). The
$B$-band internal extinction correction is based on the statistical
method of \cite{Bottinelli:1995}, which parametrizes the correction as a
function of axial ratio (and therefore inclination) and bulge-to-disc
ratio (and therefore morphological type). In exactly edge-on spiral
galaxies, the corrections used are both large (0.8--1.5\,mag) and somewhat
dubious. The $K_S$-band magnitudes include a small correction for
Galactic extinction (the median correction is 0.02\,mag) and neglect
internal extinction. The effect on the parameters of the model of
neglecting internal extinction in the $K_S$-band is discussed in
\refsec{sec:dustuncertainties}, but as far as the $B - K_S$ colour is
concerned this should introduce a systematic but small error (perhaps
0.1\,mag in the spirals and less in the S0s). We therefore conclude that
the uncertainties in the $B - K$ colours are dominated by large
uncertainties in the corrections to the $B$-band magnitude, especially
the internal correction.

The first prediction we compare our results to is the colour--$(M/L)$ relation of
\cite{Bell:2003}. This relation is derived from spectrophotometric
galaxy evolution models that use the \textsc{pegase} stellar population
model (see \citealt{Fioc:1997} and \citealt{Bell:2003}) with a Salpeter
initial mass function (IMF) modified by globally scaling down the
stellar mass by a factor of 0.7. \cite{Bell:2003} do not present
relations involving near-infrared colours, so we adopt a characteristic
$R - {K_S}$ colour of 2.2\,mag to derive the line shown in
\reffig{fig:colorml} from their $(B - R)$--$(M/L)_{K_S}$ relation,
yielding $\log(M/L)_{K_S} = 0.138 - 0.568 (B - {K_S})$. \cite{Bell:2001}
do use near-infrared colours, but their relations are based on a small
selection of ages and metallicities rather than the distribution
observed in the local universe. This leads them to derive a rather
flatter relation than the ones shown in \cite{Bell:2003}.

We show in \reffig{fig:colorml} a further comparison to an independent
set of stellar population models from \cite{Maraston:2005} that predict
$K$-band mass-to-light ratios for stellar populations with ages less
than 15\,Gyr and metallicities $-1.33 < [\mathrm{Z/H}] < 0.67$. The
models include a semi-empirical treatment of the thermally pulsing
asymptotic giant branch, to which the near-infrared luminosity is
particularly sensitive. We show models assuming a Kroupa IMF. These
tracks can be transformed to predictions for the scaled down Salpeter
IMF of \cite{Bell:2003} and the normal Salpeter by adding 0.09\,dex and
0.24\,dex respectively to the logarithm of the $(M/L)$ ratio. In this
case we show evolutionary tracks for a range of metallicities rather
than a statistical characterisations of local galaxies. 

As the figure demonstrates, the predictions of the \cite{Bell:2003}
galaxy evolution models are below our dynamical determinations in all
but four cases. A power law fit to our models, constrained to the same
slope as the \cite{Bell:2003} line and from which the outlier
ESO~240-G11 is excluded, implies a systematic offset of $\Delta
\log(M/L)_{K_S} = 0.06$\,dex. At least part of this small offset could
be due to the introduction of a systematic error in the approximate
transformation of the \cite{Bell:2003} to $B-K_S$. Our measurements are
systematically toward the upper end, if not outside, the range of values
predicted by the \cite{Maraston:2005} models. If the stellar population
models are correct, this implies systematically old stellar populations,
$\gtrsim 10$\,Gyr. These differences are intriguing and could suggest
problems in the stellar evolution models or the IMFs adopted, but we
note four reasons which lead us to argue that our findings are broadly
consistent with their predictions. Firstly, there is a random scatter of
$0.1$--\,$0.2$\,dex in the \cite{Bell:2003} relation (larger at the blue
end). Secondly, there may be systematic errors in the $(B-K_S)$ colours
adopted for our sample of edge-on galaxies due to internal extinction.
These are difficult to quantify. Thirdly, in the case of the
\cite{Maraston:2005} tracks, the mass-to-light ratios are for single
stellar population models, while many galaxies, especially spirals, are
likely to have composite stellar populations. Finally, our sample of 28
bright early-type disc galaxies is not necessarily representative of
local galaxies. We will perform a more detailed comparison on the models
and data in a future work. 

We end this discussion of the $(M/L)_{K_S}$ ratios measured by noting
the existence of a significant outlier galaxy, ESO~240-G011, which has
$(M/L)_{K_S} = 2.41\pm0.14$. This is more than four standard
deviations above the median $(M/L)_{K_S}$ of the sample. ESO~240G11 has
the latest Hubble type in the sample, and later galaxies are generally
thought to be more dark matter-dominated. However, our modelling method
aims to correctly account for dark matter so $(M/L)_{K_S}$ should be the
mass-to-light ratio of the stellar component, not the total matter
distribution. This particular galaxy therefore remains puzzling.

\subsection{Models with dark haloes: dark halo masses}
\label{sec:discusshaloes}

For those galaxies with constrained halo masses, we find a median
$M_\mathrm{DM} = 10^{12.85}\,M_\odot$ with an rms scatter of 0.7\,dex.
Using the concentration--halo mass correlation of \cite{Maccio:2008},
which our haloes are constrained to lie on, this corresponds to $c_{\rm
vir} = 7.9$ with an rms scatter of 1.2. These results should be treated
with caution, however, because most of the dark mass lies beyond radii
at which our models are constrained by either photometry or kinematics.
$M_{\rm DM}$ is therefore strongly dependent on the assumptions
described in Section~\ref{sec:halo}. 

\begin{figure}
\includegraphics[width=8.4cm]{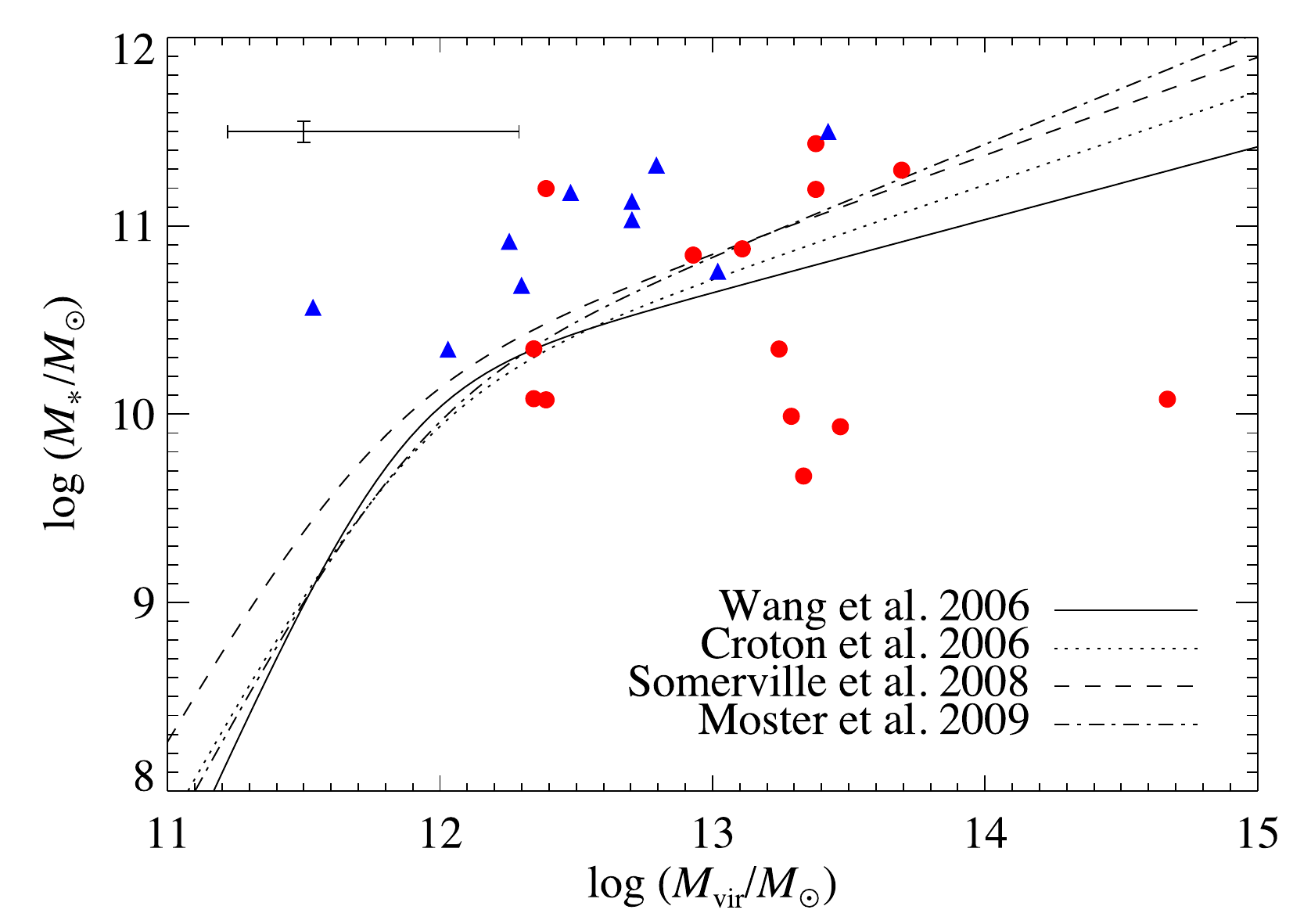}
\caption{Comparison of the stellar and halo masses of our mass models
(data points) to the independent predictions of \protect\cite{Wang:2006},
\protect\cite{Croton:2006}, \protect\cite{Somerville:2008} and
\protect\cite{Moster:2009}.
Median error bars are shown in black in the upper left corner. Symbols are
as in \reffig{fig:appmag}.}
\label{fig:mvirmstar}
\end{figure}

With this strong caveat in mind, we note, however, that the virial
masses $M_{\rm vir} = 1.2\,M_{\rm DM}$ of the models are at least
consistent with the predictions of semi-analytic and halo occupation
distribution models with respect to their stellar mass. We present a
comparison in \reffig{fig:mvirmstar}. \cite{Croton:2006} and
\cite{Somerville:2008} use semi-analytic models of galaxy formation
that, among other things, attempt to predict the relationship between
the stellar mass and dark halo mass of a galaxy by modelling relevant
physical processes such as the growth of structure, cooling, star
formation and feedback. \cite{Wang:2006} and \cite{Moster:2009} use a
halo occupation distribution approach to populate a halo catalogue with
stellar mass. The halo catalogue is taken from \cite{Springel:2005}. It
is populated to reproduce the local observed stellar mass function and
clustering function.

Our models do not agree with the predictions in a few cases, but there
is no evidence of any systematic offset. The stellar and virial masses
of our mass models do not seem to be correlated. This is not true of the
predictions, especially below $M_\mathrm{vir} = 10^{12}\,M_\odot$, which
we unfortunately do not probe. The comparison is therefore limited by
both a lack of mass range in our sample, and the significant
uncertainties in our halo mass determinations. We can only conclude that
our galaxies have stellar mass to halo mass ratios consistent with the
range predicted by the theoretical models for these halo masses.
However, the assumptions and observational constraints used to make
their predictions are quite independent from ours, so even this
relatively weak statement is not trivial. Ignoring the predictions, we
note that the spiral galaxies in our sample all have large
$M_*/M_\mathrm{vir}$ ratios relative to the average S0.

The dark matter enclosed within a sphere of radius $r$ is given by
\begin{equation}
M_{\rm DM}(r) = M_\mathrm{DM} \frac{A(r/r_\mathrm{s})}{A(c_\mathrm{vir})}.
\end{equation}
From this and the ratios of the circular velocities of the dark and
stellar components, we can calculate the dark-to-total mass fraction
$X_\mathrm{DM}$ as a function of radius. We find a mean $X_{\rm
DM}(\Reff{})$ of 15 per cent (10 per cent rms scatter) and a mean
$X_{\rm DM}(R_{25})$ of 49 per cent (15 per cent scatter). As we
mentioned in the introduction, a maximal disc is strictly defined as one
in which the stellar component contributes 75--95 per cent of the
rotational velocity at $2.2\,R_\mathrm{disc}$, i.e.
$X_\mathrm{DM}(2.2\,R_\mathrm{disc}) \le 45$ per cent
\citep{Sackett:1997} (2.2\,$R_\mathrm{disc}$ is the radius at which the
rotation curve of a thin disc reaches its maximum). The disc scale length
of edge-on, barred systems are uncertain \citep{Bureau:2006}, but our
estimates of $R_\mathrm{disc}$ suggest that $X_{\rm
DM}(2.2\,R_\mathrm{disc})$ is 27 per cent (11 per cent scatter). The
best fitting mass model is sub-maximal in only two cases [NGC~2310,
$X_\mathrm{DM}(2.2\,R_\mathrm{disc}) = 46$ per cent and PGC~44931,
$X_\mathrm{DM}(2.2\,R_\mathrm{disc}) = 50$ per cent].

We can also determine the radius at which dark and stellar matter
contribute equally to the circular velocity and mass enclosed within a
sphere. We do this in units of \Reff{} and $R_{25}$ and find $R(X_{\rm
DM} = 0.5)/\Reff{} = 4.1$ (with rms scatter of 1.3) and $R(X_{\rm DM} =
0.5)/R_{25} = 1.0$ (with rms scatter of 0.5). These results should be
treated with a little caution, however, because in several cases
$R(X_{\rm DM} = 0.5) > R_{\rm max}$, the last kinematic data point.
Beyond the kinematics, the models are obviously particularly model
dependent. 

As discussed in \refsec{sec:randomuncertainties}, $X_\mathrm{DM}$ and
$R(X_\mathrm{DM}=0.5)$ are almost unaffected by distance or photometric
calibration errors. Finally, we note that there is no evidence of any
significant difference between the dark matter content of the spirals
and S0s in our sample, at least at the radii within which the models are
constrained by the kinematic data.

\subsection{Models with dark haloes: anisotropies}
\label{sec:anisotropy}

The best-fitting anisotropies for our models are in the range $0.0
\lesssim \beta_z \lesssim 0.5$. As we discussed in \refsec{sec:jeans},
however, the constraints on these measurements are weak because the
galaxies in our sample are rotation-dominated and we do not have
integral field data. Moreover, we assume a constant anisotropy
throughout the galaxy, which is of course a significant simplification,
since many of our galaxies have a bulge, bar and disc. With these
caveats in mind, the typical anisotropy we find is broadly consistent
with the average of previous observations of ellipticals, that are
expected to have similar dynamics to the bulges ($\beta_z \lesssim 0.4$,
\citealt{Cappellari:2007,Thomas:2009}) and discs ($0.35
\lesssim \beta_z \lesssim 0.75$, \citealt{Shapiro:2003}).

\subsection{Neglecting dark matter}

As discussed in the introduction, previous studies have shown that it is
usually possible to construct mass models of bright spiral galaxies that
reproduce the general form of their rotation curves within $R_{25}$ with
no or only a sub-dominant contribution from dark matter. It is therefore
not surprising that, in some cases, the improvement in fit quality when
a dark halo added is not overwhelming. 

How many galaxies can we fit without a dark halo? Many of the second
velocity moments of the best mass models without a dark halo are, by eye
at least, satisfactory fits to the observed kinematics (compare the thin
red lines to the data points in the lower right plot of each panel in
Figures~\ref{fig:results} and \ref{fig:control}). In the formal sense,
in four cases (NGC~1886, IC~4937, ESO~240-G011, and IC~5176) the removal
of the dark halo does not improve the fit at the $3\,\sigma$ level. 

We therefore now discuss an Occam's Razor argument which states that,
since we \emph{can} fit some of these galaxies well without dark matter,
dark matter is not a significant component by mass within the radii
probed by our kinematic data (0.5--1\,$R_{25}$ or 2--3\,\Reff{}) in at
least those cases. We reject this argument because the parameters of our
best fitting mass models with dark matter are in line with independent
expectations, whereas those for the haloless models are not. For
example, the mass-to-light ratios we measure for mass models without
haloes are inconsistent with stellar population results. We find the
median $(M/L)_{K_S,\,\mathrm{nohalo}} = 1.25(M/L)_{K_S}$, where
$(M/L)_{K_S,\,\mathrm{nohalo}}$ is the stellar mass to light ratio
measured using a mass model without a dark halo. The $(M/L)_{K_S,\,\rm
nohalo}$ we measure is therefore significantly higher than the
predictions the stellar population models of \cite{Bell:2003} and
\cite{Maraston:2005}. Either the stellar population models and their
assumptions are flawed, or the stellar mass models without dark matter
have too much mass. The halo sizes (and concentrations) of the models
with dark matter also match the expectations of galaxy formation. Of
course, adding another free parameter to a model is always going to
improve the quality of the fit, but if the true halo of a galaxy is not
significant (or not present), the agreement between the halo
masses that most improves the fits and the predictions and the
expectations of $\Lambda$CDM would be a striking coincidence.

Indeed, the fact that the $(M/L)$ ratio of a mass model including only
stars systematically exceeds that of a mass model accounting for dark
matter or a stellar population estimate can be used to estimate how
`wrong' the former is. The median $(M/L)_{K_S, {\rm nohalo}}$ must be
\emph{decreased} by 20 percent [$\Delta \log (M/L) \approx -0.10$] to
match $(M/L)_{K_S}$, the dynamically determined stellar mass-to-light
ratio with dark matter. This is similar to the finding of
\cite{Cappellari:2006}, that the median $I$-band dynamical $(M/L)$ ratio
neglecting dark matter must be decreased by 30 per cent [$\Delta \log
(M/L)_I \approx -0.15$] to match the stellar population $(M/L)$ ratio.

\subsection{Uncertainties and assumptions}
\label{sec:uncertainties}

\subsubsection{Random errors}
\label{sec:randomuncertainties}

The errors presented in Table~\ref{tab:upsilons} are $3\,\sigma$ formal
fitting errors. These neglect random errors due to distance
uncertainties and photometric calibration. As shown in
Table~\ref{tab:sample}, we adopted surface brightness fluctuation
distance estimates in two cases (NGC~1381 and NGC~1596) and the Virgo
cluster distance in two others (NGC~4469 and NGC~4710). The
uncertainties on these estimates is likely to be $\approx 1$\,Mpc,
limited either by the surface brightness fluctuation method or the physical size of the Virgo
cluster. All other distance estimates are redshift-based estimates from
NED assuming a Virgocentric flow model \citep{Mould:2000}. The
uncertainties on these distances are much larger, probably $\approx 20$
per cent. Because of the calibration process described in
Appendix~\ref{app:2mass}, the error on the calibration of the $K_S$-band
images we use is of order that of the 2MASS survey
\citep{Skrutskie:2006}, which is negligible compared to the distance
uncertainties.

Random photometric and distance errors propagate linearly into the
uncertainty in total dynamical mass of the model. In the presence of a
dark halo these errors propagate into the stellar and dark parameters of
the mass model in a complex and non-linear way. However, we find
empirically that a given fractional error in photometric calibration or
distance propagates into the same random error in $(M/L)_{K_S}$ and
$M_\mathrm{DM}$ to within a factor of 2 or so. In fact, the formal
fitting errors overwhelm these observational uncertainties. This can be
seen from \reffig{fig:jamchi2}, where the $3\,\sigma$ confidence region
allows us to constrain $M_{\rm DM}$ within $\approx 1$--2\,dex only, a
much bigger uncertainty than the distance and photometry errors.

Despite the non-linear way in which photometric distance errors
propagate into the stellar and dark masses of the mass models, we note
that these errors almost cancel out in $X_\mathrm{DM}(R)$, the
dark-to-total mass fraction at radius $R$. For example, plausible
distance errors of 20 per cent result in changes in
$X_\mathrm{DM}(\Reff)$, $X_\mathrm{DM}(2.2\,R_\mathrm{disc})$ and
$X_\mathrm{DM}(R_{25})$ of less than 2 per cent.

\subsubsection{Systematic errors due to dust}
\label{sec:dustuncertainties}

We now discuss two sources of systematic error, both of which are due to
dust. Firstly, if the internal extinction at $K_S$-band in these
galaxies is significant, then the current modelled mass-to-light are
overestimates. If that absorption varies significantly across the
projected galaxies, then this would also affect the halo masses
determined. We believe that this effect leads to $\lesssim 10$ per cent
reduction in the total light detected at $K_S$-band, and probably much
less. Absorption at $K_S$-band it typically smaller than at optical
wavelengths by an order of magnitude. This is reflected in detailed
examination of the $R$-band \citep{Bureau:1999} and $K_S$-band
\citep{Bureau:2006} images of the sample. Dust lanes that are prominent
in the optical cannot usually be detected at $K_S$-band and half of the
sample are S0s, which do not have optical dust lanes at all, so are
likely to be totally unaffected by dust at near-infrared wavelengths. 

The second potential systematic effect is due to the positioning of the
slit used to measure the stellar kinematics. As described in
\cite{Bureau:1999} and \cite{Chung:2004}, in the dustiest, exactly
edge-on systems the long-slit had to be moved a little away from the
major axis. If the kinematics at this distance from the plane are
significantly different from that along the major axis, then it would
likely affect both the $(M/L)$ ratio and halo mass determinations. The
most likely result would be a systematic reduction of both parameters,
since both rotational velocity and dispersion typically fall with height
above the disc. In actual fact, however, the effect may be smaller than
this in the B/PS sample. B/PS bulges are believed to rotate
cylindrically, i.e. their kinematics do not vary with height
\citep{Jarvis:1987,Shaw:1993,Shaw:1993a,Fisher:1994,Donofrio:1999}.
If this is true then this effect should be small within the
B/PS bulges. More importantly, the sample was constructed to avoid
placing the slit above the plane. The dustiest galaxies in the sample
(e.g. NGC~5746) are in fact inclined at slightly less than $90^\circ$.
It was therefore almost always possible for the centre of the galaxy to
lie within the slit, without dust affecting observations much.

\subsubsection{Systematic errors due to model assumptions}
\label{sec:modeluncertainties}

Finally we discuss the systematic errors introduced by our modelling
technique, which assumes that the stellar mass-to-light ratio and
anisotropy are constant across each galaxy, that the halo is of the form
described in \refsec{sec:halo}, and, most importantly that the galaxy
is axisymmetric.

In the case of both $(M/L)_{K_S}$ and $\beta_z$, the best-fitting parameters
may be thought of as global averages for each galaxy. We are, however,
justified in assuming a constant value in both cases. The stellar
population mass-to-light ratio at $K_S$-band varies relatively little
between galaxies \citep{Bell:2003}, and rather less within them. As
discussed in \refsec{sec:jeans} and \refsec{sec:anisotropy}, our results
are largely insensitive to the anisotropy chosen because our galaxies
are rotation-dominated. Choosing a more sophisticated, varying
anisotropy would not change our conclusions. 

It is one of the goals of this study to offer constraints on halo masses
and dark matter fractions by \emph{assuming} a particular, theoretically
motivated, single-parameter halo model. We are effectively asking how
large haloes are, not what shape they have. If real haloes do not follow
the NFW profile then our estimates of its virial mass are unlikely to be
correct. We are encouraged, however, by the good correspondence between
the stellar masses of the mass models including a halo and the
predictions of stellar population models. This is strong circumstantial
evidence that our method correctly apportions mass to the stellar and
dark components, despite (or perhaps because of) the strong assumptions
we make about the halo.

Conclusions drawn from these axisymmetric models may be flawed if the
assumption of axisymmetry is significantly violated. In line with local
disc galaxy demographics, a majority of the galaxies in our sample are
thought to host bars (see \refsec{sec:sample}). The possible
consequences of the application of axisymmetric modelling to the present
sample (or indeed any representative local sample) are therefore a
concern. Indeed, there are hints of our modelling breaking down,
presumably due to a bar, in the inner $\approx 10$\,arcsec of some
galaxies, where the model second velocity moment overpredicts the
observations (e.g. NGC~1381, ESO~240-G011 and NGC~3957). None of the
models underpredicts the data in this region. This can easily be
explained by the sample selection. The B/PS bulges dominating our sample
are thought to be bars viewed side-on. As such, the orbits supporting
the bars are elongated perpendicular to the line-of-sight, resulting in
observed velocities smaller than the circular velocities of the
equivalent axisymmetric mass distribution. End-on bars, which lead to
the opposite effect, are systematically excluded from our sample as they
appear round. This slight but systematic failure of our models in the
central regions is therefore consistent with previous studies of this
sample \citep{Bureau:1999,Chung:2004,Bureau:2006} and others
\citep[e.g.][]{Kuijken:1995,Merrifield:1999} supporting a close
relationship between B/PS bulges and bars. 

This, however, does not mean that our axisymmetric mass models are not
appropriate. Firstly, although a bar distorts the surface brightness of
a galaxy, the potential and the shapes of the orbits themselves are
significantly less distorted \citep[see, e.g., figure 2\,(a) of][which
demonstrates for a particular mass model that the radial peak in the
axial ratio of the $x_1$ orbits supporting the bar is quite
narrow]{Bureau:1999a}. It is these orbital shapes that the observed
stellar kinematics depend upon. This probably explains why the region in
which the over-prediction occurs is much smaller than the B/PS bulge.
The effect of the bar, at least as far as our method is concerned, is
restricted to only the inner few arcseconds.

Secondly, there is no systematic difference between the fit quality of
haloless models in the inner regions of the galaxies with B/PS bulges
and the control sample of six galaxies with spheroidal bulges.
Unfortunately, we are limited by the rather size of the control sample.
Demonstrating a statistically significant correlation between $\mu_2$
being too large in the central few arcseconds and more reliable bar
diagnostics \citep[e.g.][]{Bureau:1999,Chung:2004} is therefore
unlikely.

Nevertheless, it seems that the pressing question is not so much why the
stellar kinematics of so many galaxies deviate slightly but
systematically in the central regions, but rather why Jeans modelling of
axisymmetric mass distributions reproduces the kinematics of barred
galaxies so well. The high quality of the kinematic fits perhaps
suggests that the $\mu_2$ profiles primarily trace the \emph{enclosed
masses} as a function of galactocentric radius and are fairly (but not
entirely) insensitive to the details of the dynamics. Because of this,
axisymmetric mass models that reproduce the `bulge-plateau-disc' surface
brightness profile often observed in this sample \citep[see
\reffig{fig:negfit} and][]{Bureau:2006} can also reproduce the second
velocity moment of intrinsically non-axisymmetric galaxies with the same
radial profile. Irrespective of the importance of pressure support in
our sample galaxies, $\mu_2$ therefore appears to be analogous to the
circular velocity in purely rotationally-supported systems.

We further note that there is no physically motivated evolutionary
scenario which would systematically lead to axisymmetric galaxies with
such a radial mass distribution, but such surface brightness plateaus
develop naturally in barred disc galaxies due to the angular momentum
and mass exchanges mediated by the bar \citep[see,
e.g.,][]{Bureau:2005}. In effect, observations of $\mu_2$ circularize
the intrinsic mass distribution so $\mu_2$ is simply too coarse a
measurement to properly constrain the internal dynamics of galaxies. To
do so requires knowledge of the full shape of the line-of-sight velocity
distribution. This approach was used by \cite{Chung:2004} for this
sample. 

In summary, we argue that the overall overwhelming quality of the fits
for galaxies with and without B/PS bulges provides confidence in the
reliability of the derived masses and mass-to-light ratios, despite the
likely non-axisymmetry of the inner regions of many of the galaxies. Our
results, however, are not inconsistent with observations and simulations
that demonstrate that these galaxies are barred. The application of the
\textsc{jam} technique to model galaxies of known non-axisymmetric
morphologies is of course the best way to confirm definitively the
validity of this argument.

\section{Conclusions and future work} \label{sec:conclusion}

We presented mass models for a large sample of spiral and S0 galaxies.
These models allowed us to constrain the stellar and dark matter content
of the sample galaxies. For each galaxy, the stellar mass distribution
was derived from near-infrared photometry under the assumptions of
axisymmetry and a constant stellar mass-to-light ratio. We added an NFW
dark halo and assumed a correlation between concentration and virial
mass. We solved the Jeans equations for the corresponding potential
under the assumption of constant anisotropy in the meridional plane. By
comparing the predicted second velocity moment to observed long-slit
stellar kinematics, we determined the best-fitting parameters of the
mass models. In some galaxies the observed second velocity moment rises
monotonically, in others it plateaus at small radii, and in others it
falls significantly before rising again. Despite this wide range of
observed behaviours, our simple models, with only three free
parameters (stellar mass-to-light ratio, dark halo mass and anisotropy),
are able to reproduce the observed kinematics very well. The observed
kinematics typically extend to 2--3\,\Reff{} or, equivalently,
0.5--1\,$R_{25}$, and $1 \lesssim \chi_\mathrm{red} \lesssim 2$ for all
galaxies.

For our sample of 14 spirals and 14 S0s, we find a median $(M/L)_{K_S}$ of
1.09\,$(M/L)_{K_S,\odot}$ with rms scatter 0.31. Our values are roughly
consistent with the small number of previous independent determinations
using different dynamical methods. The $K_S$-band mass-to-light ratios
that we measure are unique, however, because of the size of the sample
and the way we have attempted to correctly apportion mass between the
stellar and dark components of the galaxies, without resorting to either
a maximal disc assumption or results from stellar population models.
Because they do not depend on stellar population models, they can be
used to attempt to constrain the normalization and IMF of such models in
the near-infrared.

We also performed preliminary comparisons of our dynamical $(M/L)$
ratios to the predictions of two stellar population models: the
color--$(M/L)$ relations of \cite{Bell:2003}, which use the
\textsc{pegase} stellar population models, and evolutionary tracks in
color--$(M/L)$ space for a range of metallicities from
\cite{Maraston:2005}. The \cite{Bell:2003} prediction is offset from our
models by a small but systematic amount and some of our galaxies are
just outside the range expected by the \cite{Maraston:2005} models.
These differences could be due to systematic errors introduced in our
comparison, but may also hint at problems with the stellar population
models or the IMFs they assume. In a future work we will extend this
comparison by determining absorption line strength indices for the
present sample, allowing us to directly compare dynamical and stellar
population estimates of $(M/L)$ for individual galaxies.

Once accurately known, $K_S$-band mass-to-light ratios are particularly
useful for constraining the stellar mass budget of the universe.
Firstly, $K_S$ (or $K$) is the waveband at which the effects of dust on
observations of light from stars is minimized. Shorter wavelengths are
subject to absorption and longer wavelengths have contributions from hot
dust seen in emission. Secondly, $K$-band is dominated by light from the
sub-solar mass main sequence stars that dominate the total mass budget
(and trace the smooth potential) of galaxies. It therefore comes as no
surprise that, as found dynamically in this work and using stellar
population models \citep[e.g.][]{Bell:2003}, $(M/L)_{K_S}$ is a
relatively constant quantity in the local universe compared to the
$(M/L)$ ratio at $B$-band. We both find that $(M/L)_{K_S}$ varies by
$\lesssim$ 0.3\,dex (a factor of 2) across our samples. This is much
less than the variations observed at shorter wavelengths. The $K_S$-band
luminosity function can therefore be used as a reliable proxy (subject
to the small variation in $(M/L)_{K_S}$) for the stellar mass density
function \citep[e.g.][]{Bell:2003,Arnouts:2007,Devereux:2009}.

Our best-fitting mass models include NFW haloes with a median dark mass
within the virial radius of $M_{\rm DM} = 10^{12.85}\,M_\odot$ with rms
scatter of 0.7\,dex. With our adopted concentration--halo mass
correlation, this corresponds to a concentration $c_{\rm vir} = 7.9$
with rms scatter of 1.2. These parameters of the best-fitting dark
haloes are defined in terms of their behaviour out to the virial
radius, which is well beyond our kinematic constraints. They are
therefore model dependent and should be treated with caution. The
dark-to-total mass fraction within the galaxy is, however, a
well-constrained quantity. We find that, on average, the haloes
contribute around 15 per cent by mass within \Reff{} and 49 per cent
within $R_{25}$. All but two galaxies are consistent with the maximal
disk assumption defined in \citealt{Sackett:1997}. Models without dark
matter are able to satisfactorily reproduce the observed kinematics in
most cases, although there are problems at large radii in several
galaxies. The improvement when a halo is added is statistically
significant in all but four cases and the stellar mass-to-light ratios
of mass models with dark haloes match the independent expectations of
stellar population models better.

There is no systematic difference between the dark matter content of the
S0, Sa and Sb galaxies in the sample. This hints at a homology between
S0s and spirals. \cite{Emsellem:2007} and \cite{Cappellari:2007} show
that elliptical and S0 galaxies exhibiting large-scale rotation (which
they call fast-rotators) constitute a homogeneous class in terms of
their shape, stellar kinematics and photometric properties. Assuming
this direct link between S0s and fast-rotating ellipticals, our
constraints on the dark matter content of Sa, Sb and S0 galaxies can in
turn be applied to fast-rotating ellipticals, implying a continuum of
properties from Sb spirals to fast-rotating ellipticals.
\cite{Bertola:1993} made a particularly interesting pioneering
suggestion when, with limited data, they suggested that spirals and
ellipticals share a common scale, 1.2\,\Reff{}, within which luminous
and dark matter contribute an equal amount to the total mass. For our
larger sample, which contains no elliptical galaxies, we find that this
radius is, on average, 4.1\,\Reff{} or 1.0\,$R_{25}$. 

The method we have used is generally applicable to axisymmetric (and
spherically symmetric) galaxies and the code we have used is public.
Although we have applied it to edge-on galaxies, it can equally be
applied to any galaxy with a well-constrained inclination. We assumed
that the galaxies are axisymmetric but argued that even in galaxies that
are probably barred the method gives sensible results. Application of
the method to simulated barred galaxies is needed to demonstrate this
definitively, however. Other limitations of our work include our
inability to constrain anisotropy and the well-motivated but
model-dependent way in which we eliminate one of the halo parameters.
Kinematics above and below the major axis of the galaxy could allow us
to lift these degeneracies without making assumptions, which would
provide even more robust observational tests of galaxy formation models
and simulations. We have acquired long slit spectroscopy at multiple
heights above the equatorial plane for a subset of the galaxies in
sample, which will allow us to try this idea, but the ideal method is of
course integral field spectroscopy from an instrument whose field of
view and spatial resolution are optimized to reach the radii at which
dark matter becomes important \citep[e.g.][]{Verheijen:2007}.

\section*{Acknowledgements}

We thank Aeree Chung for giving us access to the stellar kinematics of
the sample and Giuseppe Aronica for giving us access to the $K$n-band
images. We also thank Liam Cook and Davor Krajnovi\'c for participating
in an early feasibility study for this project and Victor Debattista,
Richard Ellis, Susan Kassin and John Magorrian for valuable comments. We
thank the anonymous referee for his/her helpful comments, which
encouraged to consider the role of dark matter in more detail. MJW is
supported by an STFC Postgraduate Studentship, MB by the STFC rolling
grant `Astrophysics at Oxford' (PP/E001114/1) and MC by an STFC Advanced
Fellowship (PP/D005574/1). This publication makes use of data products
from the Two Micron All Sky Survey, which is a joint project of the
University of Massachusetts and the Infrared Processing and Analysis
Center/California Institute of Technology, funded by the National
Aeronautics and Space Administration and the National Science
Foundation. We also acknowledge the use of use of the HYPERLEDA database
(\url{http://leda.univ-lyon1.fr}) and the NASA/IPAC Extragalactic
Database (NED) which is operated by the Jet Propulsion Laboratory,
California Institute of Technology, under contract with the National
Aeronautics and Space Administration. Numerical integration was
performed using Craig Markwardt's \textsc{qpint1d} numerical quadrature
code.

\appendix

\section{MGE model parameters}
\label{app:mge}

For each galaxy, the parameters of the best-fitting MGE parametrizations
of the projected light are presented in Table~\ref{tab:mge}.

\begin{table*}
\caption{MGE parameters for the deconvoled $K_S$-band surface brightness.}
\label{tab:mge}
\begin{tabular*}{\textwidth}{r@{\hspace{0.5em}}ccc@{\hspace{2em}}r@{\hspace{0.5em}}ccc@{\hspace{2em}}r@{\hspace{0.5em}}ccc@{\hspace{2em}}r@{\hspace{0.5em}}ccc}
\hline
Sign & $\log L_i$ & $\log\sigma_i$ & $q_i$ & Sign & $\log L_i$ & $\log\sigma_i$ & $q_i$ & Sign & $\log L_i$ & $\log\sigma_i$ & $q_i$ & Sign & $\log L_i$ & $\log\sigma_i$ & $q_i$ \\
& $(L_{K_S,\odot})$ & (arcsec) & & & $(L_{K_S,\odot})$ & (arcsec) & & & $(L_{K_S,\odot})$ & (arcsec) & & & $(L_{K_S,\odot})$ & (arcsec) & \\[1pt]
\hline
\multicolumn{16}{c}{B/PS sample} \\
\\
           \multicolumn{4}{c}{NGC 128} &       \multicolumn{4}{c}{ESO 151-G004} &           \multicolumn{4}{c}{NGC 1381} &           \multicolumn{4}{c}{NGC 1596} \\
\\
  $+$ &    4.334 &    0.182 &    0.596 &   $+$ &    4.093 &    0.341 &    0.368 &   $+$ &    4.624 &    0.012 &    0.766 &   $+$ &    4.782 &    0.067 &    0.598 \\
  $+$ &    3.866 &    0.709 &    0.596 &   $+$ &    4.977 &    0.820 &    0.291 &   $+$ &    4.143 &    0.480 &    0.766 &   $+$ &    4.127 &    0.481 &    0.598 \\
  $+$ &    4.026 &    1.199 &    0.421 &   $-$ &    4.976 &    0.825 &    0.286 &   $+$ &    3.946 &    0.970 &    0.505 &   $+$ &    5.774 &    0.781 &    0.437 \\
  $+$ &    4.790 &    1.280 &    0.367 &   $+$ &    3.170 &    1.125 &    0.154 &   $+$ &    2.434 &    1.078 &    1.000 &   $-$ &    5.794 &    0.783 &    0.430 \\
  $-$ &    5.097 &    1.286 &    0.355 &   $+$ &    2.818 &    1.198 &    0.303 &   $-$ &    3.925 &    1.109 &    0.305 &   $+$ &    4.518 &    0.808 &    0.359 \\
  $+$ &    4.746 &    1.303 &    0.335 &       &          &          &          &   $+$ &    3.702 &    1.303 &    0.168 &   $+$ &    3.540 &    0.974 &    0.685 \\
  $+$ &    1.986 &    1.651 &    0.400 &       &          &          &          &   $+$ &    3.765 &    1.445 &    0.351 &   $+$ &    3.214 &    1.360 &    0.208 \\
      &          &          &          &       &          &          &          &   $-$ &    3.917 &    1.446 &    0.338 &   $+$ &    2.791 &    1.636 &    0.219 \\
      &          &          &          &       &          &          &          &   $+$ &    3.455 &    1.479 &    0.286 & \\
      &          &          &          &       &          &          &          &   $+$ &    1.110 &    1.803 &    0.479 & \\
\\
          \multicolumn{4}{c}{NGC 1886} &           \multicolumn{4}{c}{NGC 2310} &       \multicolumn{4}{c}{ESO 311-G012} &           \multicolumn{4}{c}{NGC 3203} \\
\\
  $+$ &    4.392 &    0.016 &    0.331 &   $+$ &    4.823 &   -0.143 &    0.736 &   $+$ &    4.838 &   -0.023 &    0.681 &   $+$ &    4.724 &   -0.175 &    0.724 \\
  $+$ &    4.035 &    0.614 &    0.331 &   $+$ &    3.886 &    0.398 &    0.736 &   $-$ &    4.478 &    0.131 &    0.681 &   $+$ &    4.134 &    0.252 &    0.724 \\
  $-$ &    4.792 &    0.685 &    0.198 &   $+$ &    3.715 &    0.936 &    0.382 &   $+$ &    4.372 &    0.181 &    0.681 &   $+$ &    4.493 &    0.723 &    0.422 \\
  $+$ &    4.721 &    0.696 &    0.181 &   $-$ &    3.716 &    1.109 &    0.243 &   $+$ &    4.206 &    0.409 &    0.681 &   $-$ &    4.441 &    0.751 &    0.391 \\
  $+$ &    4.061 &    1.080 &    0.282 &   $+$ &    3.067 &    1.158 &    0.448 &   $+$ &    4.522 &    0.864 &    0.384 &   $+$ &    3.276 &    1.027 &    0.446 \\
  $-$ &    4.470 &    1.160 &    0.203 &   $+$ &    3.256 &    1.445 &    0.138 &   $-$ &    4.472 &    0.906 &    0.333 &   $+$ &    3.116 &    1.387 &    0.149 \\
  $+$ &    4.266 &    1.223 &    0.165 &   $+$ &    2.772 &    1.725 &    0.178 &   $+$ &    2.309 &    1.140 &    1.000 &   $+$ &    2.594 &    1.565 &    0.229 \\
  $+$ &    3.358 &    1.297 &    0.131 &       &          &          &          &   $+$ &    3.618 &    1.159 &    0.406 & \\
  $+$ &    2.400 &    1.717 &    0.138 &       &          &          &          &   $+$ &    3.211 &    1.517 &    0.100 & \\
      &          &          &          &       &          &          &          &   $+$ &    2.918 &    1.733 &    0.158 & \\
\\
          \multicolumn{4}{c}{NGC 3390} &           \multicolumn{4}{c}{NGC 4469} &           \multicolumn{4}{c}{NGC 4710} &          \multicolumn{4}{c}{PGC 44931} \\
\\
  $+$ &    4.640 &   -0.154 &    0.578 &   $+$ &    4.728 &   -0.014 &    0.176 &   $+$ &    4.664 &   -0.199 &    0.393 &   $+$ &    4.786 &   -0.174 &    0.311 \\
  $+$ &    4.125 &    0.375 &    0.578 &   $+$ &    4.230 &    0.400 &    0.176 &   $+$ &    5.728 &    0.374 &    0.393 &   $+$ &    5.023 &    0.444 &    0.311 \\
  $+$ &    4.678 &    0.869 &    0.338 &   $+$ &    3.717 &    0.442 &    0.786 &   $-$ &    5.730 &    0.381 &    0.382 &   $-$ &    5.009 &    0.456 &    0.311 \\
  $-$ &    4.683 &    0.882 &    0.318 &   $+$ &    3.446 &    0.837 &    0.622 &   $+$ &    3.967 &    0.733 &    0.100 &   $+$ &    3.690 &    0.608 &    0.422 \\
  $+$ &    3.244 &    1.031 &    0.617 &   $+$ &    4.031 &    1.365 &    0.284 &   $+$ &    4.258 &    0.767 &    0.448 &   $+$ &    3.114 &    1.048 &    0.369 \\
  $+$ &    3.420 &    1.322 &    0.111 &   $-$ &    4.173 &    1.424 &    0.230 &   $-$ &    4.039 &    0.843 &    0.353 &   $-$ &    3.404 &    1.068 &    0.187 \\
  $+$ &    3.088 &    1.553 &    0.146 &   $+$ &    3.706 &    1.537 &    0.170 &   $+$ &    4.175 &    1.109 &    0.345 &   $+$ &    3.211 &    1.388 &    0.100 \\
  $+$ &    1.969 &    1.866 &    0.176 &   $+$ &    2.864 &    1.626 &    0.327 &   $-$ &    4.173 &    1.198 &    0.288 &   $+$ &    2.264 &    1.713 &    0.155 \\
      &          &          &          &       &          &          &          &   $+$ &    3.428 &    1.365 &    0.286 & \\
      &          &          &          &       &          &          &          &   $+$ &    2.067 &    1.374 &    0.770 & \\
      &          &          &          &       &          &          &          &   $+$ &    3.426 &    1.596 &    0.118 & \\
      &          &          &          &       &          &          &          &   $+$ &    2.770 &    1.739 &    0.232 & \\
\\
      \multicolumn{4}{c}{ESO 443-G042} &           \multicolumn{4}{c}{NGC 5746} &            \multicolumn{4}{c}{IC 4767} &           \multicolumn{4}{c}{NGC 6722} \\
\\
  $+$ &    4.605 &   -0.217 &    0.652 &   $+$ &    4.639 &    0.025 &    0.526 &   $+$ &    4.230 &   -0.021 &    0.559 &   $+$ &    4.518 &   -0.095 &    0.673 \\
  $+$ &    3.522 &    0.352 &    0.652 &   $+$ &    4.335 &    0.494 &    0.526 &   $+$ &    3.721 &    0.379 &    0.559 &   $+$ &    3.995 &    0.430 &    0.673 \\
  $+$ &    3.858 &    1.116 &    0.193 &   $+$ &    3.720 &    0.874 &    0.640 &   $-$ &    3.102 &    0.772 &    0.305 &   $+$ &    4.347 &    0.966 &    0.386 \\
  $-$ &    4.234 &    1.227 &    0.146 &   $-$ &    2.798 &    1.332 &    0.100 &   $+$ &    5.194 &    0.949 &    0.296 &   $-$ &    4.283 &    0.998 &    0.353 \\
  $+$ &    3.136 &    1.274 &    0.221 &   $+$ &    3.416 &    1.350 &    0.580 &   $-$ &    5.190 &    0.952 &    0.294 &   $+$ &    2.968 &    1.064 &    0.583 \\
  $+$ &    4.001 &    1.327 &    0.112 &   $+$ &    3.248 &    1.942 &    0.109 &   $+$ &    3.133 &    1.216 &    0.233 &   $+$ &    3.140 &    1.522 &    0.111 \\
  $+$ &    2.156 &    1.674 &    0.280 &       &          &          &          &   $+$ &    2.157 &    1.495 &    0.407 & \\
\\
          \multicolumn{4}{c}{NGC 6771} &       \multicolumn{4}{c}{ESO 185-G053} &            \multicolumn{4}{c}{IC 4937} &       \multicolumn{4}{c}{ESO 597-G036} \\
\\
  $+$ &    4.546 &   -0.139 &    0.437 &   $+$ &    4.624 &   -0.214 &    0.691 &   $+$ &    4.548 &   -0.316 &    0.558 &   $+$ &    4.412 &   -0.242 &    0.600 \\
  $+$ &    4.246 &    0.421 &    0.437 &   $+$ &    3.837 &    0.253 &    0.691 &   $+$ &    3.805 &    0.266 &    0.558 &   $+$ &    3.775 &    0.284 &    0.600 \\
  $+$ &    4.725 &    0.822 &    0.386 &   $+$ &    3.509 &    0.672 &    0.575 &   $+$ &    4.431 &    0.786 &    0.299 &   $-$ &    3.585 &    0.300 &    0.291 \\
  $-$ &    4.684 &    0.838 &    0.375 &   $-$ &    4.769 &    1.074 &    0.356 &   $-$ &    4.421 &    0.796 &    0.277 &   $+$ &    5.661 &    0.884 &    0.268 \\
  $+$ &    3.319 &    1.251 &    0.306 &   $+$ &    4.773 &    1.079 &    0.353 &   $-$ &    4.395 &    1.046 &    0.280 &   $-$ &    5.780 &    0.893 &    0.265 \\
  $-$ &    3.094 &    1.624 &    0.281 &   $+$ &    1.873 &    1.208 &    0.737 &   $+$ &    4.379 &    1.049 &    0.294 &   $+$ &    5.162 &    0.920 &    0.257 \\
  $+$ &    3.101 &    1.642 &    0.278 &       &          &          &          &   $+$ &    3.195 &    1.098 &    0.100 &   $+$ &    3.273 &    1.166 &    0.109 \\
      &          &          &          &       &          &          &          &   $+$ &    3.282 &    1.322 &    0.124 &   $+$ &    2.185 &    1.512 &    0.246 \\
      &          &          &          &       &          &          &          &   $-$ &    3.053 &    1.383 &    0.193 & \\
      &          &          &          &       &          &          &          &   $+$ &    2.617 &    1.621 &    0.182 & \\
\end{tabular*}
\end{table*}

\begin{table*}
\addtocounter{table}{-1}
\caption{ --- continued}
\begin{tabular*}{\textwidth}{r@{\hspace{0.5em}}ccc@{\hspace{2em}}r@{\hspace{0.5em}}ccc@{\hspace{2em}}r@{\hspace{0.5em}}ccc@{\hspace{2em}}r@{\hspace{0.5em}}ccc}
\hline
Sign & $\log L_i$ & $\log\sigma_i$ & $q_i$ & Sign & $\log L_i$ & $\log\sigma_i$ & $q_i$ & Sign & $\log L_i$ & $\log\sigma_i$ & $q_i$ & Sign & $\log L_i$ & $\log\sigma_i$ & $q_i$ \\
& $(L_{K_S,\odot})$ & (arcsec) & & & $(L_{K_S,\odot})$ & (arcsec) & & & $(L_{K_S,\odot})$ & (arcsec) & & & $(L_{K_S,\odot})$ & (arcsec) & \\[1pt]
\multicolumn{16}{c}{B/PS sample (continued)} \\
\\
           \multicolumn{4}{c}{IC 5096} &       \multicolumn{4}{c}{ESO 240-G011} & \\
\\
  $+$ &    4.829 &   -0.270 &    0.349 &   $-$ &    6.391 &   -0.219 &    0.704 \\
  $+$ &    4.253 &    0.155 &    0.349 &   $+$ &    6.394 &   -0.216 &    0.704 \\
  $+$ &    3.868 &    0.394 &    0.682 &   $+$ &    3.729 &    0.548 &    0.704 \\
  $+$ &    4.895 &    0.821 &    0.439 &   $+$ &    2.996 &    0.978 &    0.608 \\
  $-$ &    4.887 &    0.827 &    0.427 &   $-$ &    6.088 &    1.134 &    0.111 \\
  $+$ &    3.542 &    0.860 &    0.772 &   $+$ &    6.088 &    1.134 &    0.111 \\
  $+$ &    3.543 &    1.254 &    0.111 &   $-$ &    2.533 &    1.396 &    0.226 \\
  $-$ &    3.927 &    1.318 &    0.134 &   $+$ &    3.325 &    1.572 &    0.100 \\
  $+$ &    3.892 &    1.395 &    0.117 &   $+$ &    2.398 &    1.881 &    0.100 \\
  $+$ &    2.653 &    1.667 &    0.142 &   $+$ &    1.565 &    1.881 &    0.254 \\
\\
\multicolumn{16}{c}{Control sample} \\
\\
          \multicolumn{4}{c}{NGC 1032} &           \multicolumn{4}{c}{NGC 3957} &           \multicolumn{4}{c}{NGC 4703} &           \multicolumn{4}{c}{NGC 5084} \\
\\
  $+$ &    4.776 &   -0.131 &    0.758 &   $+$ &    4.783 &   -0.334 &    0.336 &   $+$ &    4.413 &   -0.004 &    0.914 &   $+$ &    4.822 &    0.345 &    0.689 \\
  $+$ &    4.110 &    0.448 &    0.758 &   $+$ &    4.566 &   -0.238 &    0.336 &   $+$ &    3.627 &    0.416 &    0.914 &   $-$ &    4.377 &    0.348 &    0.100 \\
  $+$ &    3.584 &    0.854 &    0.670 &   $+$ &    4.210 &    0.305 &    0.336 &   $+$ &    3.810 &    0.584 &    0.360 &   $+$ &    4.073 &    0.840 &    0.599 \\
  $+$ &    2.423 &    1.157 &    0.373 &   $+$ &    3.594 &    0.477 &    0.713 &   $+$ &    3.276 &    0.986 &    0.327 &   $+$ &    3.161 &    1.269 &    0.735 \\
  $+$ &    3.057 &    1.334 &    0.529 &   $+$ &    4.630 &    0.685 &    0.105 &   $-$ &    3.711 &    0.992 &    0.193 &   $+$ &    3.533 &    1.359 &    0.109 \\
  $+$ &    3.524 &    1.418 &    0.367 &   $-$ &    4.595 &    0.702 &    0.112 &   $+$ &    3.103 &    1.081 &    0.470 &   $+$ &    2.980 &    1.728 &    0.131 \\
  $-$ &    4.474 &    1.499 &    0.330 &   $+$ &    3.316 &    0.990 &    0.541 &   $+$ &    3.519 &    1.159 &    0.118 &   $+$ &    2.308 &    1.869 &    0.360 \\
  $+$ &    4.427 &    1.511 &    0.321 &   $+$ &    3.400 &    1.388 &    0.131 &   $-$ &    4.536 &    1.481 &    0.123 & \\
      &          &          &          &   $-$ &    3.872 &    1.640 &    0.184 &   $+$ &    4.539 &    1.485 &    0.121 & \\
      &          &          &          &   $+$ &    3.883 &    1.649 &    0.181 &   $+$ &    2.356 &    1.719 &    0.173 & \\
\\
          \multicolumn{4}{c}{NGC 7123} &            \multicolumn{4}{c}{IC 5176} & \\
\\
  $+$ &    4.812 &   -0.289 &    0.794 &   $+$ &    3.629 &   -0.206 &    0.746 \\
  $+$ &    4.177 &    0.268 &    0.794 &   $+$ &    3.480 &    0.181 &    0.746 \\
  $+$ &    3.884 &    0.655 &    0.634 &   $+$ &    2.960 &    0.466 &    0.746 \\
  $+$ &    3.379 &    0.905 &    0.640 &   $+$ &    2.979 &    0.761 &    0.640 \\
  $-$ &    4.335 &    1.071 &    0.114 &   $+$ &    3.176 &    1.126 &    0.111 \\
  $+$ &    4.376 &    1.107 &    0.100 &   $+$ &    3.346 &    1.314 &    0.103 \\
  $+$ &    2.215 &    1.217 &    1.000 &   $+$ &    3.321 &    1.333 &    0.268 \\
  $+$ &    2.671 &    1.224 &    0.504 &   $-$ &    3.803 &    1.425 &    0.181 \\
  $+$ &    3.069 &    1.439 &    0.100 &   $+$ &    3.765 &    1.453 &    0.156 \\
  $+$ &    2.255 &    1.764 &    0.144 &   $+$ &    2.242 &    1.771 &    0.165 \\
\hline
\end{tabular*}
\begin{minipage}{\textwidth}
\emph{Notes.} Column (i) Sign of term in the Gaussian sum. (ii) Logarithm of
the Gaussian amplitude. (iii) Logarithm of the Gaussian width. (iv) Axial ratio
of the Gaussian. See \refeq{eqn:sb}.
\end{minipage}
\end{table*}

\section{Photometric recalibration with the 2MASS Extended Source
Catalog}
\label{app:2mass}

%
%
\begin{table}
\caption{$K_S$-band calibration corrections applied to the surface
brightness of the images presented by {\protect \cite{Bureau:2006}}.}
\label{tab:2mass}
\begin{center}
\begin{tabular}{rcrc}
\hline
Galaxy & $\Delta$ & Galaxy & $\Delta$ \\
 & (mag arcsec$^{-2}$) & & (mag arcsec$^{-2}$) \\
\hline
\multicolumn{2}{c}{B/PS bulges} & \multicolumn{2}{c}{Control sample}\\  
\\
      NGC 128 & -0.86 &    NGC 1032 & -0.87 \\
 ESO 151-G004 & -0.80 &    NGC 3957 & -0.73 \\
     NGC 1381 & -0.75 &    NGC 4703 & -0.78 \\
     NGC 1596 & -0.84 &    NGC 5084 & -0.97 \\
     NGC 1886 & -0.80 &    NGC 7123 & -0.92 \\
     NGC 2310 & -0.96 &     IC 5176 & -0.82 \\
 ESO 311-G012 & -0.83 & \ldots & \ldots \\
     NGC 3203 & -0.78 & \ldots & \ldots \\
     NGC 3390 & -0.83 & \ldots & \ldots \\
     NGC 4469 & -0.82 & \ldots & \ldots \\
     NGC 4710 & -0.85 & \ldots & \ldots \\
    PGC 44931 & -1.08 & \ldots & \ldots \\
 ESO 443-G042 & -0.75 & \ldots & \ldots \\
     NGC 5746 & -0.85 & \ldots & \ldots \\
      IC 4767 & -0.76 & \ldots & \ldots \\
     NGC 6722 & -1.29 & \ldots & \ldots \\
     NGC 6771 & -0.82 & \ldots & \ldots \\
 ESO 185-G053 & -0.70 & \ldots & \ldots \\
      IC 4937 & -1.21 & \ldots & \ldots \\
 ESO 597-G036 & -0.64 & \ldots & \ldots \\
      IC 5096 & -0.88 & \ldots & \ldots \\
 ESO 240-G011 & -0.82 & \ldots & \ldots \\
\hline
\end{tabular}
\end{center}
\label{lastpage}
\end{table}

During the course of this work, we discovered that the photometric
calibration zero points of \cite{Bureau:2006} were incorrect, so we
recalibrated them using $K_S$-band images of the same objects taken from
the Two Micron All Sky Survey (2MASS) Extended Source Catalog
\citep{Skrutskie:2006}.

We did this by measuring the light in matching elliptical apertures for
each pair of images, and shifting the CASPIR image by a constant zero
point offset such that its radial profile coincided with that of the
2MASS image. This offset gives a new, corrected $K_S$-band surface
brightness zero point for each CASPIR image. 

Ensuring truly corresponding elliptical apertures enclosing
identical locations on the sky is not trivial. The problem is that
measurements of ellipticity and position angle based on two images of
different depths will not necessarily yield identical results. Here the
shallower 2MASS image does not reveal much of the faint discs, so
important in determining the position angle and ellipticity.
Corresponding apertures can trivially be ensured by using circular
apertures. However, because circular apertures of increasing radii more
quickly include noisy contributions from outside the galaxy (especially
in edge on systems), they do not use as much of either galaxy image as
possible, decreasing the reliability of the recalibration. 

We therefore chose to impose the ellipticity measured for each CASPIR
image on the corresponding 2MASS image and determined the position angle
independently for both images by using an initial image truncated at the
same approximate surface brightness. In truncating both images, we
temporarily remove information from the deeper image so that it is
missing as much of the disc as the shallower image. 

Before use, we correct the 2MASS images for the effects foreground
Galactic extinction using the dust maps of \cite{Schlegel:1998}. This
ensures that our recalibration incorporates this correction. The
calibration also involves an incidental and very slight colour
transformation from $K_S$ to $K$n. Much like the $K_S$ filter used for
2MASS, the purpose of the $K$n-band filter is to reject the thermal
background admitted by a standard $K$-band filter at its long wavelength
end \citep{Skrutskie:2006}. We adopt a value for the absolute magnitude
of the Sun at $K_S$-band of $M_{K_S,\odot} = 3.29$ \citep{Blanton:2007}.
The result of this recalibration is presented in Table~\ref{tab:2mass}.
The quantity $\Delta$ is the recalibration constant which should be
added to the incorrectly calibrated $K$n-band images presented by
\cite{Bureau:2006}, which were too faint, typically by around 0.8 mag
arcsec$^{-2}$.

\end{document}